\documentclass[
	a4paper, 
	10pt, 
	unnumberedsections, 
	twoside, 
]{LTJournalArticle}

\usepackage{xspace}
\usepackage{subfigure}
\usepackage{multirow}
\usepackage{hyperref}
\usepackage{bm}
\usepackage{float}
\usepackage[figuresright]{rotating}

\newcommand{\fos}[1]{\textit{#1}\xspace}
\newcommand{\ie}{\emph{i.e.,}\xspace}
\newcommand{\eg}{\emph{e.g.,}\xspace}

\runninghead{}

\footertext{}

\title{Research Explosion: \\More Effort to Climb onto Shoulders of the Giant} 

\author{%
	Guoxiu He\textsuperscript{1*}, 
        Aixin Sun\textsuperscript{2}\thanks{Corresponding authors. Guoxiu He: gxhe@fem.ecnu.edu.cn, Aixin Sun: axsun@ntu.edu.sg}, and
        Wei Lu\textsuperscript{3}
}

\date{
\footnotesize\textsuperscript{\textbf{1}}Faculty of Economics and Management, East China Normal University, Shanghai, 200062, China\\ 
\textsuperscript{\textbf{2}}School of Computer Science and Engineering, Nanyang Technological University, Singapore, 639798, Singapore\\
\textsuperscript{\textbf{3}}School of Information Management, Wuhan University, Wuhan, 430072, China
}

\renewcommand{\maketitlehookd}{%
	\begin{abstract}
		\noindent Fast-growing scientific publications present challenges to the scientific community.  In this paper, we describe their implications to researchers. As references form explicit foundations for researchers to conduct a study, we investigate the evolution in reference patterns based on 60.8 million papers published from 1960 to 2015. The results demonstrate that recent papers contain more references than older ones, especially the well-cited papers compared with other papers. Well-cited papers receive 10 or more citations within 5 years of publication. Their references cover a longer period from  classic research to very recent studies. Authors of well-cited papers are also farsighted to discover the reference papers with good potential to receive high citation numbers in near future. We also discover that the number of accumulative publications has a negative impact on next-5-year citation count for most fields except Chemistry, Materials science, Environmental science, Biology, and Engineering. Our findings suggest that researchers are expected to devote more effort to producing impactful research. Based on all these findings, we strongly advise against judging researchers simply based on the number of their publications. On the other hand, authors and reviewers should ensure that published papers contain adequate contributions. The code for our analysis is on GitHub: \url{https://github.com/ECNU-Text-Computing/Research-Explosion}.
	\end{abstract}
}

\begin{document}

\maketitle 


\section{Introduction}

As of December 31, 2020, number of scientific publications has exceeded 200 million, based on data maintained by Microsoft Academic Graph (MAG). If we only consider the publications with at least one reference, the number is over 81 million, which is 1.31 times of the number in 2015, 3.75 times of 2000, and 63.32 times of 1960. For researchers, the giant of science they are facing now has grown taller and fatter. Accordingly, researchers now have to face a much larger search space to identify cutting-edge topics or to locate for related work to support their study. Chu and Evans~\cite{chu2021slowed} claim that the canonical progress of papers becomes slow when the field of science is growing large. In this study, through analysis of reference patterns, we show that much more effort is needed for scientists to climb onto shoulders of the giant, to produce high-quality publications. 

\begin{figure}[th]
\centering
\includegraphics[width=0.8\columnwidth]{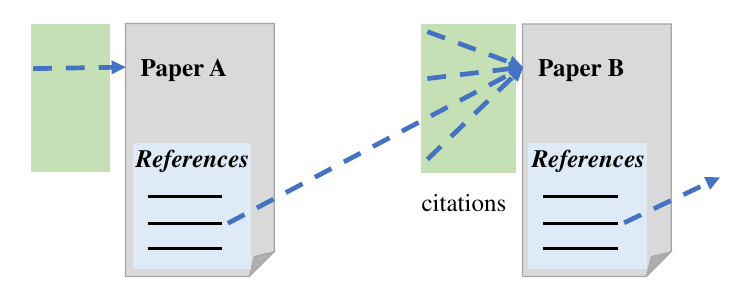}
\caption{A citing paper (Paper A) lists a cited paper (Paper B) in its references.}
\label{fig:referenceCitation}
\end{figure}

Reference pattern is a collection of measures characterising the references in publications. Fig.~\ref{fig:referenceCitation} illustrates the relationship between references and citations. The selection of references is a proactive action of authors~\cite{zhang2017citationa,zhang2017citationb}. References in a paper may reflect authors' knowledge, the trend of research topic, and even the degree of difficulty in accessing past publications. On the contrary, the acquisition of citations is a passive result produced by academic peers~\cite{bollen2009principal}. Though arguable, citation count remains a widely employed measure of scientific impact~\cite{wang2013quantifying,tahamtan2019citation}.

\begin{table*}[th]
  \centering
  \caption{Number of publications in each field of study, in descending order}
  \label{tab:dataOverall}%
  
    \begin{tabular}{l|r|rrr}
      \toprule
      \multirow{2}{2cm}{Field of study} & \multicolumn{4}{c}{Publications used in this study [1960 - 2015]} \\ 
      \cmidrule{2-5}
      & publications & non-cited (citation$=$0) & cited (1$\leq$citation$<$10) & well-cited (citation$\geq$10) \\
      \midrule
    Medicine & 13,140,385 & 3,046,062 (23.18\%) & 5,903,230 (44.92\%) & 4,191,093 (31.89\%) \\
    Engineering & 12,416,110 & 4,126,581 (33.24\%) & 6,603,158 (53.18\%) & 1,686,371 (13.58\%) \\
    Biology & 11,716,161 & 2,030,940 (17.33\%) & 5,205,567 (44.43\%) & 4,479,654 (38.23\%) \\
    Chemistry & 10,836,398 & 2,715,581 (25.06\%) & 5,278,076 (48.71\%) & 2,842,741 (26.23\%) \\
    Materi. sci. & 9,956,308 & 3,253,255 (32.68\%) & 5,152,934 (51.76\%) & 1,550,119 (15.57\%) \\
    Comput. sci. & 9,576,227 & 2,917,097 (30.46\%) & 4,931,373 (51.50\%) & 1,727,757 (18.04\%) \\
    Physics & 8,820,910 & 2,665,708 (30.22\%) & 4,531,382 (51.37\%) & 1,623,820 (18.41\%) \\
    Mathematics & 6,107,484 & 1,911,132 (31.29\%) & 3,178,463 (52.04\%) & 1,017,889 (16.67\%) \\
    Psychology & 3,632,814 & 976,351 (26.88\%) & 1,587,558 (43.70\%) & 1,068,905 (29.42\%) \\
    Economics & 2,739,908 & 953,040 (34.78\%) & 1,273,492 (46.48\%) & 513,376 (18.74\%) \\
    Geology & 2,692,794 & 801,580 (29.77\%) & 1,327,358 (49.29\%) & 563,856 (20.94\%) \\
    Sociology & 2,000,325 & 724,519 (36.22\%) & 919,543 (45.97\%) & 356,263 (17.81\%) \\
    Environ. sci. & 1,652,148 & 528,715 (32.00\%) & 780,286 (47.23\%) & 343,147 (20.77\%) \\
    Business & 1,564,470 & 570,408 (36.46\%) & 707,345 (45.21\%) & 286,717 (18.33\%) \\
    Geography & 1,508,306 & 562,629 (37.30\%) & 682,744 (45.27\%) & 262,933 (17.43\%) \\
    Polit. sci. & 1,291,658 & 570,136 (44.14\%) & 571,898 (44.28\%) & 149,624 (11.58\%) \\
    Population & 1,100,468 & 187,644 (17.05\%) & 484,899 (44.06\%) & 427,925 (38.89\%) \\
    Philosophy & 1,061,120 & 533,626 (50.29\%) & 433,732 (40.87\%) & 93,762 (8.84\%) \\
    Art & 801,264 & 494,244 (61.68\%) & 279,103 (34.83\%) & 27,917 (3.48\%) \\
    History & 575,584 & 330,788 (57.47\%) & 216,236 (37.57\%) & 28,560 (4.96\%) \\
    \midrule
    All publications & 60,803,629 & 17,876,729 (29.40\%) & 29,457,881 (48.45\%) & 13,469,019 (22.15\%) \\
    \bottomrule
    \end{tabular}%
\end{table*}%

The fast development of online databases has made it much easier for scholars to access both historical and the most up-to-date publications. However, the cost of selecting appropriate references to support their research may not be diminished. How scholars select references has been studied for several decades. This is an important research question in both bibliometrics and science of science. Previous work has approached this question from many perspectives including number of references~\cite{shearer1979citation,costas2012referencing,ucar2014growth}, time span or age of references~\cite{line1974synchronous,stinson1987synchronous,milojevic2012academic}, citation patterns of references \cite{hargens2000using,bethard2010should}, \textit{etc}. Along this line, through investigating these reference patterns, we attempt to probe how much effort scientists have to make with the rapid growth of scholarly publications from 1960 to 2015. Specifically, we study publications in three groups, non-cited, cited, and well-cited, based on the number of citations a paper receives within 5 years after publication. A paper is considered \textit{well-cited} if it receives at least 10 citations, and \textit{cited} if it receives 1 to 9 citations. 

By comparing reference patterns of the three groups, we observe that: 1) From 1960 to 2015, scholars have to reference more papers to support their findings. 2) References in a paper show an increasingly longer time span, covering both classic and recent findings. 3) In addition to including highly cited papers, authors have to identify promising research from not-well cited papers. In other words, authors must demonstrate a good sense or taste of selecting high-quality recent publications when no citations are available. These observations are more intense for well-cited papers. 

We then perform an OLS regression model with paper's citation count in 5 years as the dependent variable. Independent variables are current number of publications and reference patterns. In 15 out of 20 fields of study, the increasing number of publications prevents new findings from gaining more citations. The science explosion pushes researchers to keep more comprehensive references for high-quality outcomes. 

\section{Dataset}
\label{sec:data}

This study is conducted on Microsoft Academic Graph (MAG), released on April 26, 2021. MAG~\cite{sinha2015overview, shen2018web} is a Web entity graph about scholarly publications. Publications are connected with citing / cited publications, and have attributes \eg published year, fields of study. To the best of our knowledge, MAG dataset is the most comprehensive and the largest academic collection, that is publicly accessible. 

MAG maintains more than 200 million publications, and they are tagged with 700K fields of study. These fields are organized in a six-level hierarchy. There are 19 top-level fields (\eg \fos{Medicine}, \fos{Engineering}), and 292 second-level fields (\eg \fos{Traditional Medicine}, \fos{Nuclear Engineering}). The hierarchy is not a tree structure. A child field could have more than one parent field, as the result of interdisciplinary research. In this study, we are interested in top-level fields. We consider that a paper $p$ belongs to a top-level field $F$ as long as $p$ is tagged with $F$ or any of $F$'s second-level child fields. We note a special field named \fos{Population}, which does not belong to any top-level field. As its number of publications is comparable to some top-level fields, we consider \fos{Population} also a top-level field in this study. Table~\ref{tab:dataOverall} lists the number of publications in the 20 top-level fields. As a publication may be assigned to more than one field, and a low-level field may have multiple parents, there are overlaps between different fields. The last row in Table~\ref{tab:dataOverall} shows the number of unique publications. 

To study reference pattern and its relationship with citations, in this work, we only select publications with references. \footnote{In MAG, publications are not limited to research papers. Because we only consider the publications that contain references, the selected ones are mostly papers. We hereby use ``publication'' and ``paper'' interchangeably in our discussion.}
As citation grows along time, for a fair comparison, we consider the citation count a paper receives in 5 years. Given the dataset was released in April 2021, to ensure every paper included in this study has 5 years to grow citations, we extract all papers that are published between 1960 and 2015. The total number of publications included in this study is about 60.8 million. Table~\ref{tab:dataOverall} lists the papers included in this study, in total and in each of the 20 fields. Fig.~\ref{sfig:fieldsCluster} further illustrates the overlapping between the 20 top-level fields, measured by Jaccard similarity. In this plot, size of a node (field of study) is proportional to its number of publications. Width of an edge is proportional to Jaccard similarity $J = \frac{|F_a\cap F_b|}{|F_a \cup F_b|}$ between two fields $F_a$ and $F_b$. In Fig.~\ref{sfig:fieldsCluster}, the 20 fields are grouped into 5 clusters; fields in the same cluster share more publications than fields in different clusters. As expected, \fos{Medicine} and \fos{Biology} share more papers than \fos{Medicine} and \fos{Geology}. The clustering of fields serves as a complementary aspect to understanding the data composition in this study.

These 60.8 million papers are divided into three citation groups: without any citations within 5 years (non-cited), having 1 to 9 citations (cited), and having 10 or more citations (well-cited). The threshold (10 citations) between cited and well-cited refers to i10 index in Google Scholar, and the threshold (5 years) refers to h5 index in Google Scholar Metrics.\footnote{\url{https://scholar.google.com/}} The distribution of papers in three groups are reported in Table~\ref{tab:dataOverall}, with percentages specific to each field.

\section{Methods}
\label{sec:methods}

\begin{figure*}[th] 
\centering
\subfigure[Five clusters of fields of study in different colors \label{sfig:fieldsCluster}] {\includegraphics[width=0.3665\textwidth]{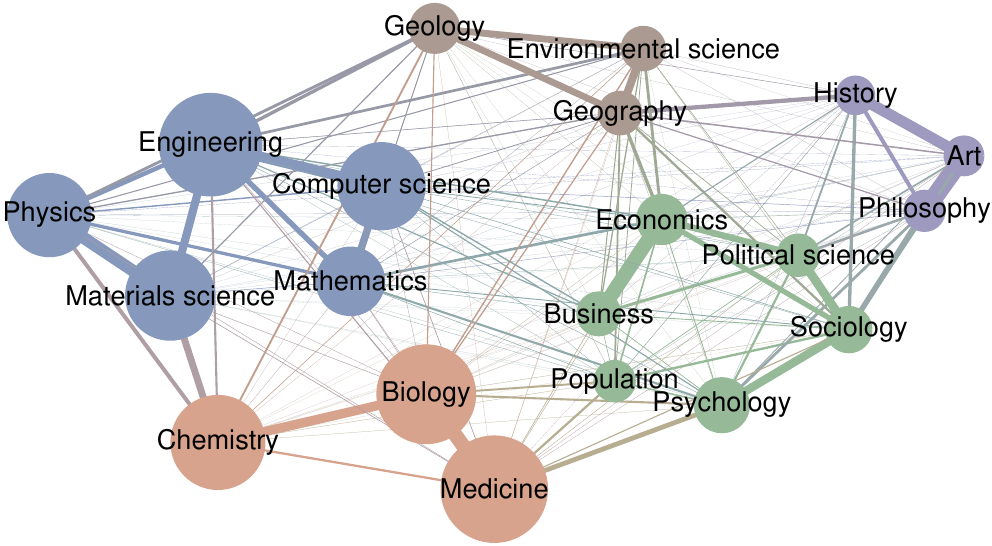}}\quad
\subfigure[Number of accumulative publications (million) \label{sfig:historicalYearCount}] {\includegraphics[width=0.3822\textwidth]{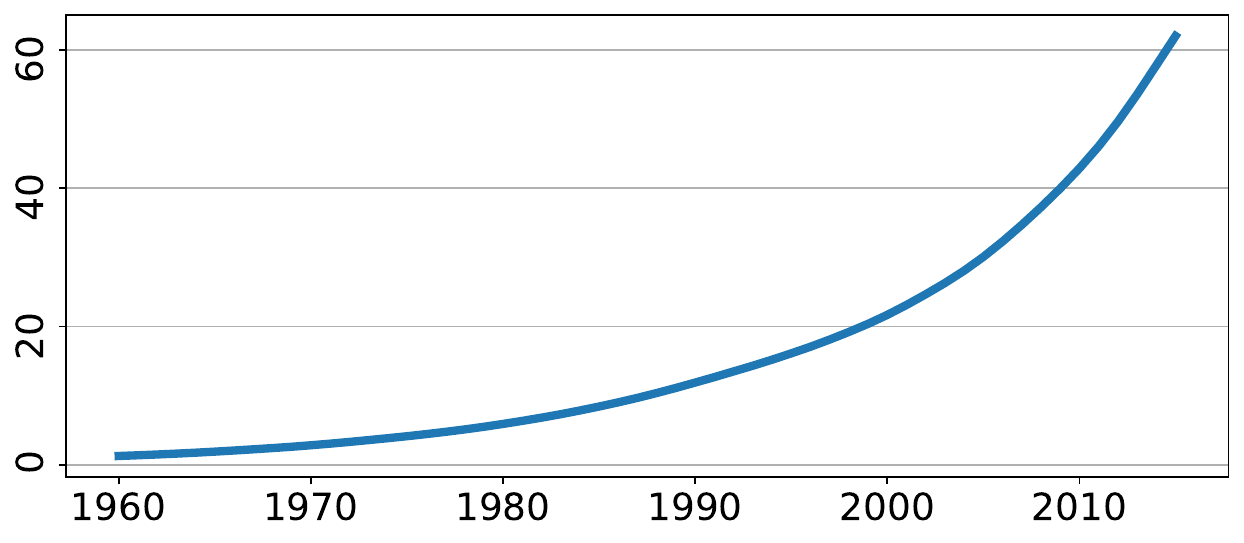}}
\\
\subfigure[Number of papers published in each year (million)\label{sfig:growYearCount}] {\includegraphics[width=0.3822\textwidth]{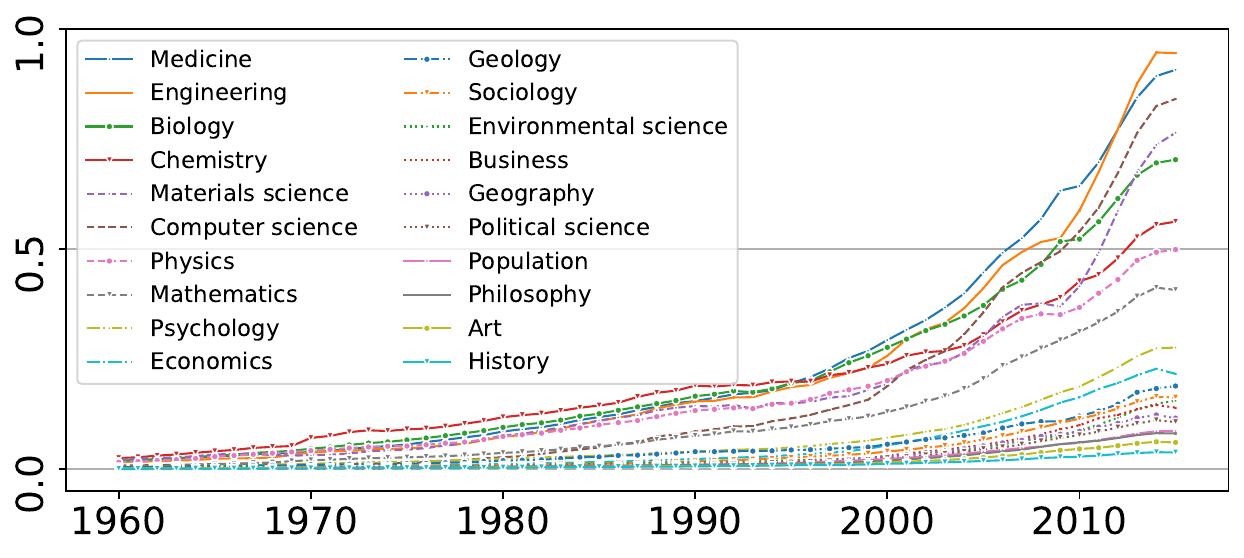}}
\quad
\subfigure[Papers published in each year by citation group \label{sfig:growYearCountSplit}] {\includegraphics[width=0.36\textwidth]{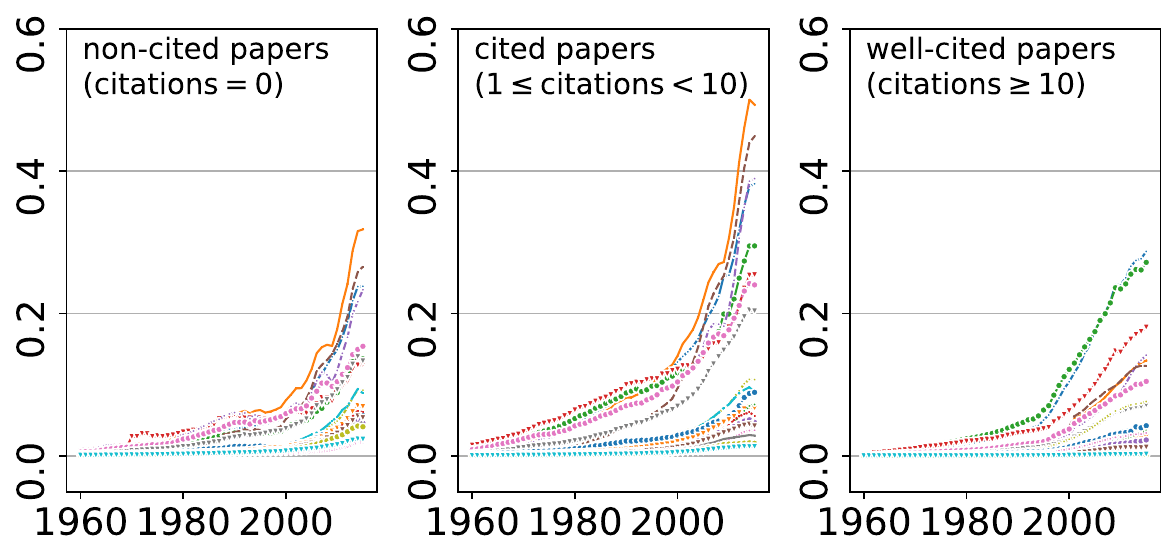}}
\caption{Statistics of publications used in this study. (a) Relationships between the 20 fields of study. Size of node is proportional to the number of publications in the field, and width of edge is proportional to the overlap between two fields. The clusters are generated by Gephi (\url{https://gephi.org/}). (b) Number of accumulative publications (in millions) from 1960 to 2015. (c) Number of publications (in millions) published in each year from 1960 to 2015, in each field. (d) Number of publications (in millions) published in each year, in each filed, in three citation groups. The citation count is collected in 5-year time after publication.}
\label{fig:yearCount}
\end{figure*}

With the extracted dataset, we first show the growth of scientific publications in accumulation, annual growth in the 20 fields, and by citation groups. Figs.~\ref{sfig:historicalYearCount}, \ref{sfig:growYearCount}, and \ref{sfig:growYearCountSplit} plot the growth, where x-axis indicates the year and y-axis is the number of papers in million. 

We explore reference pattern from 3 perspectives: number of references, age of reference, and citation of reference~\cite{schubert1993cognitive,allen1994persuasive,egghe1997price,milojevic2012academic}. For reference number, we calculate the median number of references per year from 1960 to 2015, plotted in Fig.~\ref{sfig:refNum}. Reference number partially reflects the amount of literature a scholar needs to read to support a new publication. 

Age of a reference in a paper is the difference between the paper's and the reference's publication years. We consider the following indicators: (i) Maximum age of all references in a paper, which reflects how old the literature scholars trace, (ii) Variation ratio $c_v=\frac{\sigma}{\mu}$ of ages excluding top 10\% oldest references, \ie age distribution of the remaining 90\% references, (iii) Ratio of 1-year-old references, which reflects how scholars chase the latest achievements, and (iv) Absolute number of 1-year-old references. We plot the median value of each indicator for papers published in each year, in Fig.~\ref{sfig:refAge}. 

We only consider citations of a reference received before the citing paper's publication time. Hence, the same reference may have different citation counts in different citing papers. Similar to age of reference, we calculate the maximum and variation ratio of references' citation counts in a paper. In addition, we probe the ratio as well as the absolute number of non-cited references in a paper. A non-cited reference has not received any citation, at the time of the citing paper's publication. Citing non-cited references partially reflects a scholar's taste, in identifying high-quality research. Then we further trace how many citations these non-cited references subsequently receive in 5 years after the citing paper's publication year. Fig.~\ref{sfig:refCitation} plots the median value of these 5 indicators in each year.

All plots in Fig.~\ref{fig:referencePattern} are by paper's citation group (\ie non-cited, cited, well-cited), visualizing the scholars' effort to conduct research at different levels. The reference pattern of each field of study is in Appendix, Figs. \ref{fig:referencePatternMedicine} to \ref{fig:referencePatternHistory}, respectively. 

We then estimate the effect of the size of literature body using the following specification:
$$c = \alpha + \beta a + \gamma R + \epsilon$$
where $c$ is paper's citation count in 5 years, $a$ is number of accumulated publications, and $R$ stands for aforementioned indicators of reference patterns. Among all variables, paper's citation count, number of accumulated publications, and maximum reference citation are $log_{10}$-scaled to make all variables in comparable data range. 

We use the 60.8 million publications to find out the evidence of the effect. For the same publication year, the number of accumulated publications is the same for all papers. Take 2015 as an example year of publication. Theoretically, all scholars can refer to the same collection of papers that are published before 2015. If two papers in different years share similar reference pattern, then the number of accumulated publications in the two different years becomes a factor. The coefficient $\beta$ of this variable indicates the global effect of publication numbers on potential citation count $c$. We also look for empirical evidence of the effect employing publications in each of the 20 fields of study. Here, number of accumulated publications remains to be all papers published in each year, and not the papers specific to each field. It is common for research in one field to cite papers from other fields. The OLS regression results are shown in Appendix, Tables \ref{tab:olsRegResChemistry} to \ref{tab:olsRegResBusiness}.

Finally, Fig.~\ref{sfig:logEffect} presents the effect on citation count ($log_{10}$) by the number of accumulated publications ($log_{10}$) with respect to the coefficient for each field. To better understand this plot, we visualize the relationship between $log_{10}(citations)$ and $citations$ (\ie without log scale) in Fig.~\ref{sfig:absoluteEffect}, as its x-axis and y-axis respectively.

\section{Results}
\label{sec:results}

\subsection{Research Explosion} 
\label{subsec:research-explosion}

Table~\ref{tab:dataOverall} lists the number of publications, by fields of study maintained in MAG. Included in this study are the 60 million publications which contain references and are published from 1960 to 2015. By publication number, the largest field is \fos{Medicine} with 13 million papers, while the smallest field \fos{History} has 0.5 million papers. 

From 1960 to 2015, Fig.~\ref{sfig:historicalYearCount} shows an exponential growth of publications, particularly after year 2000. Clearly, the giant of scientific research had grown up swiftly in recent years. If we look into the number of papers published in each year, by the 20 fields of study in Fig.~\ref{sfig:growYearCount}, almost all fields show significant growth from 1960 to 2015, particularly after the year 2000. Based on the plot of yearly publications in Fig.~\ref{sfig:growYearCount}, these 20 fields can be roughly divided into 2 groups, a fast-growth group and a slow-growth group. The divider between the two groups is the gap between \fos{Mathematics} and \fos{Psychology}. The fast-growth group includes the top 8 fields listed in Table~\ref{tab:dataOverall}, which also corresponds to the two clusters plotted on the left hand side of Fig.~\ref{sfig:fieldsCluster}. The slow-growth group includes the last 12 fields listed in Table~\ref{tab:dataOverall}, and the three clusters plotted on the right hand side of Fig.~\ref{sfig:fieldsCluster}.

\begin{figure*}[ht] 
\centering
\subfigure[Number \label{sfig:refNum}] {\includegraphics[width=0.1233\textwidth]{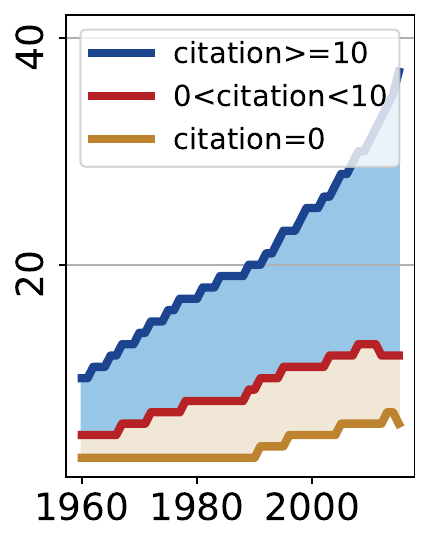}}
\quad
\subfigure[Age \label{sfig:refAge}]{\includegraphics[width=0.3605\textwidth]{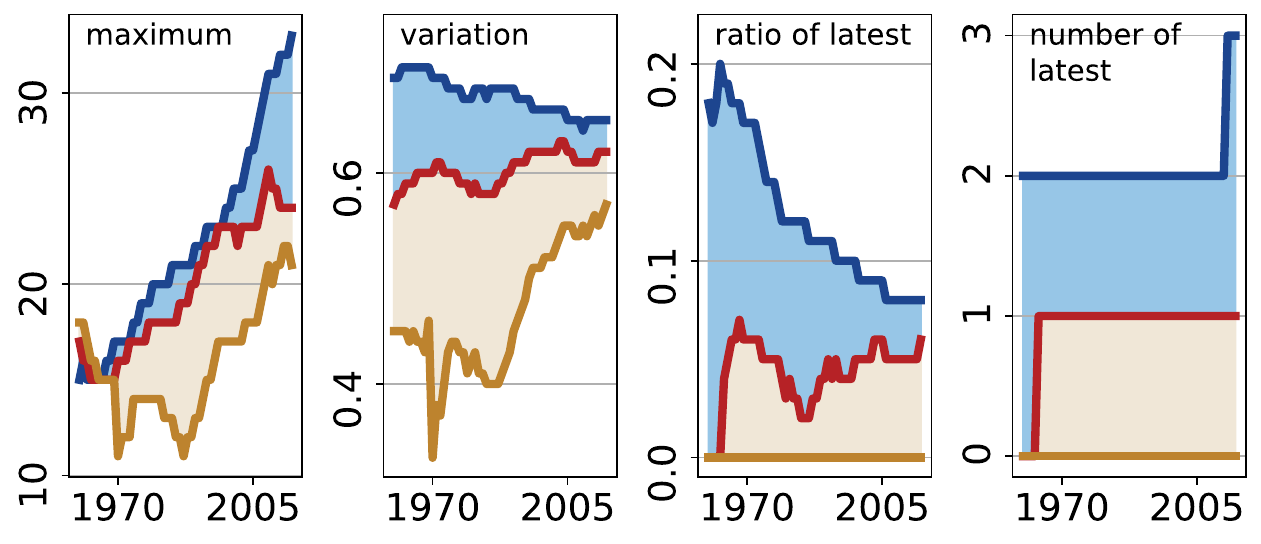}}
\quad
\subfigure[Citation count \label{sfig:refCitation}] {\includegraphics[width=0.45\textwidth]{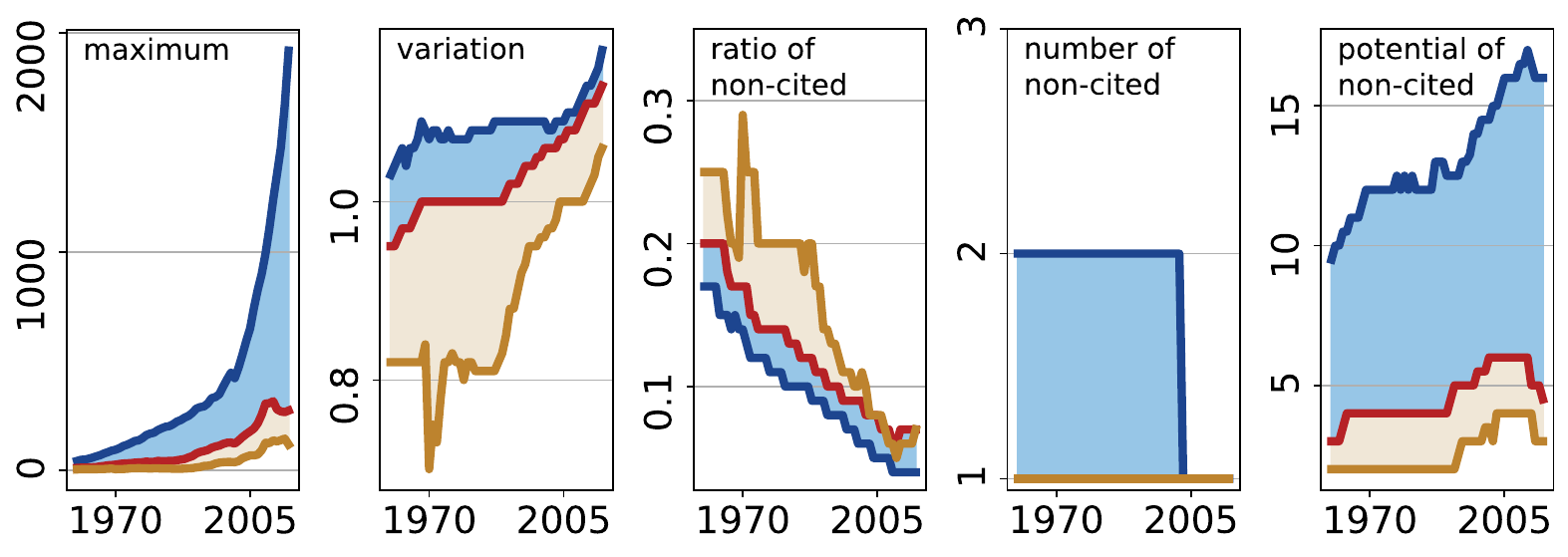}}
\caption{The indicators of reference pattern from 1960 to 2015. (a) indicates the median number of references per year, from 1960 to 2015. (b) presents the reference age patterns year by year from 1960 to 2015. Specifically, the 4 sub-figures in (b) plot the maximum reference age, variation ratio of age, ratio of latest papers in reference, and the absolute number of latest papers in references, respectively. (c) presents the citation count of references. Particularly, the 5 sub-figures in (c) show maximum of citation counts, variation ratio of citation counts, ratio of non-cited publications in references, absolute number of non-cited publications in references, and the potential (number of citations) of the non-cited references, respectively.}
\label{fig:referencePattern}
\end{figure*}

The rapid growth of publications can be a double-edged sword. Ostensibly, it is a strong indication that all fields in sciences are collecting more and more findings, and scholars can stand higher. On the other hand, it is also well known that not all publications bring in the same impact. Scholars hence have to be able to distinguish papers that are worth spending more time on. If we consider citation to be an indicator of research impact, Fig.~\ref{sfig:growYearCountSplit} reports the yearly publications in the 20 fields by citation group: non-cited, cited, and well-cited in 5-year time. The cited ($1\le citation <10$) group receives the most rapid growth, followed by non-cited ($citation = 0$), then well-cited ($citation \ge 10$). Collectively, the cited group accounts for majority (48.45\%) of publications, while the number of well-cited group is only 22.15\%. The non-cited group makes up 29.40\% of all publications. Specifically, the percentage of non-cited papers are 61.68\%, 57.47\%, 23.18\%, and 17.33\% respectively for \fos{Art}, \fos{History}, \fos{Medicine}, and \fos{Biology}. Interestingly, \fos{Art} and \fos{history} are among the smallest groups by number of publications, and \fos{Medicine} and \fos{Biology} are among the largest groups. Although there are relatively fewer papers to choose from, the chance of reading a non-cited paper in \fos{Art} and \fos{history} is much higher than that in \fos{Medicine} and \fos{Biology}. Observe that the absolute number of non-cited publications in \fos{Medicine} and \fos{Biology} are also much higher than that in most other fields of study. In short, different fields of study demonstrate different patterns in paper growth and in citation patterns.

The observations made above call for a deeper look into how authors reference (or spend time on) existing publications when conducting research and what effect of the research explosion on potential citations. Next, we study the reference pattern of papers and try to establish the connections between reference patterns and citations, as a way to reflect the effort researchers need to make with the fast-growing publications. In specific, we characterise papers in different citation groups, \ie well-cited, cited, and non-cited publications. 

\subsection{More Effort}
\label{subsec:more-effort}

In this section, we investigate the reference patterns of papers by citation groups, non-cited, cited, and well-cited. 

Consistent with many other studies, in general, there is an increasing trend in the number of references per paper from 1960 to 2015, as shown in Fig.~\ref{sfig:refNum}. In particular, reference number of well-cited publications has a higher growth rate than cited and non-cited publications. Consider the number of references as a simple measure. Authors of more impactful (or well-cited) papers put more effort to source for supporting documents from the continuously growing large body of research. Non-cited papers, on the other hand, are not well supported by many references. Nevertheless, a simple number of references does not reveal much about authors' effort in searching for these references. Next, we look at the age of the references. 

Fig.~\ref{sfig:refAge} show 4 subfigures, all related to reference age, computed as the difference between publication years of the citing paper and cited paper (\ie the reference). Again, we observe an increasing trend in maximum of reference age, in all citation groups, well-cited, cite, and non-cited, in the first subfigure on maximum age. All three groups demonstrate a similar growth rate. Yet, the averaged maximum reference age of well-cited paper is much larger than that of cite, and that of non-cited. That is, authors of impactful work search for older references to support their study. We may also interpret that these authors have a more thorough understanding of the development of a research area. It is interesting to observe that, after the year 2000, the growth rate of well-cited publications suddenly became much higher than the previous years. It is a signal that the authors of impactful research turn to seek more classical references, probably with the help of more convenient online scientific databases. 

Next, we explore the temporal variation of reference ages in a paper. As shown in the second subfigure of Fig.~\ref{sfig:refAge}, while the variation of reference age of well-cited papers becomes lower from 1960 to 2015, the ratio remains higher than cited and non-cited publications. Both cited and non-cited papers show an increasing trend in recent years, along with the increase in maximum age of references, as discussed earlier. The increasing maximum age and the large variation of reference age from well-cited papers suggest that scholars have to hunt for the right support documents spreading in different time periods, to support their findings. 

Next, we look at the use of the latest publications as references. The third subfigure, ratio of latest publications in the reference list, shows a similar pattern as the variation of reference age. While the ratio of well-cited publications decreases along time, it remains higher than cited and non-cited publications overall. Interestingly, a close to zero ratio is plotted for non-cited papers. The reduction in ratio does not mean that papers use fewer latest papers as references, because the overall number of references goes up with a stead slop, see Fig.~\ref{sfig:refNum}. Hence, in the last subfigure, we show the absolute number of latest publications. In general, there is an increasing trend for well-cited papers, from 2 to 3 latest references. The fact that scholars have to follow most up-to-date achievements had not changed much. Taking both Figs.~\ref{sfig:refNum} and~\ref{sfig:refCitation} into consideration, impactful research is supported by increasing number of references, with more classic as well as most up-to-date results, and other results that are well-spread over the years. Yet, these references are selected from the ever-growing body of literature. The immediate question is: are the selection of such references affected by reference papers' citations? Again, we assume citation is accepted by researchers as an indication of research impact. 

After studying number and age of references, we now investigate citations of references in a paper, plotted in Fig.~\ref{sfig:refCitation}. There are 5 subfigures in Fig.~\ref{sfig:refCitation}. The first subfigure shows the maximum citation count among references, in a paper, published from 1960 to 2015. We see an exponential growth in well-cited papers, particularly after the year 2000. There is also an increasing trend for cited and non-cited papers. The large difference between well-cited and these two is a strong indicator that authors of well-cited papers often cite other papers which are well cited. It is not comprehensive to barely refer to the very highly cited publications. The second subfigure shows the variation of reference citations. There is an increasing trend for all the three citation groups, with well-cited papers remaining to have a high variation of reference citations, \ie a good coverage of papers with very large and also very low (or even zero) citations. 

We then check if scholars would cite papers that do not have any citations at the time of paper writing. \footnote{We use the time of paper publication instead, as time of paper writing is not recorded. Nevertheless, if a reference has no citations when a paper is published, then the reference has no citations when the paper was written.} The third subfigure in Fig.~\ref{sfig:refCitation} shows that the ratio of non-cited references decreases for papers in all the three citation groups, largely due to the increase of reference numbers as shown in Fig.~\ref{sfig:refNum}. Interestingly, both non-cited and cited papers show higher ratio of non-cited references, compared to well-cited papers. However, as the ratio of non-cited references depends on the total number of references in a paper, we then look at the absolute number of non-cited references in the fourth subfigure of Fig.~\ref{sfig:refCitation}. In terms of number of non-cited references, well-cited papers used to have 2, and now have 1 on average. Considering the increasing trend in number of references, scholars become very careful in selecting non-cited references.

The next question is: what happens to these non-cited references, once they start receiving citations. The last subfigure in Fig.~\ref{sfig:refCitation} plots the number of citations a non-cited reference receives in 5 years after the citing paper's publication. The non-cited references chosen by well-cited papers show significantly larger number of citations, compared to that of cited and non-cited papers. This plot suggests that the authors of well-cited papers demonstrated better ability in selecting the potentially impactful papers, when there is no clear evidence on the impact of these papers. This may also translate into more effort in reading and selecting papers. 

Overall, we argue that scholars have to make more effort along the research explosion, reflected by the explicit (number) and implicit (age and citation) growth in reading references to support their research. 

\subsection{Negative Effect}
\label{subsec:negative-effect}

\begin{figure*}[ht] 
\centering
\subfigure[Coefficient of number of accumulative publications \label{sfig:logEffect}] {\includegraphics[width=0.3864\textwidth]{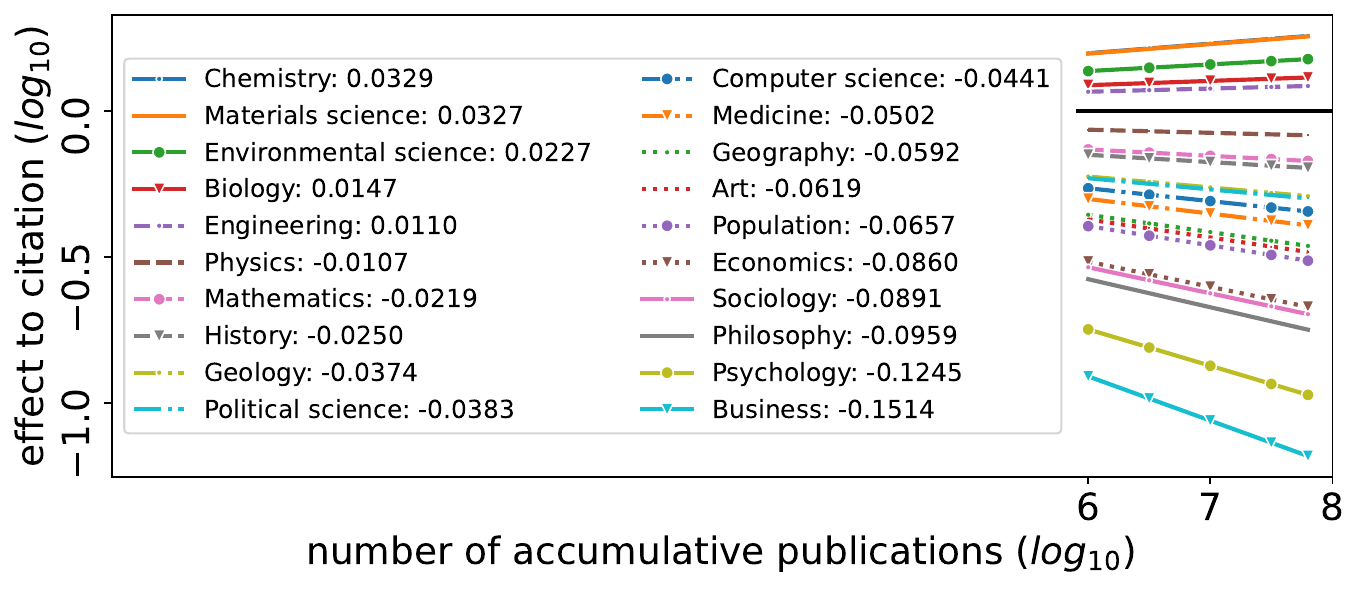}} 
\quad
\subfigure[Effects on citations, from $log_{10}$ scale to linear scale \label{sfig:absoluteEffect}]{\includegraphics[width=0.171\textwidth]{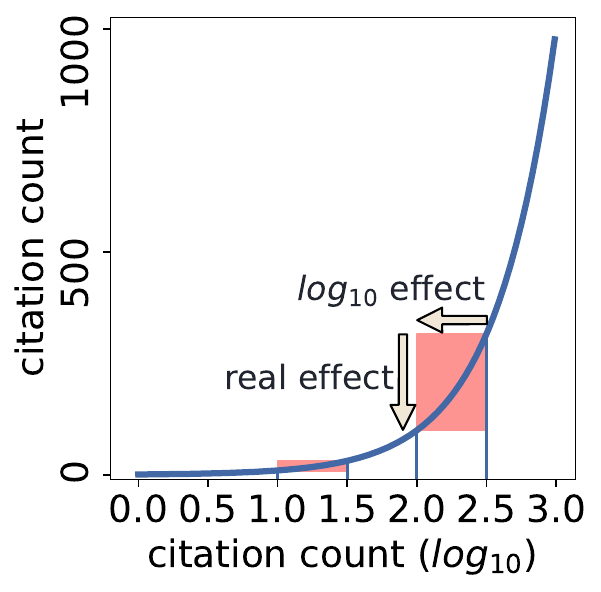}}
\caption{Effect of accumulation number. (a) presents the coefficient of the accumulation number for all fields of study. The x-axis is the $log_{10}$ accumulation number, the y-axis is the $log_{10}$ citation count, the slopes are the coefficients of accumulation number for all fields, and the biases of these plots are set to 0. The legend presents the coefficient of each field. This sub-figure controls other independent variables to remain unchanged. (b) presents the real effect on citations. In other words, the x-axis is the possible change of $log_{10}$ citation from (a), and the y-axis is the real effect on potential citation counts within 5 years.}
\label{fig:citationEffect}
\end{figure*}

The analysis of results in Fig.~\ref{fig:referencePattern} shows a clear difference in reference patterns of well-cited, cited, and non-cited papers. As references in each publication are selected from the existing work at the time of writing, the plots do not show the impact of research explosion along time. Given that both number of published papers and reference patterns change along time, we aim to estimate the impact of the number of accumulative publications on papers' potential citation count. We use a total of 60.8 million publications, and the publications in each of the 20 fields to find out the evidence of the impact.

The coefficient for the accumulated number of publications from the 60 million publications is $-0.0279^*$ ($p-value \leq 0.001$), a negative value. Plotted in Fig.~\ref{sfig:logEffect}, the coefficients which are collected from publications in each field are mostly negative, except for five fields \fos{Chemistry}, \fos{Materials science}, \fos{Environmental science}, \fos{Biology}, and \fos{Engineering}. Moreover, the range of the caused positive citation (in $log_{10}$ scale) effect is from 0.07 to 0.26. The range of the caused negative citation effect is from -0.06 to -1.20. In other words, the ever-growing publication numbers slightly benefit these five fields for higher citations, while for all the remaining 15 fields the increase in publication numbers leads to a negative impact on citations. Fig.~\ref{sfig:absoluteEffect} plots the mapping between $log_{10}$ scale and linear scale, \ie the real value $y=10^x$ of the citation changes in $log_{10}$ scale. Nevertheless, papers' citation counts are dependent on other independent variables comprising indicators of reference pattern.

Citation count remains adopted as one of the most direct indicators of research impact. From our analysis, well-cited papers, or papers with more citations, in general are results of more research effort, \eg more number of references, more carefully selected references with wide spread coverage of classic and research results, and good taste in selecting references with great potential. With respect to the fast growth in the overall number of publications, scholars may need to put more effort into getting impactful results. Here, the coefficient of number of papers can be regarded as a direct indicator of the effect on potential citation. Out of the 20 fields, only 5 fields observe positive impact of growth in number of papers to citation increase. Among them, the most positive effect is observed for \fos{Chemistry}, and the most negative effect is for \fos{Business}. The large number of publications may push the development in \fos{Chemistry}, but may have hurt the development of in \fos{Business} field. 

\section{Discussion}
\label{sec:discussion}

This study demonstrates that scholars have to devote more effort to climbing onto shoulders of the giant due to research explosion. The evolution on reference patterns clearly shows well-cited papers are supported by well selected references and non-cited papers may unnecessarily increase the effort of scholars in selecting these references. Reflected by the negative coefficient, the fast growth of publication numbers may not benefit scholarly research in general. There could be many reasons behind the fast-growing publication numbers. One possible reason is that scholars are evaluated by number of publications for grant application or promotion. Hence scholars are under pressure to produce more publications within a limited time period. Indicated by our study on reference patterns, in general, researchers need to make more effort to produce well-cited papers, but citation may only come after a number of years. Yet, there is a high chance that papers may not be well-cited. Our study calls for a better way to evaluate research output. Another possible reason is the paper review process. As about 29.4\% of papers have no citations within 5 years, a good number of papers in this group may not be worth publishing. This calls for a better review system to reduce the papers that do not bring in new contributions. Researchers might also need to put more effort in order to produce more recognizable results, exampled by well-cited papers. 


\printbibliography 


\newpage

\section{Appendix} 

\begin{figure*}[ht]
\centering
\subfigure[Number \label{sfig:refNumMedicine}] {\includegraphics[width=0.1233\textwidth]{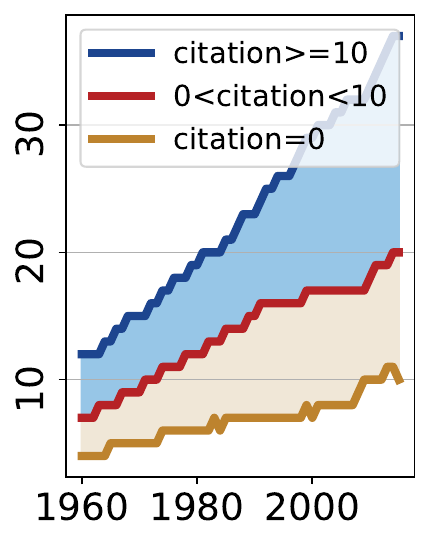}}
\quad
\subfigure[Age \label{sfig:refAgeMedicine}]{\includegraphics[width=0.3605\textwidth]{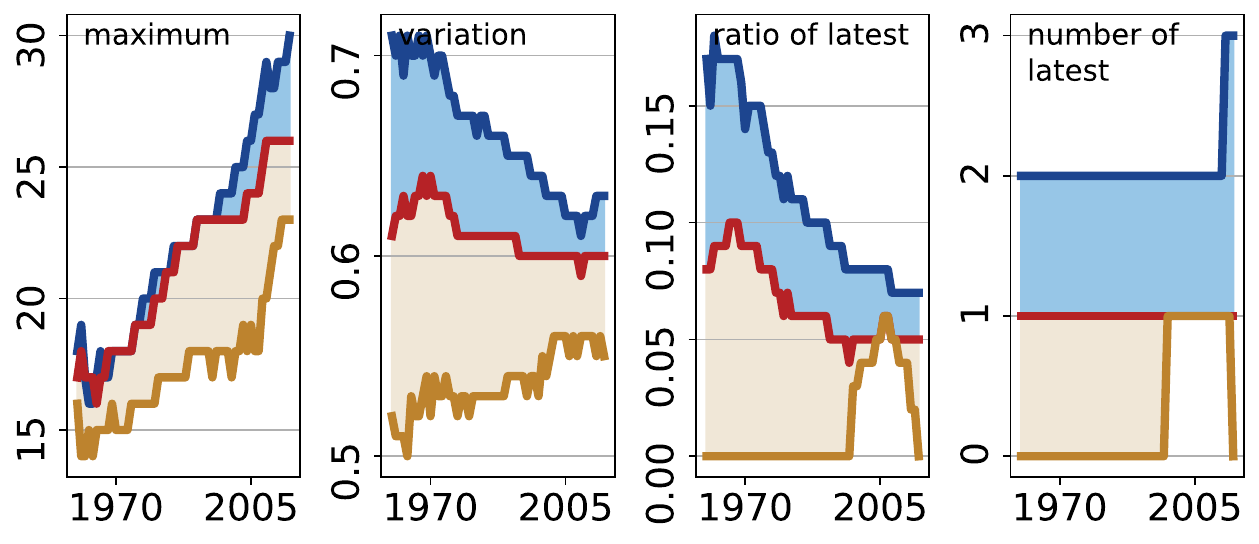}}
\quad
\subfigure[Citation count \label{sfig:refCitationMedicine}] {\includegraphics[width=0.45\textwidth]{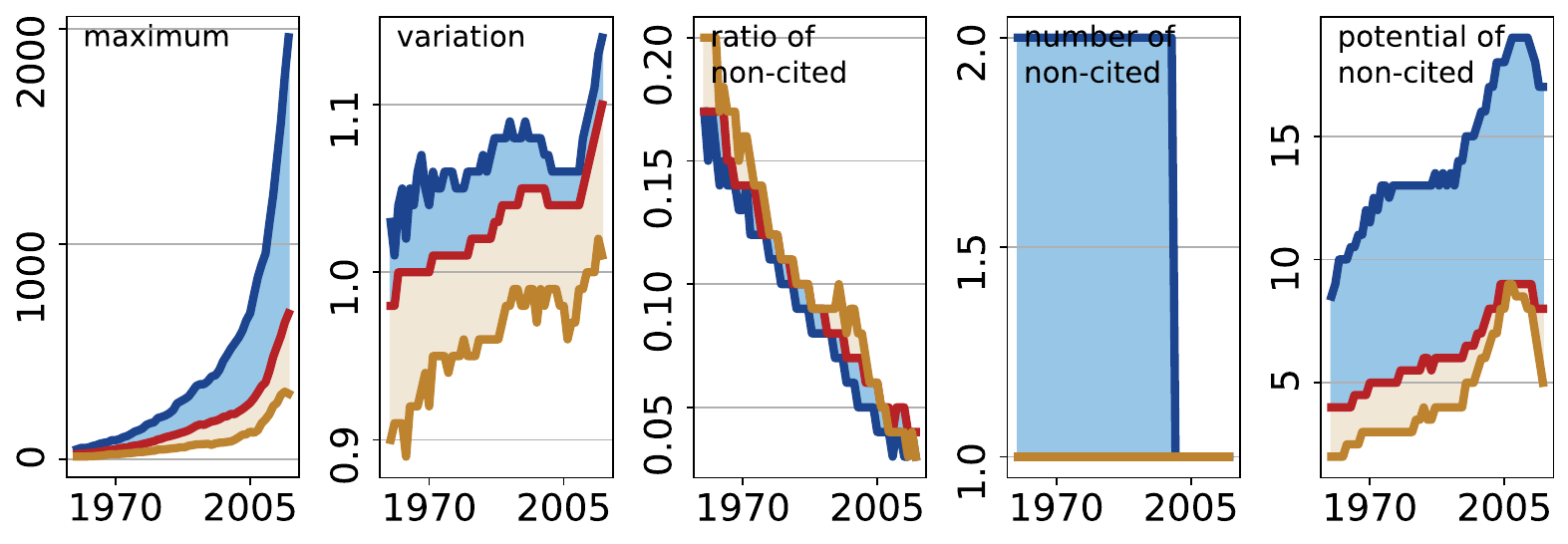}}
\caption{The indicators of reference patten of Medicine from 1960 to 2015. (a) indicates the median number of references per year, from 1960 to 2015. (b) presents the reference age patterns year by year from 1960 to 2015. Specifically, the 4 sub-figures in (b) plot the maximum reference age, variation ratio of age, ratio of latest papers in reference, and the absolute number of latest papers in references, respectively. (c) presents the citation count of references. Particularly, the 5 sub-figures in (c) show maximum of citation counts, variation ratio of citation counts, ratio of non-cited publications in references, absolute number of non-cited publications in references, and the potential (number of citations) of the non-cited references, respectively.}
\label{fig:referencePatternMedicine}
\end{figure*}

\begin{figure*}[ht]
\centering
\subfigure[Number \label{sfig:refNumEngineering}] {\includegraphics[width=0.1233\textwidth]{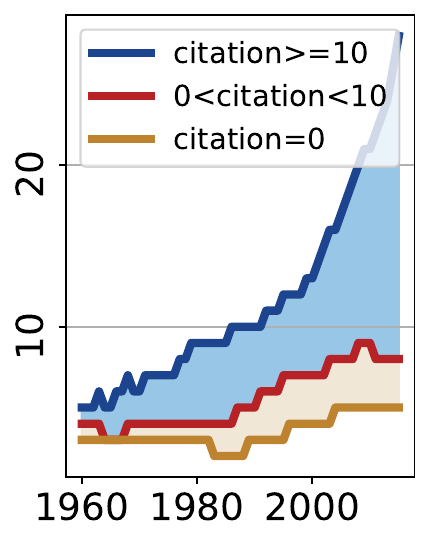}}
\quad
\subfigure[Age \label{sfig:refAgeEngineering}]{\includegraphics[width=0.3605\textwidth]{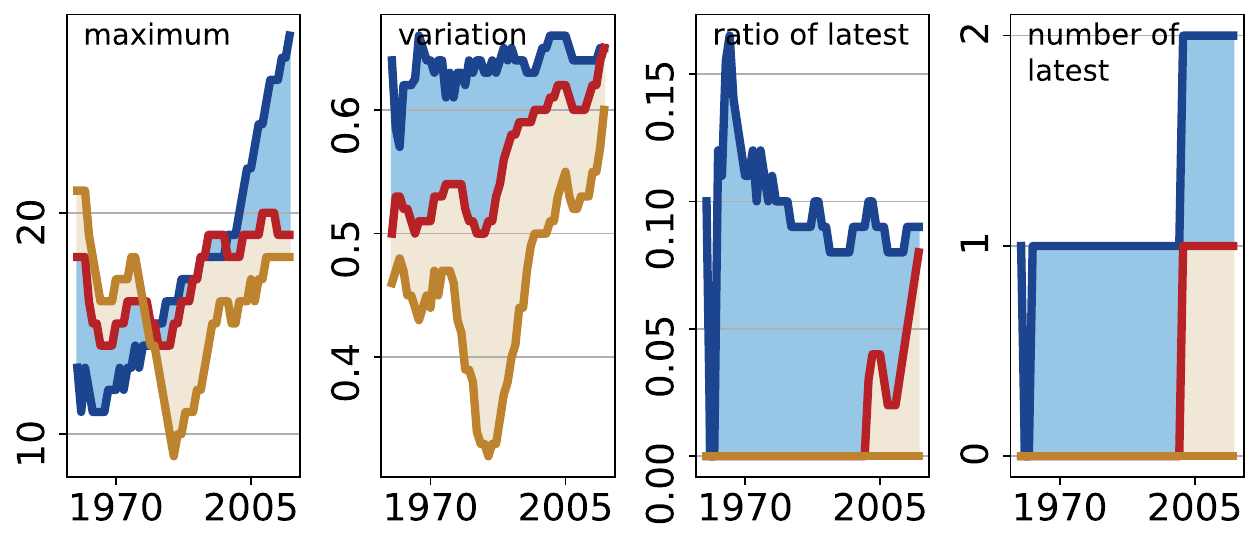}}
\quad
\subfigure[Citation count \label{sfig:refCitationEngineering}] {\includegraphics[width=0.45\textwidth]{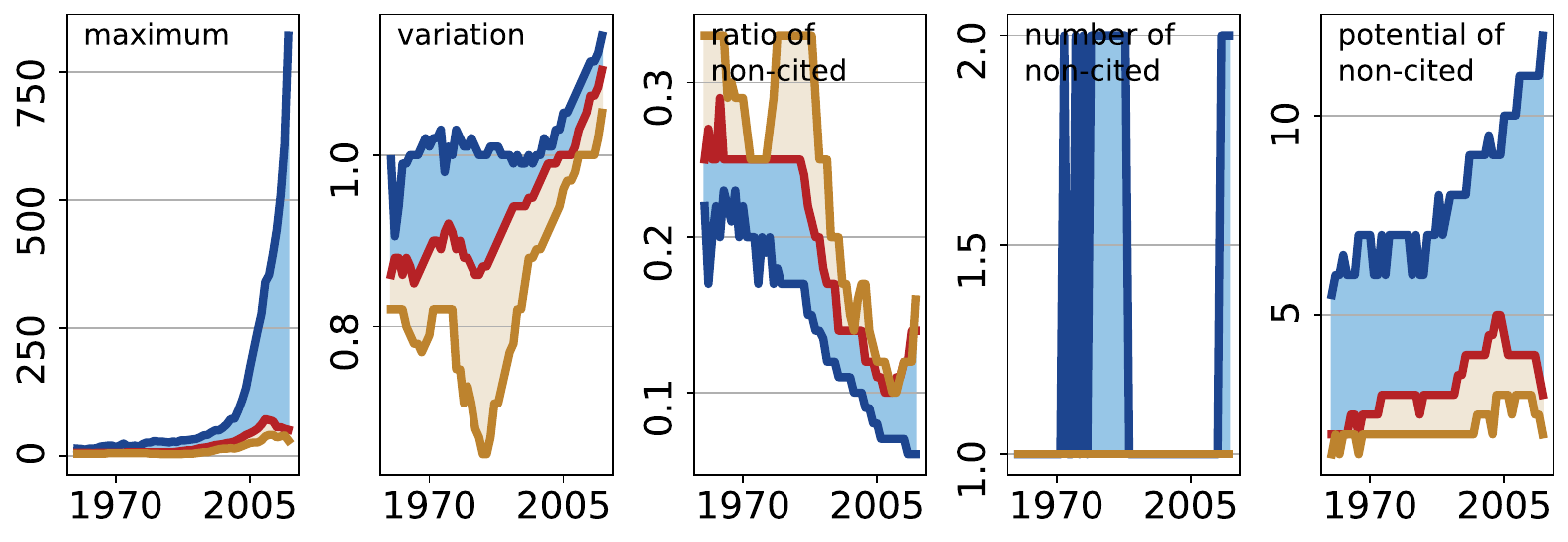}}

\caption{The indicators of reference patten of Engineering from 1960 to 2015.}
\label{fig:referencePatternEngineering}
\end{figure*}

\begin{figure*}[ht]
\centering
\subfigure[Number \label{sfig:refNumBiology}] {\includegraphics[width=0.1233\textwidth]{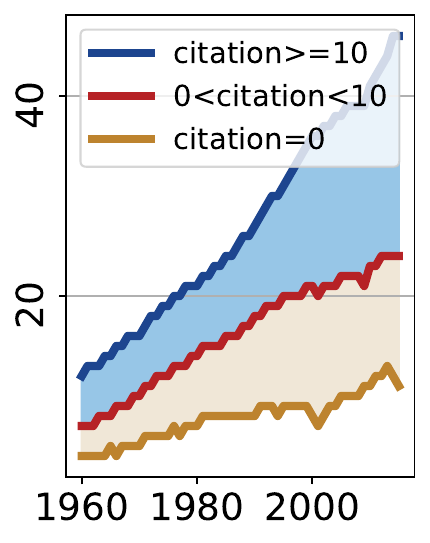}}
\quad
\subfigure[Age \label{sfig:refAgeBiology}]{\includegraphics[width=0.3605\textwidth]{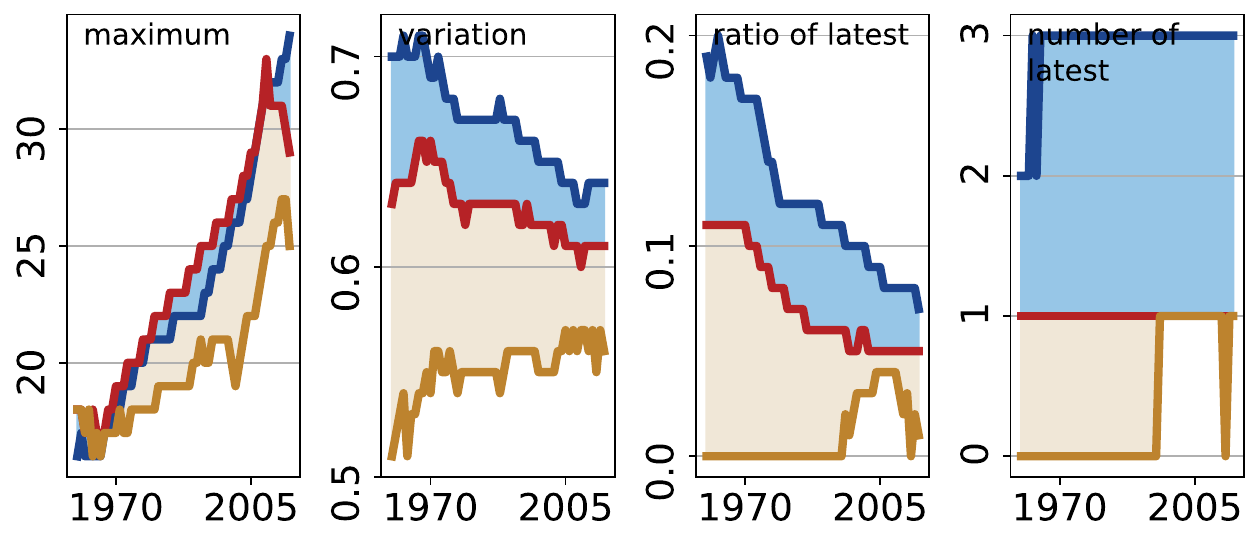}}
\quad
\subfigure[Citation count \label{sfig:refCitationBiology}] {\includegraphics[width=0.45\textwidth]{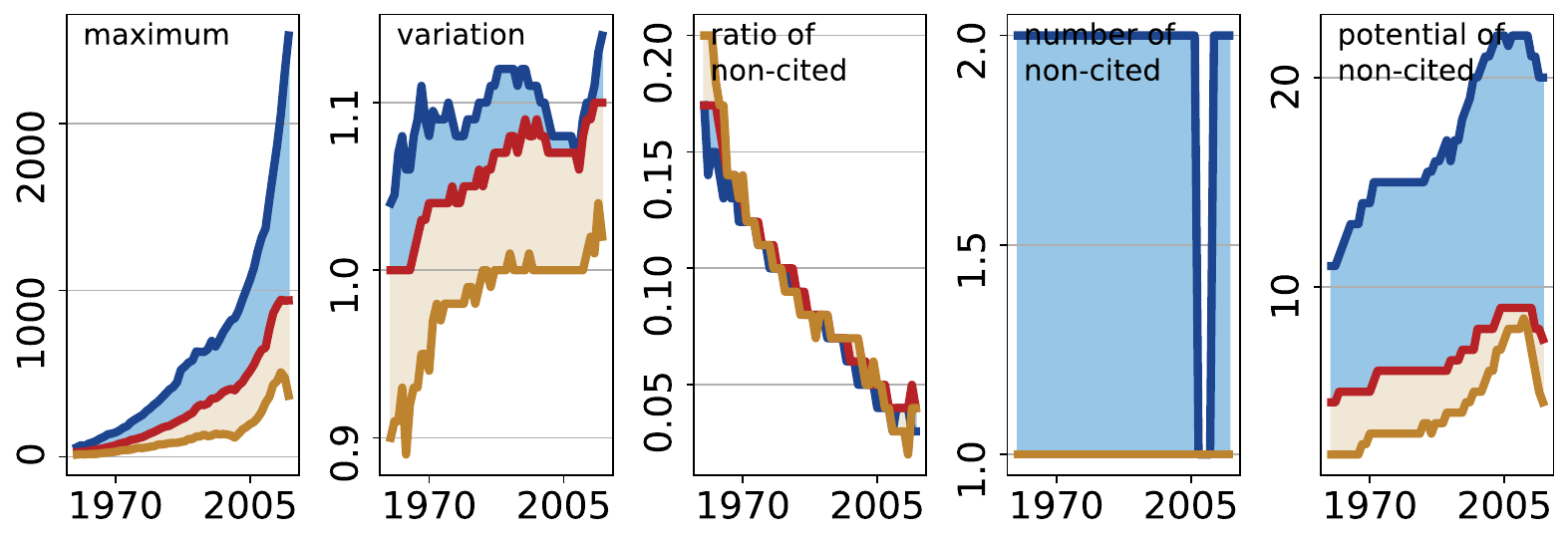}}

\caption{The indicators of reference patten of Biology from 1960 to 2015.}
\label{fig:referencePatternBiology}
\end{figure*}

\begin{figure*}[ht]
\centering
\subfigure[Number \label{sfig:refNumChemistry}] {\includegraphics[width=0.1233\textwidth]{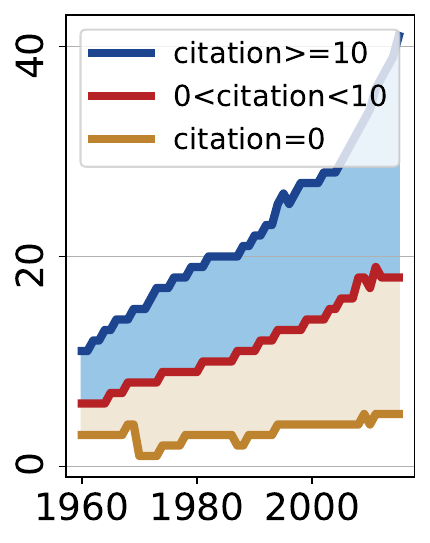}}
\quad
\subfigure[Age \label{sfig:refAgeChemistry}]{\includegraphics[width=0.3605\textwidth]{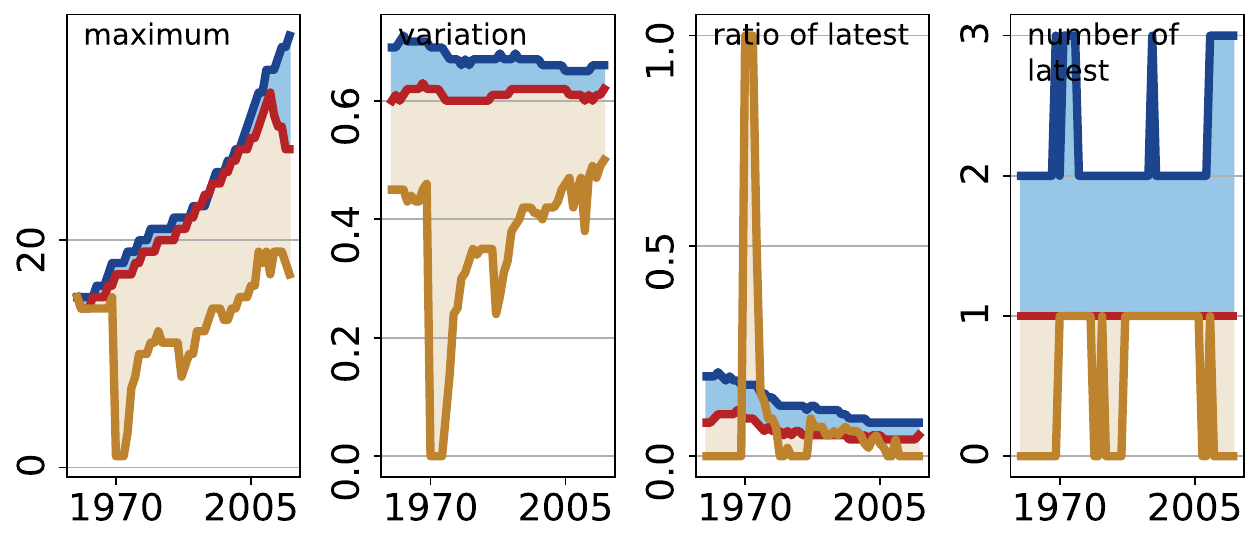}}
\quad
\subfigure[Citation count \label{sfig:refCitationChemistry}] {\includegraphics[width=0.45\textwidth]{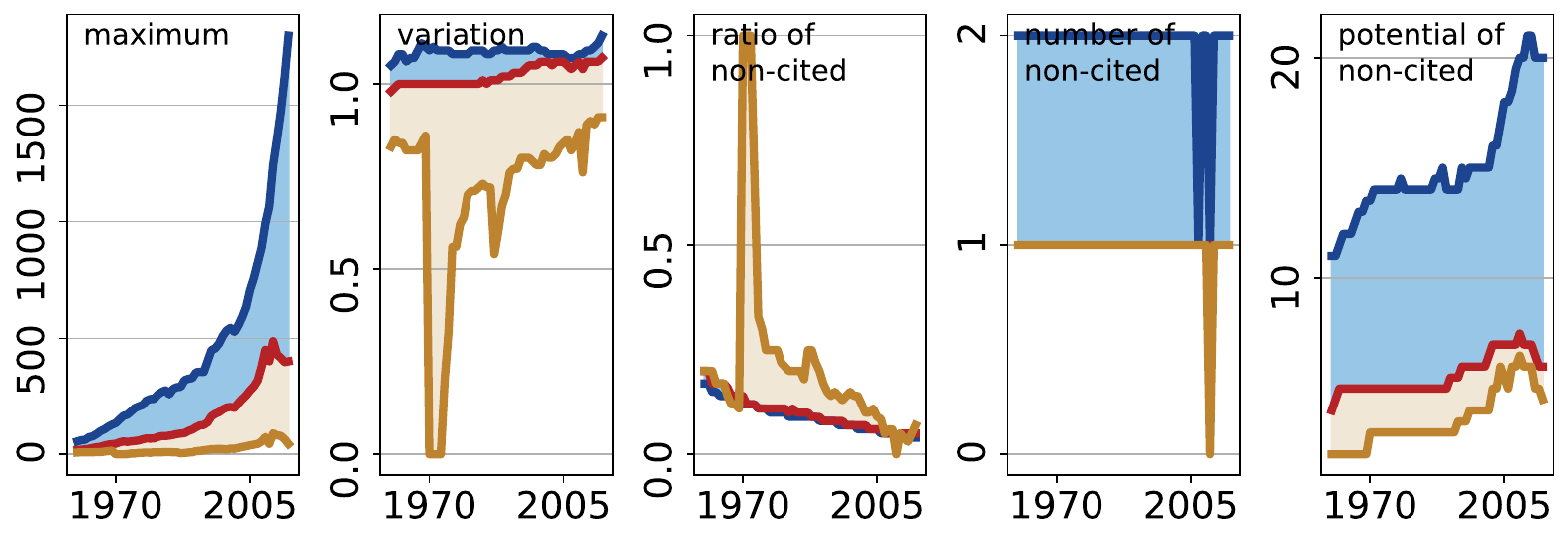}}
\caption{The indicators of reference patten of Chemistry from 1960 to 2015.}
\label{fig:referencePatternChemistry}
\end{figure*}

\begin{figure*}[ht]
\centering
\subfigure[Number \label{sfig:refNumMaterials}] {\includegraphics[width=0.1233\textwidth]{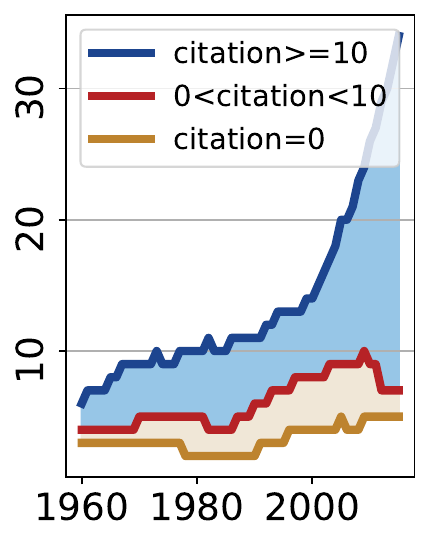}}
\quad
\subfigure[Age \label{sfig:refAgeMaterials}]{\includegraphics[width=0.3605\textwidth]{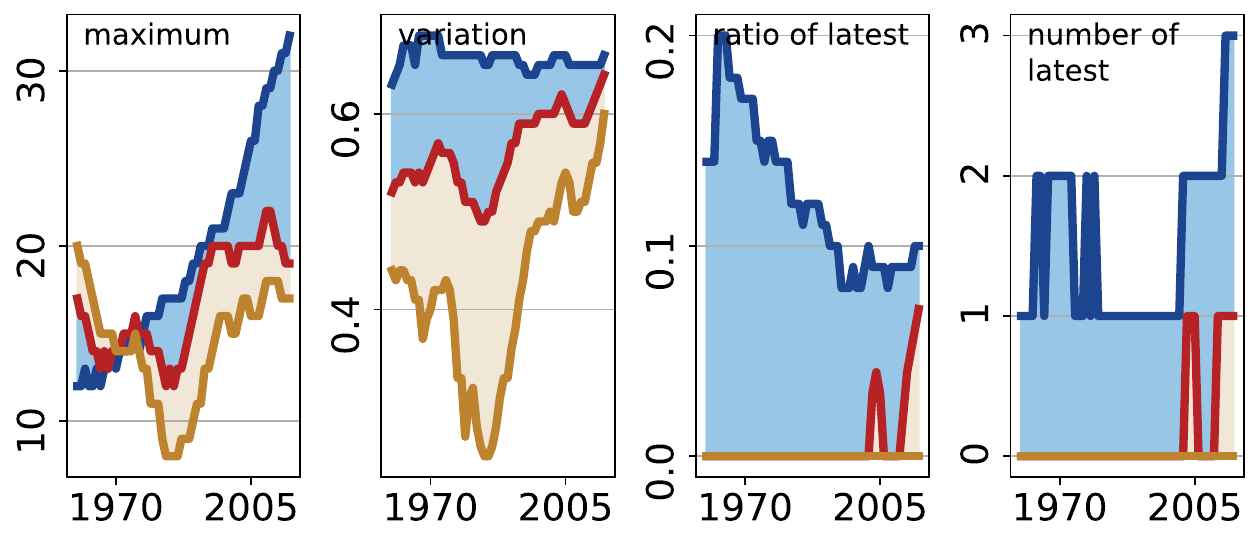}}
\quad
\subfigure[Citation count \label{sfig:refCitationMaterials}] {\includegraphics[width=0.45\textwidth]{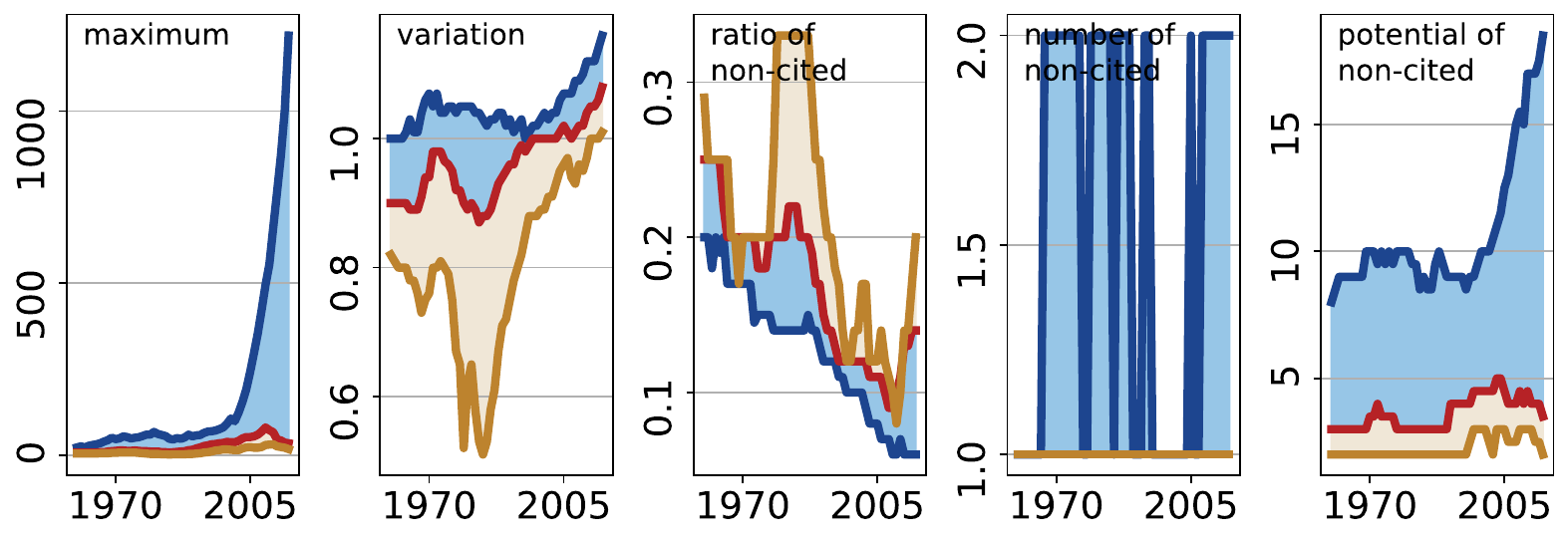}}

\caption{The indicators of reference patten of Materials science from 1960 to 2015. (a) indicates the median number of references per year, from 1960 to 2015. (b) presents the reference age patterns year by year from 1960 to 2015. Specifically, the 4 sub-figures in (b) plot the maximum reference age, variation ratio of age, ratio of latest papers in reference, and the absolute number of latest papers in references, respectively. (c) presents the citation count of references. Particularly, the 5 sub-figures in (c) show maximum of citation counts, variation ratio of citation counts, ratio of non-cited publications in references, absolute number of non-cited publications in references, and the potential (number of citations) of the non-cited references, respectively.}
\label{fig:referencePatternMaterialsScience}
\end{figure*}

\begin{figure*}[ht]
\centering
\subfigure[Number \label{sfig:refNumComputer}] {\includegraphics[width=0.1233\textwidth]{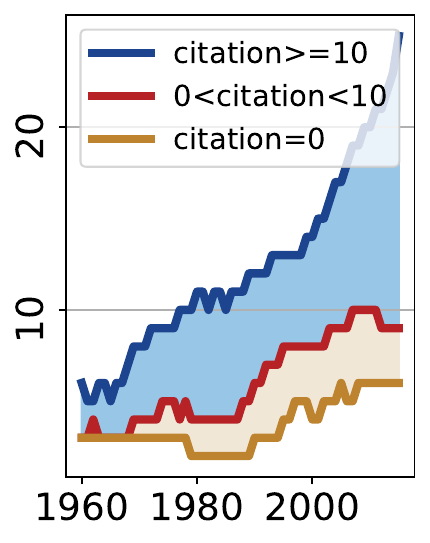}}
\quad
\subfigure[Age \label{sfig:refAgeComputer}]{\includegraphics[width=0.3605\textwidth]{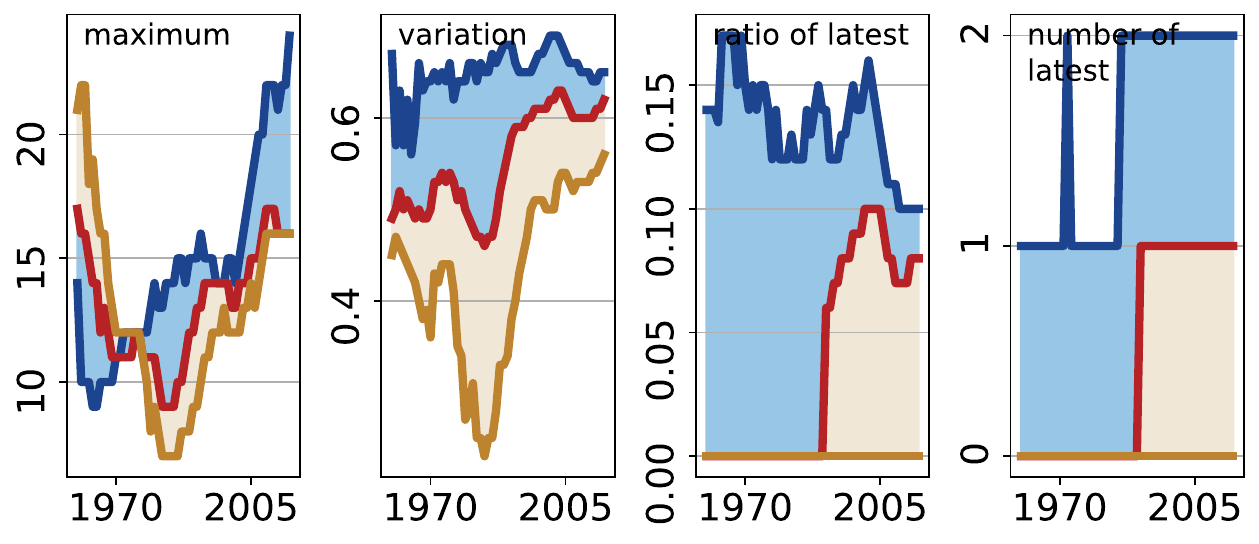}}
\quad
\subfigure[Citation count \label{sfig:refCitationComputer}] {\includegraphics[width=0.45\textwidth]{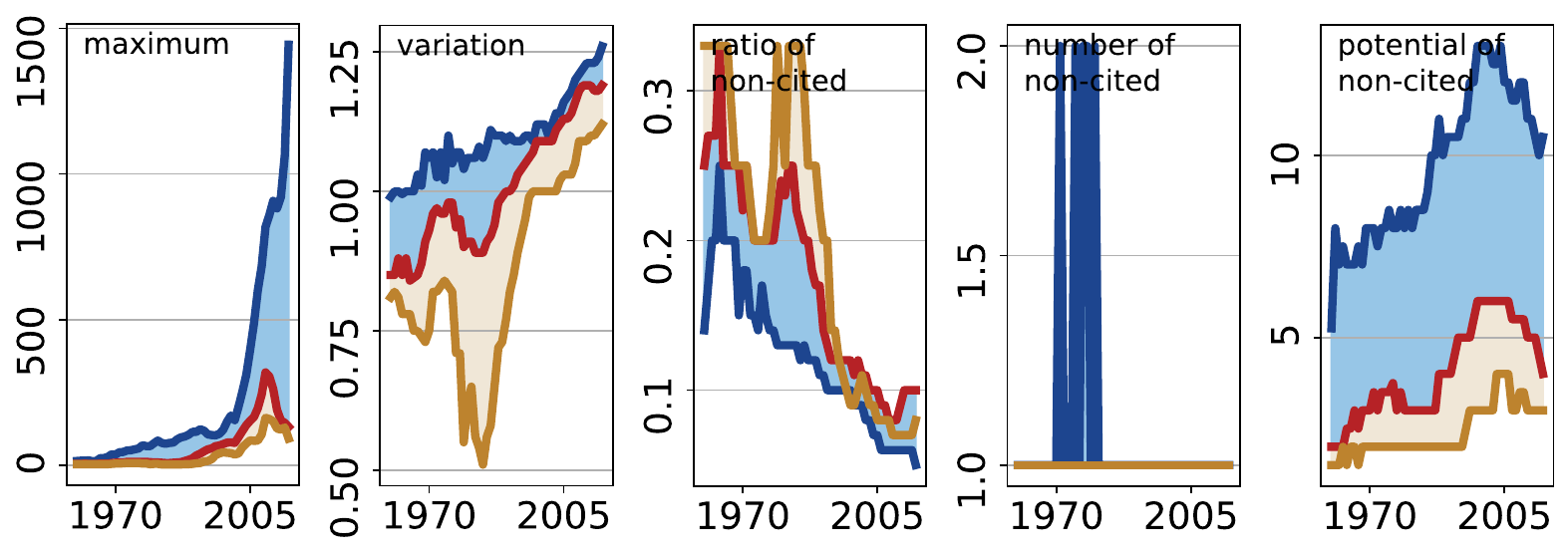}}

\caption{The indicators of reference patten of Computer science from 1960 to 2015.}
\label{fig:referencePatternComputerScience}
\end{figure*}

\begin{figure*}[ht]
\centering
\subfigure[Number \label{sfig:refNumPhysics}] {\includegraphics[width=0.1233\textwidth]{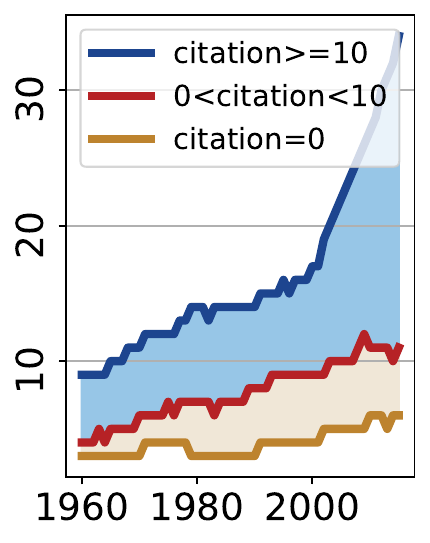}}
\quad
\subfigure[Age \label{sfig:refAgePhysics}]{\includegraphics[width=0.3605\textwidth]{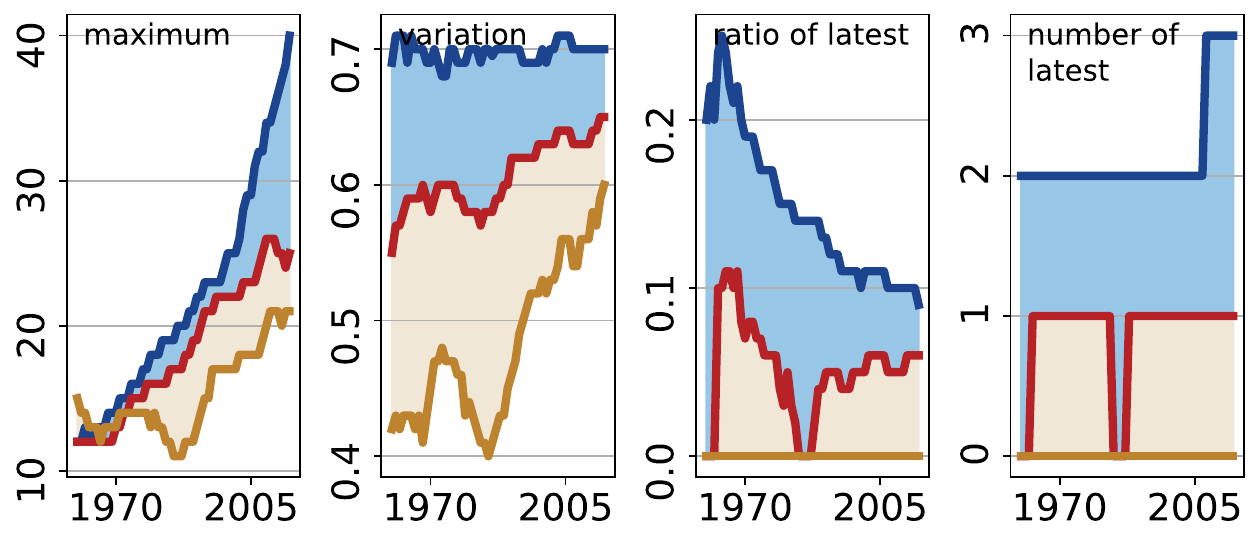}}
\quad
\subfigure[Citation count \label{sfig:refCitationPhysics}] {\includegraphics[width=0.45\textwidth]{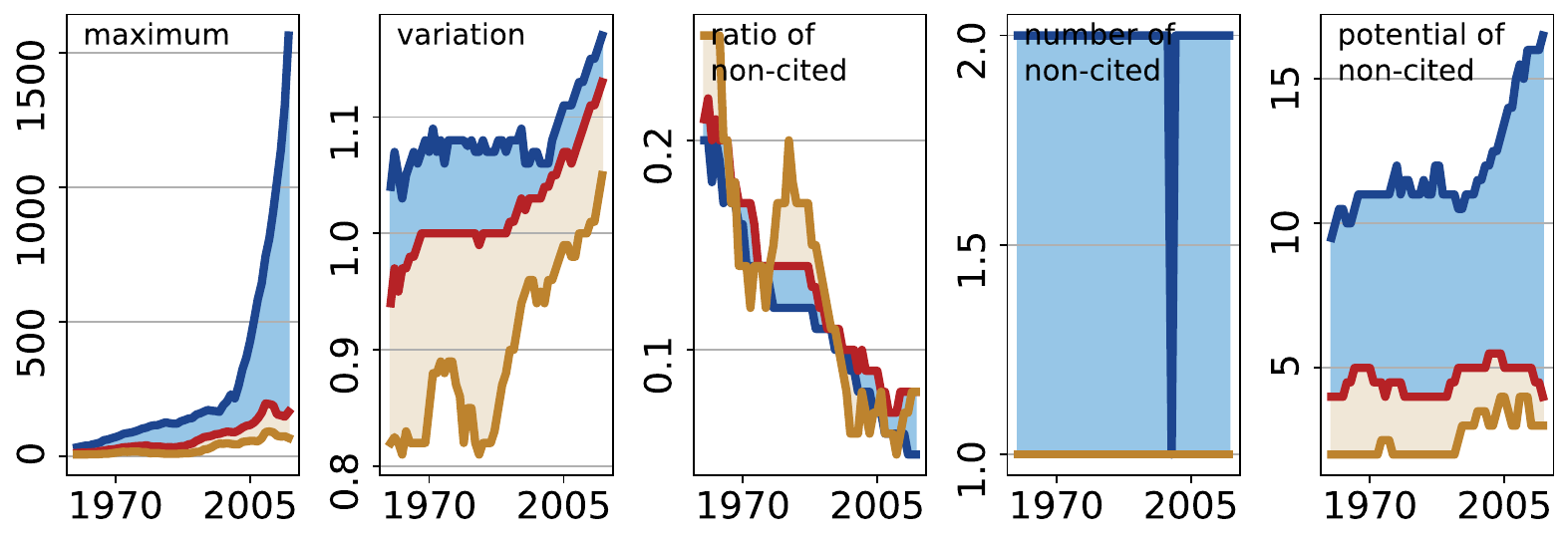}}

\caption{The indicators of reference patten of Physics from 1960 to 2015.}
\label{fig:referencePatternPhysics}
\end{figure*}

\begin{figure*}[ht]
\centering
\subfigure[Number \label{sfig:refNumMathematics}] {\includegraphics[width=0.1233\textwidth]{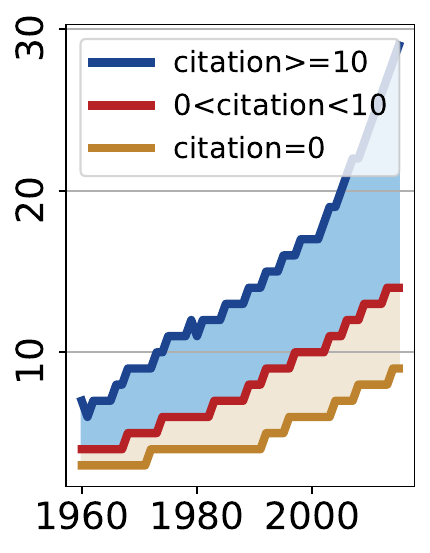}}
\quad
\subfigure[Age \label{sfig:refAgeMathematics}]{\includegraphics[width=0.3605\textwidth]{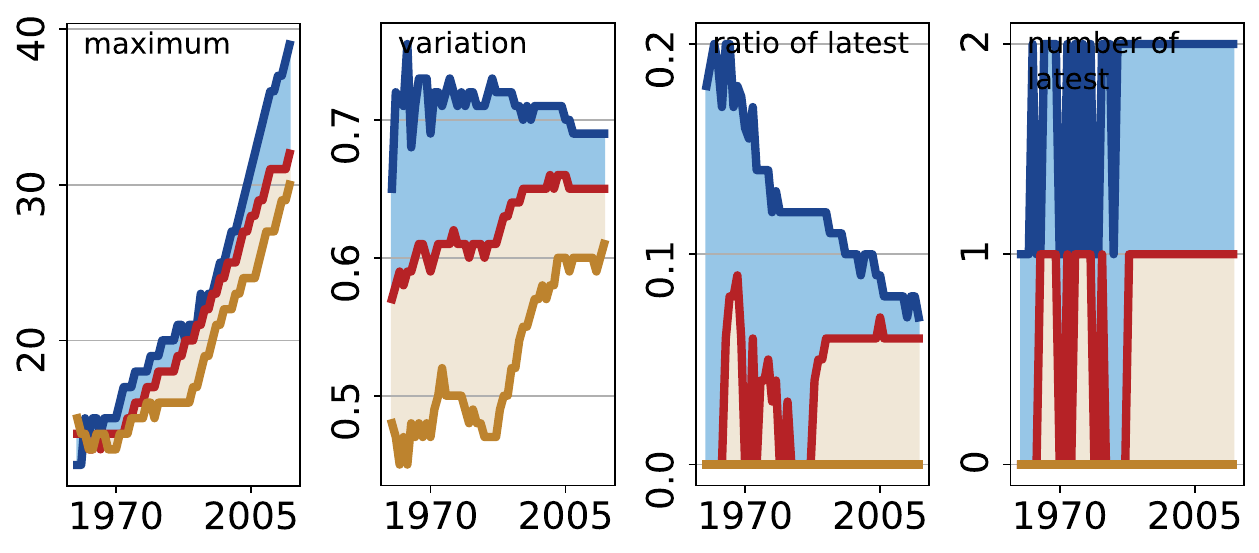}}
\quad
\subfigure[Citation count \label{sfig:refCitationMathematics}] {\includegraphics[width=0.45\textwidth]{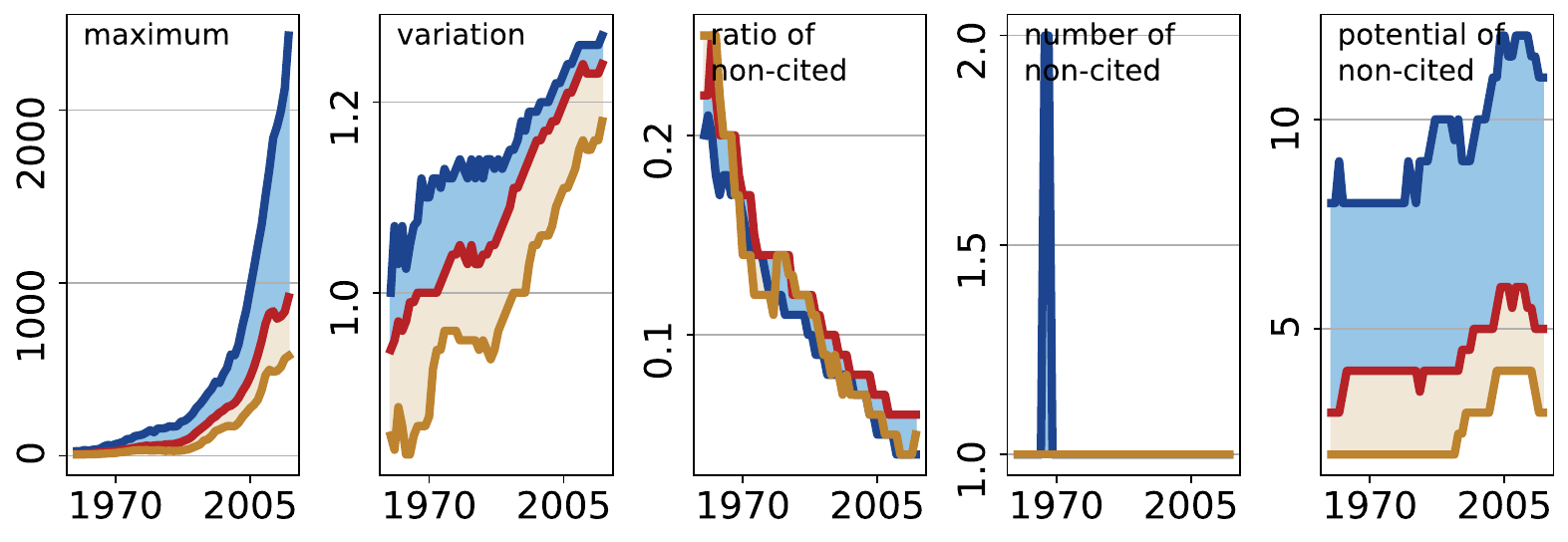}}

\caption{The indicators of reference patten of Mathematics from 1960 to 2015.}
\label{fig:referencePatternMathematics}
\end{figure*}

\begin{figure*}[ht]
\centering
\subfigure[Number \label{sfig:refNumPsychology}] {\includegraphics[width=0.1233\textwidth]{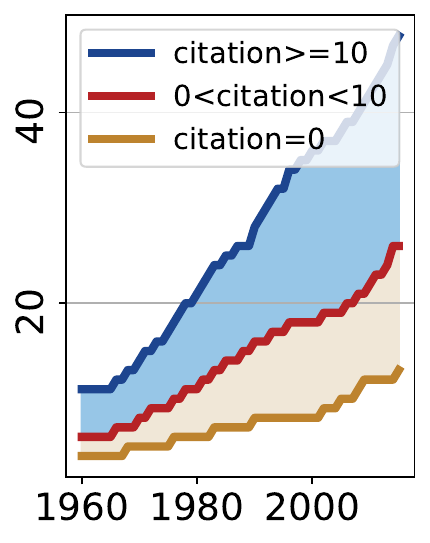}}
\quad
\subfigure[Age \label{sfig:refAgePsychology}]{\includegraphics[width=0.3605\textwidth]{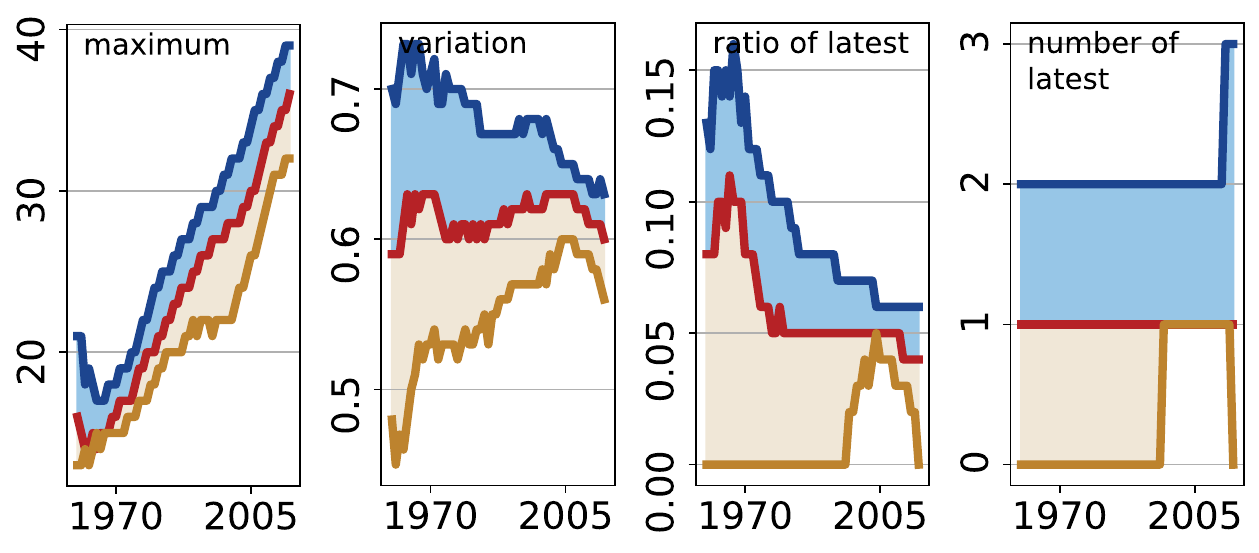}}
\quad
\subfigure[Citation count \label{sfig:refCitationPsychology}] {\includegraphics[width=0.45\textwidth]{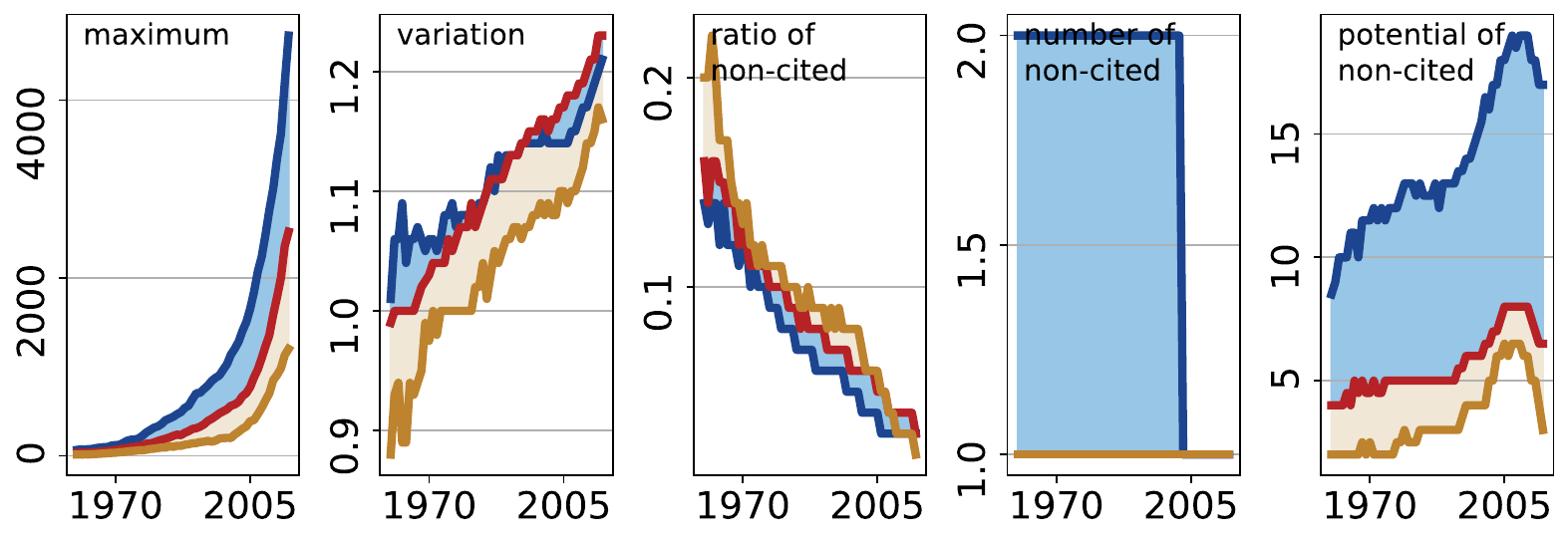}}

\caption{The indicators of reference patten of Psychology from 1960 to 2015. (a) indicates the median number of references per year, from 1960 to 2015. (b) presents the reference age patterns year by year from 1960 to 2015. Specifically, the 4 sub-figures in (b) plot the maximum reference age, variation ratio of age, ratio of latest papers in reference, and the absolute number of latest papers in references, respectively. (c) presents the citation count of references. Particularly, the 5 sub-figures in (c) show maximum of citation counts, variation ratio of citation counts, ratio of non-cited publications in references, absolute number of non-cited publications in references, and the potential (number of citations) of the non-cited references, respectively.}
\label{fig:referencePatternPsychology}
\end{figure*}

\begin{figure*}[ht]
\centering
\subfigure[Number \label{sfig:refNumEconomics}] {\includegraphics[width=0.1233\textwidth]{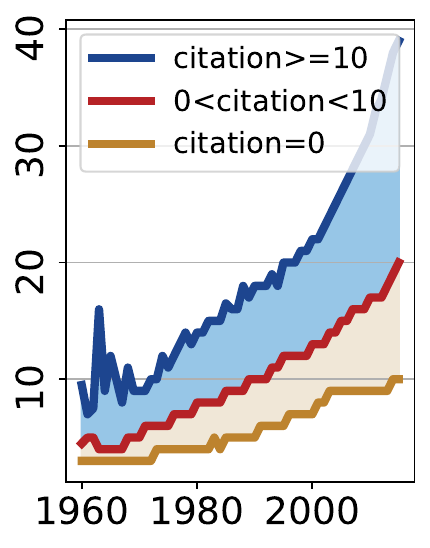}}
\quad
\subfigure[Age \label{sfig:refAgeEconomics}]{\includegraphics[width=0.3605\textwidth]{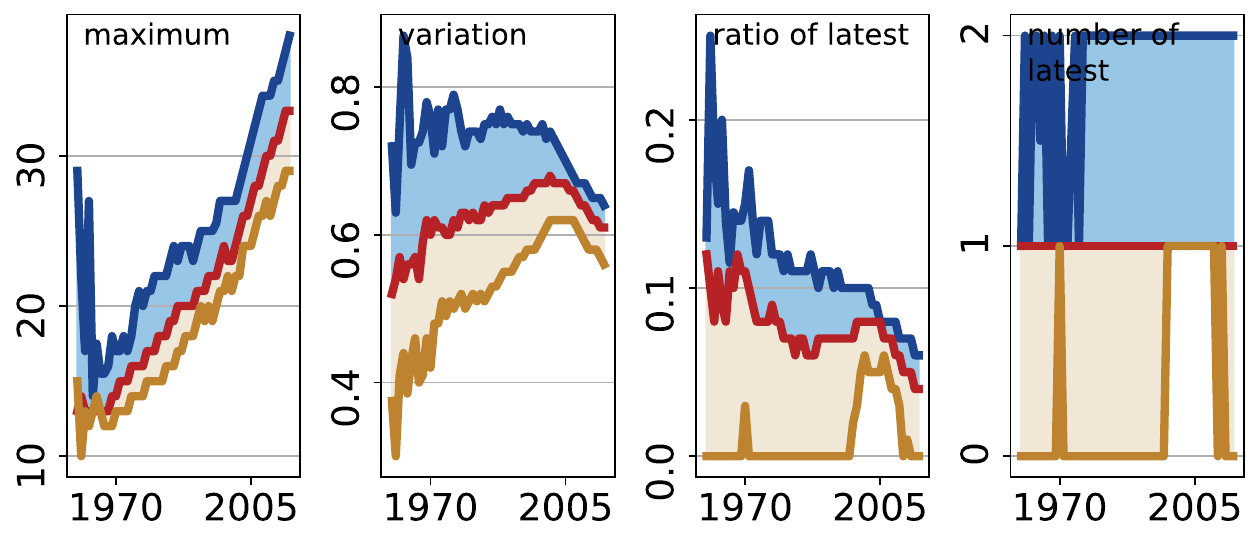}}
\quad
\subfigure[Citation count \label{sfig:refCitationEconomics}] {\includegraphics[width=0.45\textwidth]{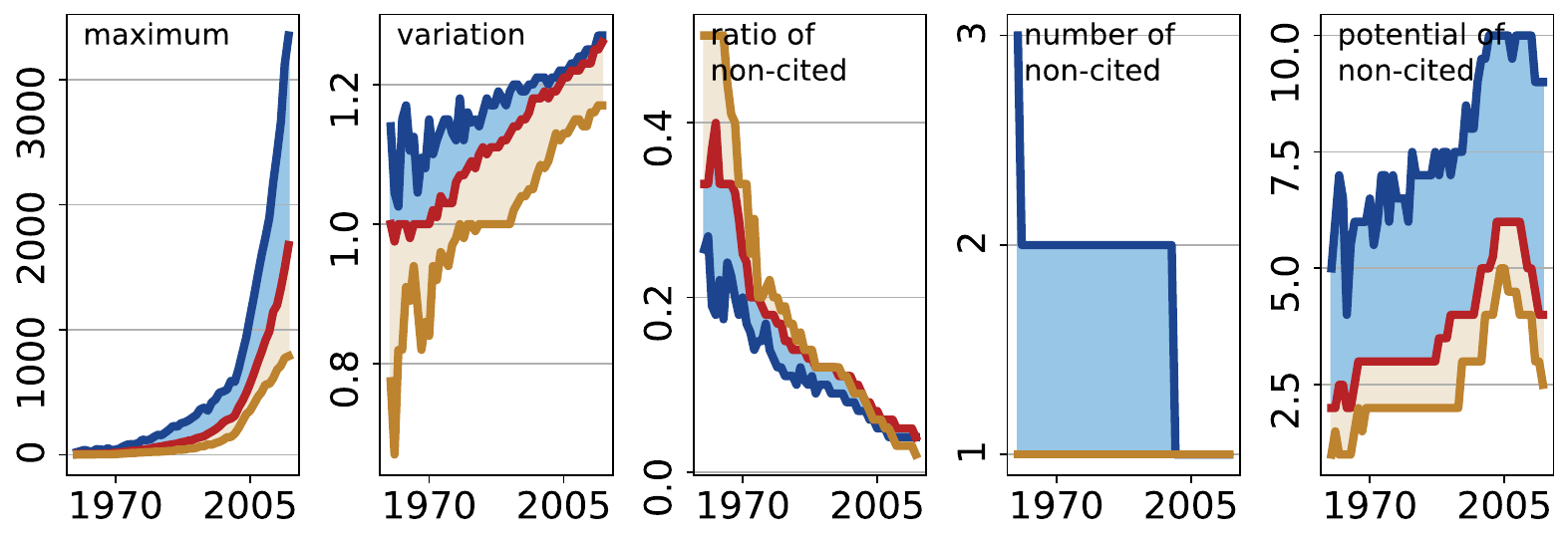}}

\caption{The indicators of reference patten of Economics from 1960 to 2015.}
\label{fig:referencePatternEconomics}
\end{figure*}

\begin{figure*}[ht]
\centering
\subfigure[Number \label{sfig:refNumGeology}] {\includegraphics[width=0.1233\textwidth]{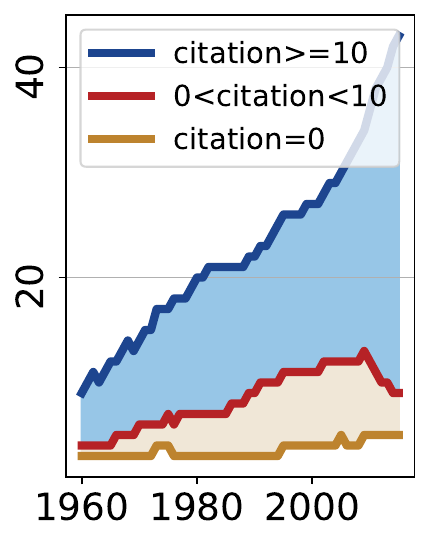}}
\quad
\subfigure[Age \label{sfig:refAgeGeology}]{\includegraphics[width=0.3605\textwidth]{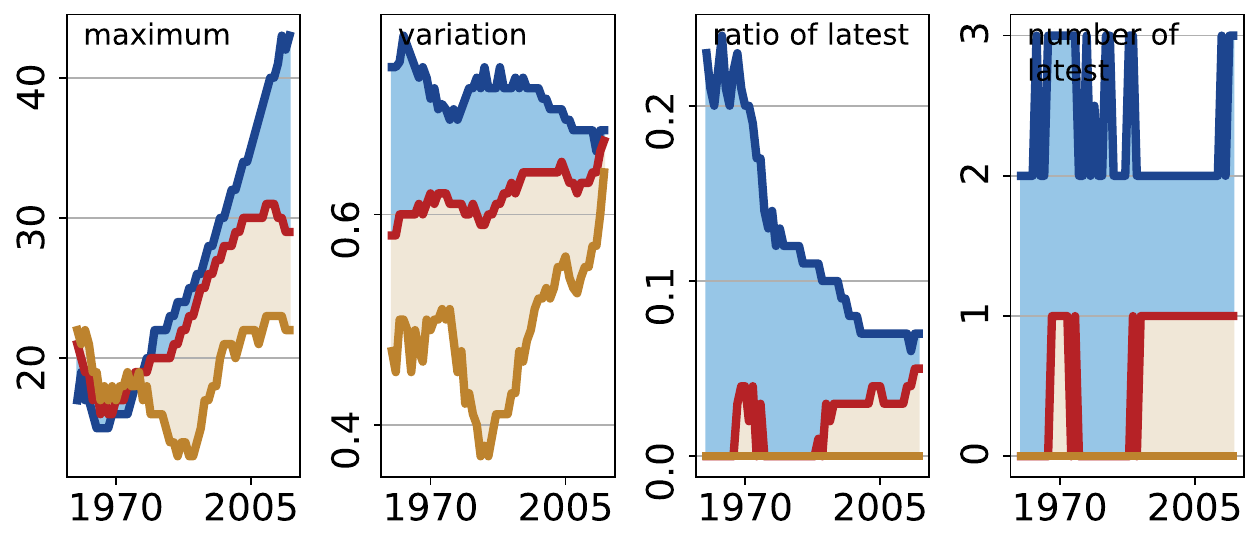}}
\quad
\subfigure[Citation count \label{sfig:refCitationGeology}] {\includegraphics[width=0.45\textwidth]{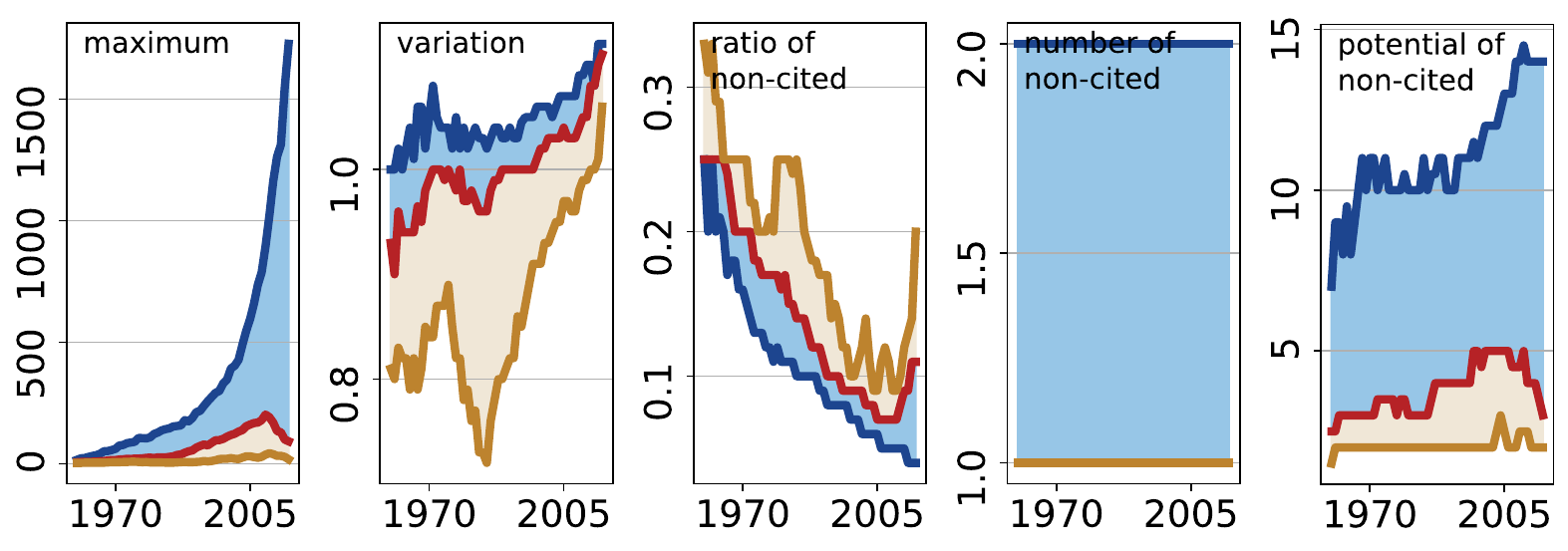}}

\caption{The indicators of reference patten of Geology from 1960 to 2015.}
\label{fig:referencePatternGeology}
\end{figure*}

\begin{figure*}[ht]
\centering
\subfigure[Number \label{sfig:refNumSociology}] {\includegraphics[width=0.1233\textwidth]{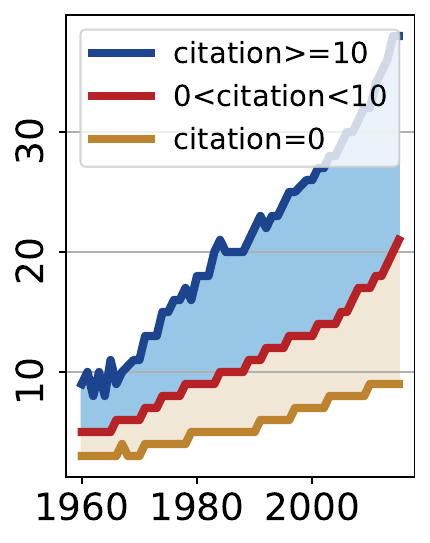}}
\quad
\subfigure[Age \label{sfig:refAgeSociology}]{\includegraphics[width=0.3605\textwidth]{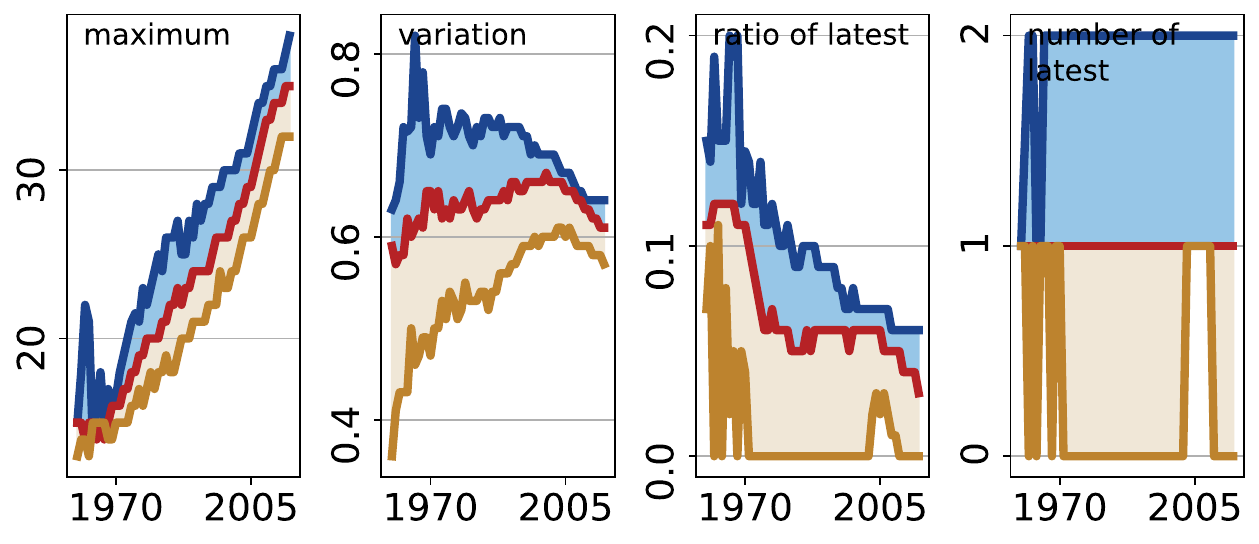}}
\quad
\subfigure[Citation count \label{sfig:refCitationSociology}] {\includegraphics[width=0.45\textwidth]{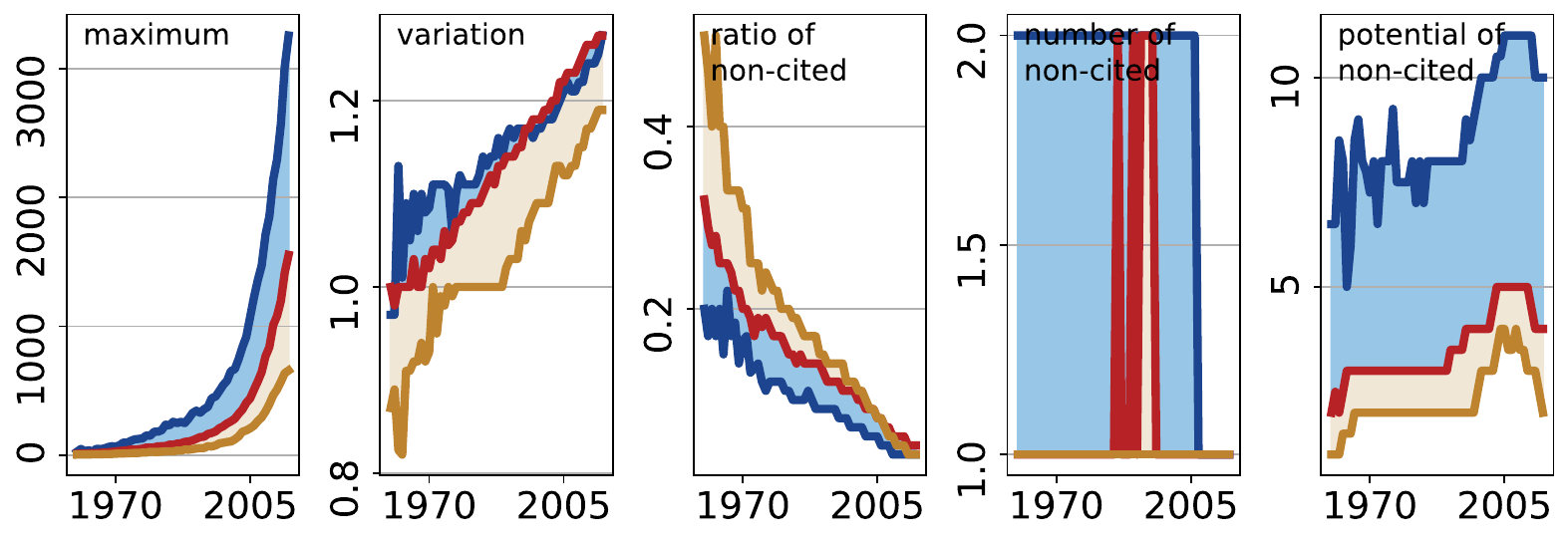}}

\caption{The indicators of reference patten of Sociology from 1960 to 2015.}
\label{fig:referencePatternSociology}
\end{figure*}

\begin{figure*}[ht]
\centering
\subfigure[Number \label{sfig:refNumEnvironmental}] {\includegraphics[width=0.1233\textwidth]{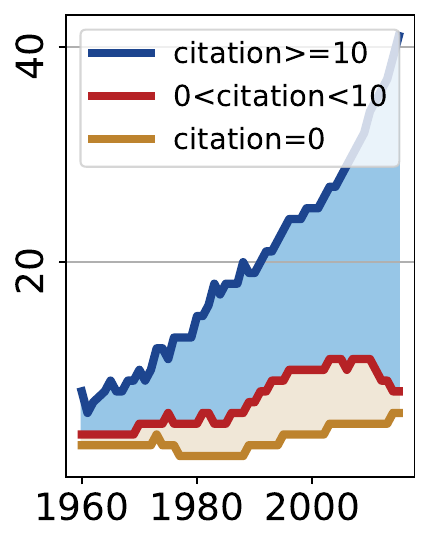}}
\quad
\subfigure[Age \label{sfig:refAgeEnvironmental}]{\includegraphics[width=0.3605\textwidth]{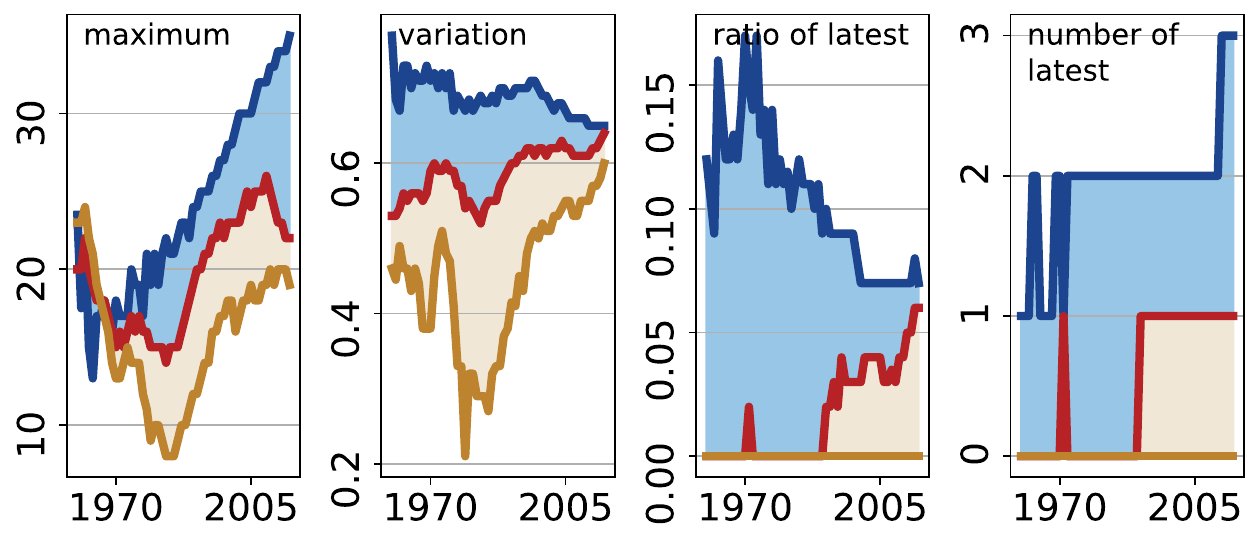}}
\quad
\subfigure[Citation count \label{sfig:refCitationEnvironmental}] {\includegraphics[width=0.45\textwidth]{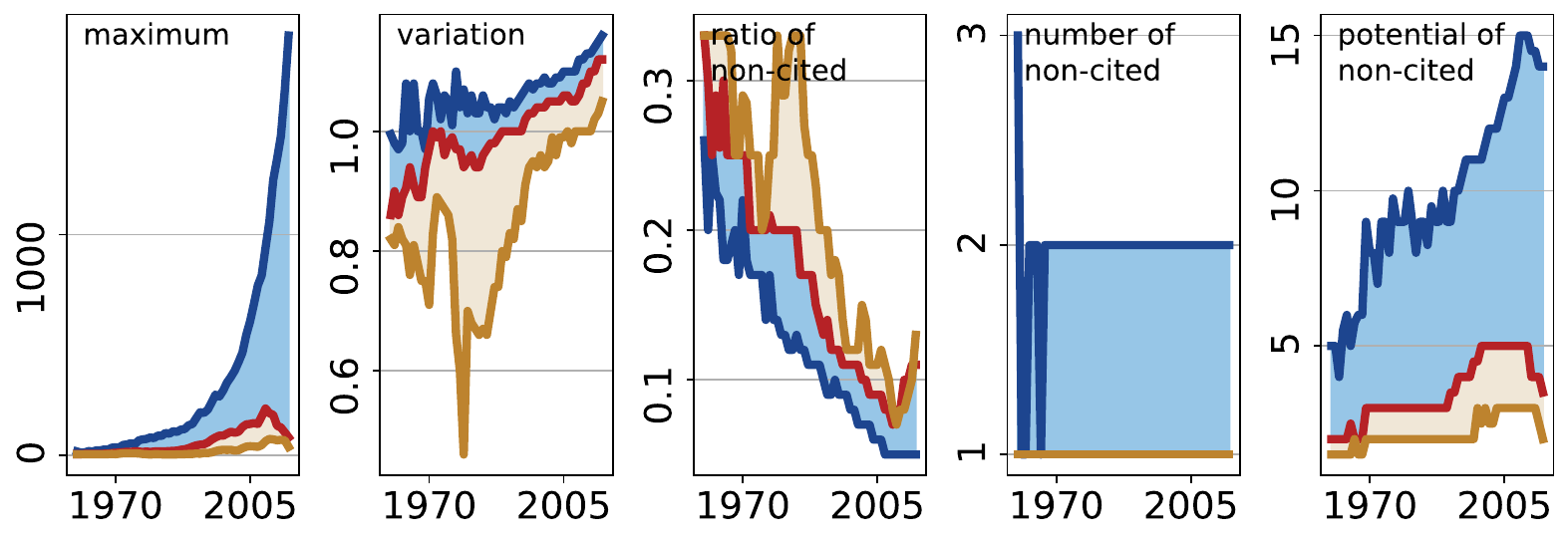}}

\caption{The indicators of reference patten of Environmental science from 1960 to 2015. (a) indicates the median number of references per year, from 1960 to 2015. (b) presents the reference age patterns year by year from 1960 to 2015. Specifically, the 4 sub-figures in (b) plot the maximum reference age, variation ratio of age, ratio of latest papers in reference, and the absolute number of latest papers in references, respectively. (c) presents the citation count of references. Particularly, the 5 sub-figures in (c) show maximum of citation counts, variation ratio of citation counts, ratio of non-cited publications in references, absolute number of non-cited publications in references, and the potential (number of citations) of the non-cited references, respectively.}
\label{fig:referencePatternEnvironmentalScience}
\end{figure*}

\begin{figure*}[ht]
\centering
\subfigure[Number \label{sfig:refNumBusiness}] {\includegraphics[width=0.1233\textwidth]{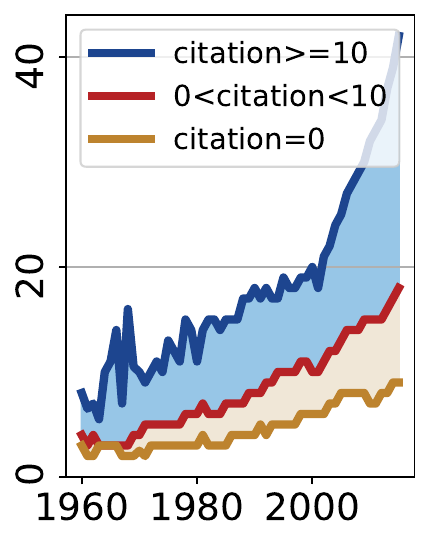}}
\quad
\subfigure[Age \label{sfig:refAgeBusiness}]{\includegraphics[width=0.3605\textwidth]{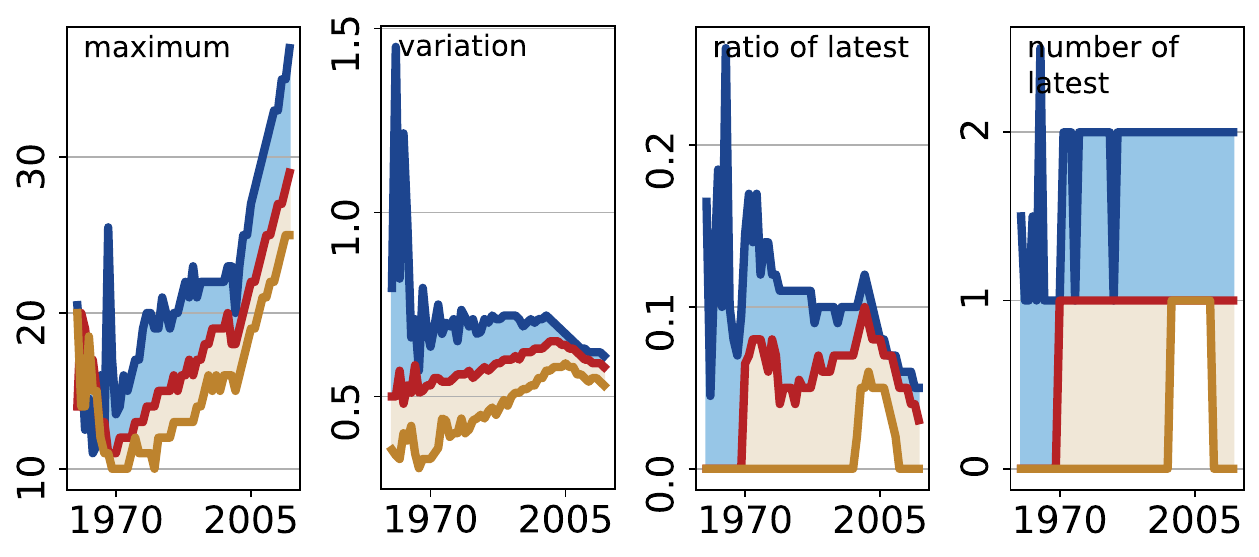}}
\quad
\subfigure[Citation count \label{sfig:refCitationBusiness}] {\includegraphics[width=0.45\textwidth]{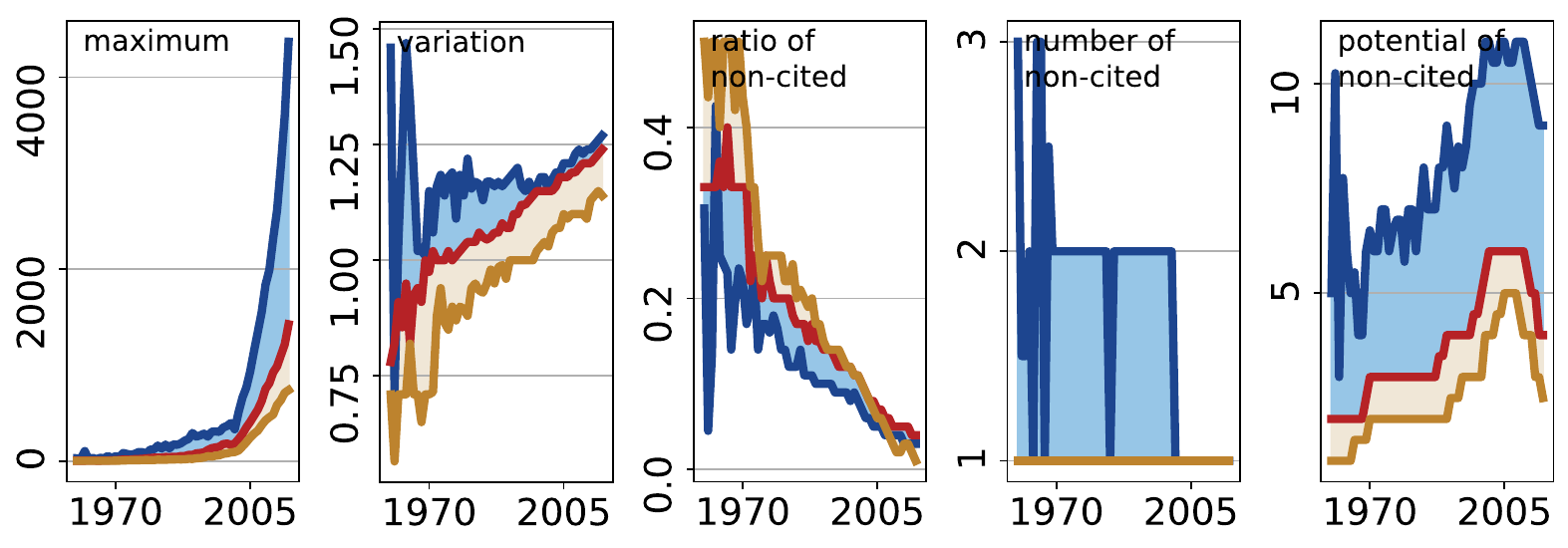}}
\caption{The indicators of reference pattern of Business from 1960 to 2015.}
\label{fig:referencePatternBusiness}
\end{figure*}

\begin{figure*}[ht]
\centering
\subfigure[Number \label{sfig:refNumGeography}] {\includegraphics[width=0.1233\textwidth]{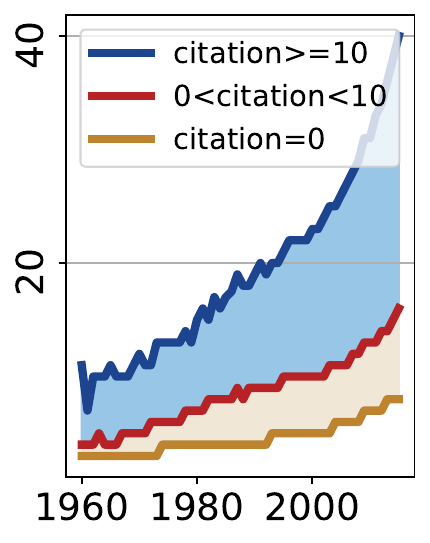}}
\quad
\subfigure[Age \label{sfig:refAgeGeography}]{\includegraphics[width=0.3605\textwidth]{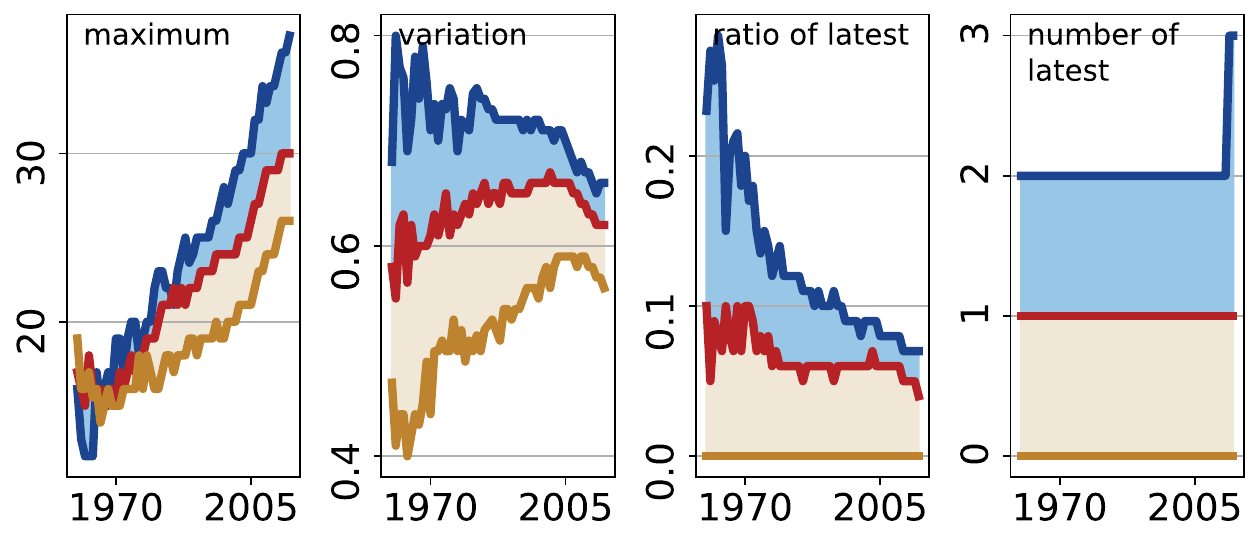}}
\quad
\subfigure[Citation count \label{sfig:refCitationGeography}] {\includegraphics[width=0.45\textwidth]{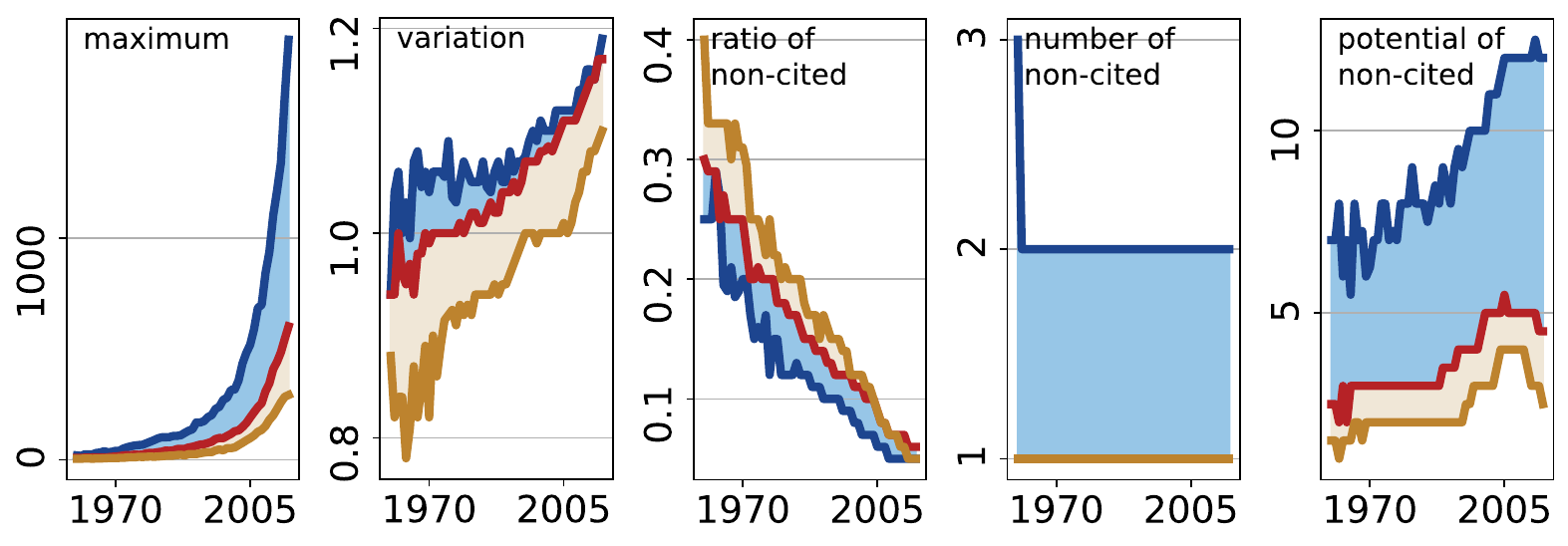}}
\caption{The indicators of reference pattern of Geography from 1960 to 2015.}
\label{fig:referencePatternGeography}
\end{figure*}

\begin{figure*}[ht]
\centering
\subfigure[Number \label{sfig:refNumPolitical}] {\includegraphics[width=0.1233\textwidth]{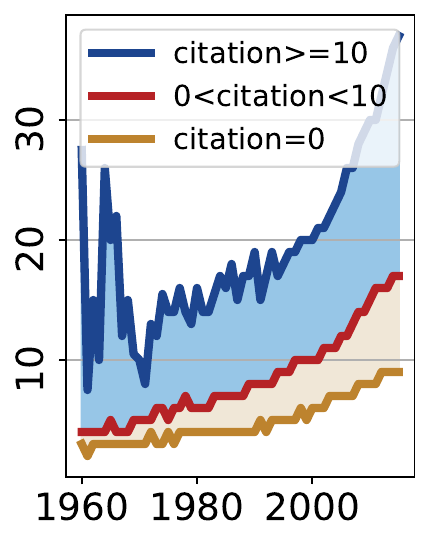}}
\quad
\subfigure[Age \label{sfig:refAgePolitical}]{\includegraphics[width=0.3605\textwidth]{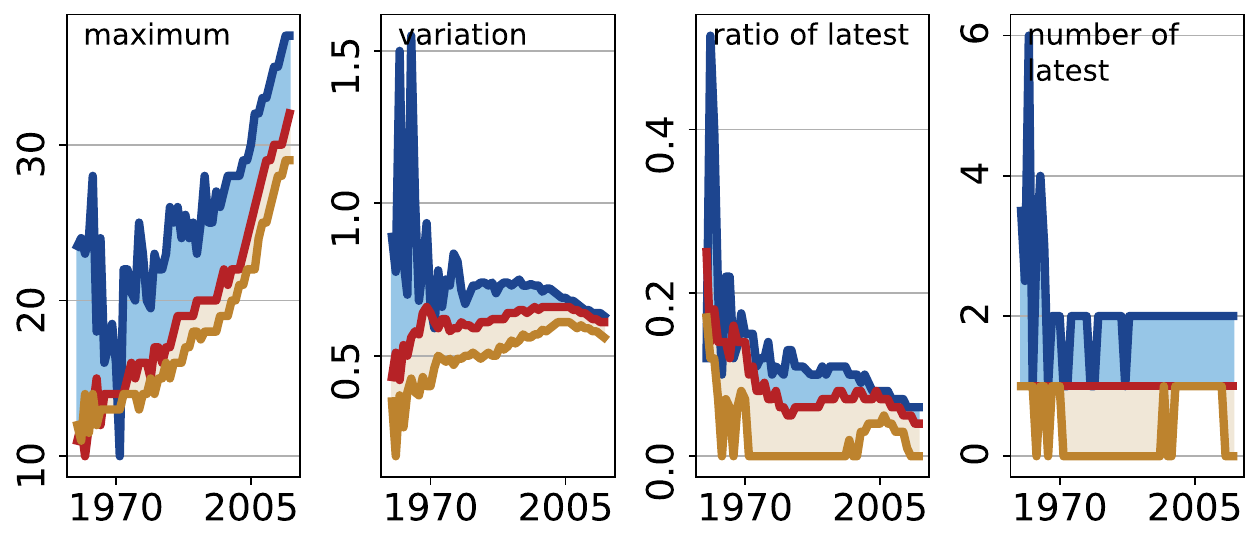}}
\quad
\subfigure[Citation count \label{sfig:refCitationPolitical}] {\includegraphics[width=0.45\textwidth]{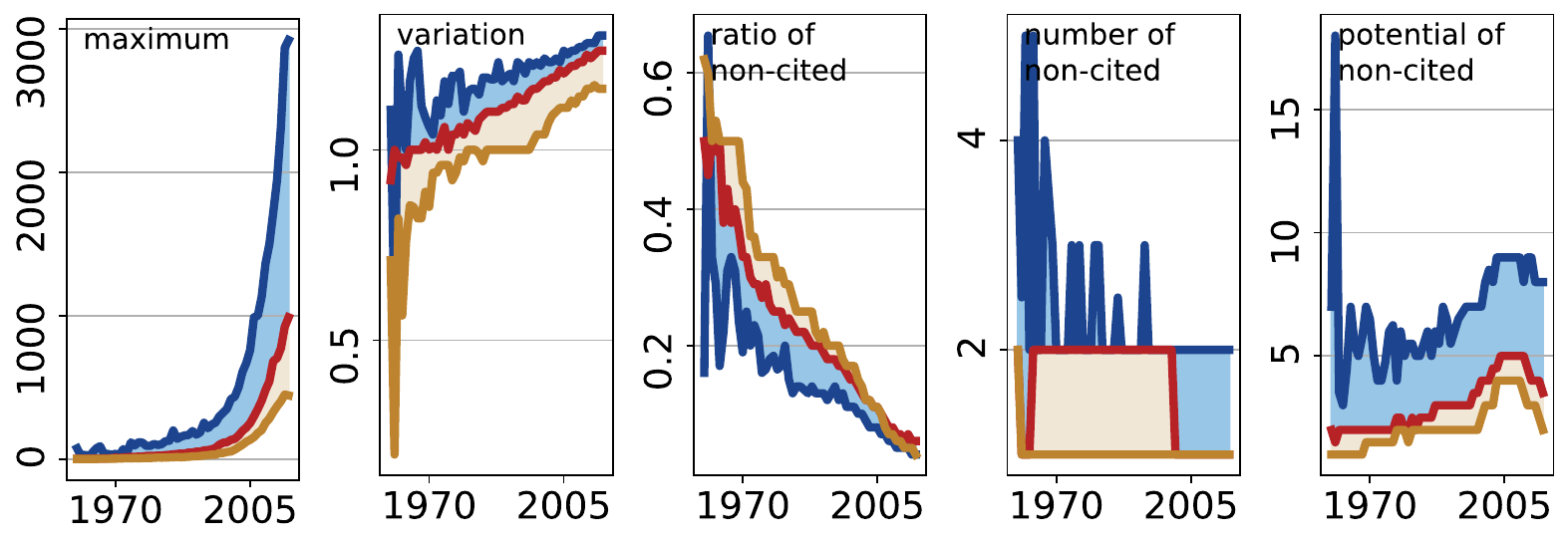}}
\caption{The indicators of reference pattern of Political science from 1960 to 2015.}
\label{fig:referencePatternPoliticalScience}
\end{figure*}

\begin{figure*}[ht]
\centering
\subfigure[Number \label{sfig:refNumPopulation}] {\includegraphics[width=0.1233\textwidth]{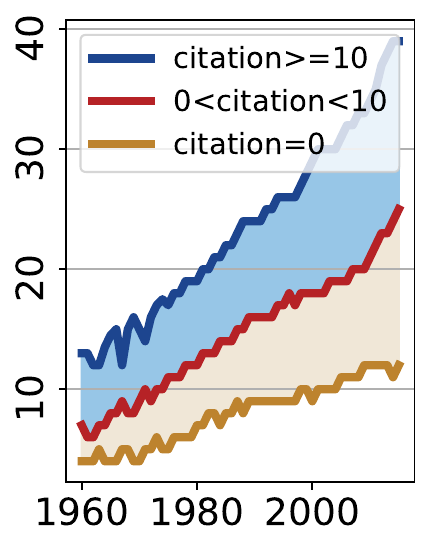}}
\quad
\subfigure[Age \label{sfig:refAgePopulation}]{\includegraphics[width=0.3605\textwidth]{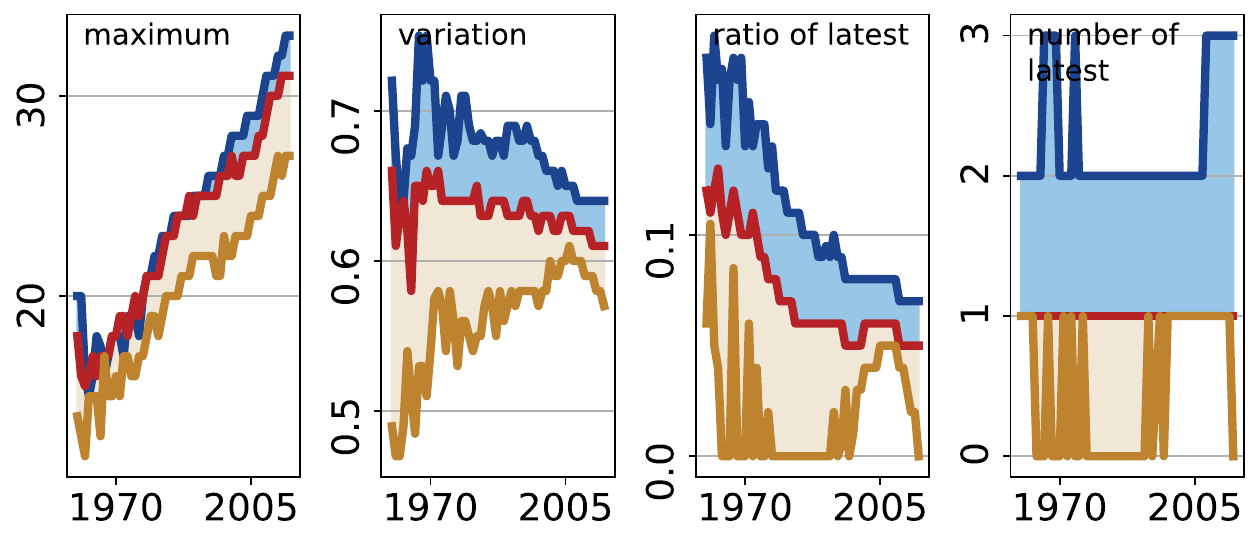}}
\quad
\subfigure[Citation count \label{sfig:refCitationPopulation}] {\includegraphics[width=0.45\textwidth]{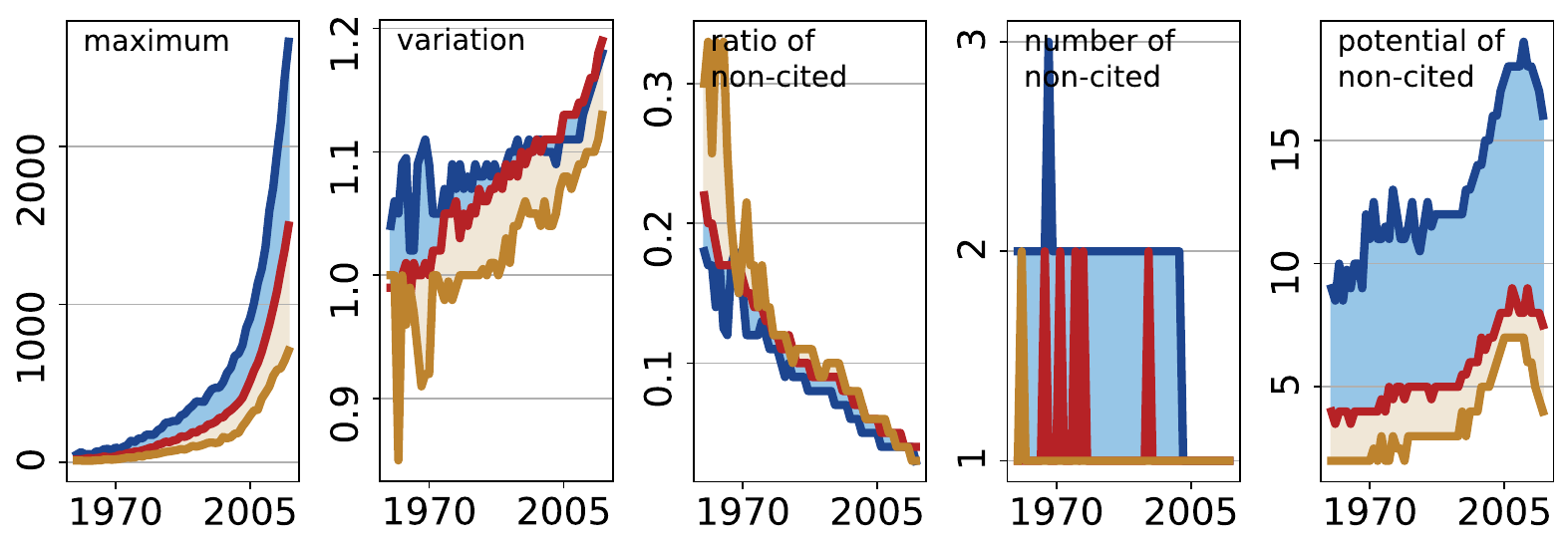}}
\caption{The indicators of reference pattern of Population from 1960 to 2015. (a) indicates the median number of references per year, from 1960 to 2015. (b) presents the reference age patterns year by year from 1960 to 2015. Specifically, the 4 sub-figures in (b) plot the maximum reference age, variation ratio of age, ratio of latest papers in reference, and the absolute number of latest papers in references, respectively. (c) presents the citation count of references. Particularly, the 5 sub-figures in (c) show maximum of citation counts, variation ratio of citation counts, ratio of non-cited publications in references, absolute number of non-cited publications in references, and the potential (number of citations) of the non-cited references, respectively.}
\label{fig:referencePatternPopulation}
\end{figure*}

\begin{figure*}[ht]
\centering
\subfigure[Number \label{sfig:refNumPhilosophy}] {\includegraphics[width=0.1233\textwidth]{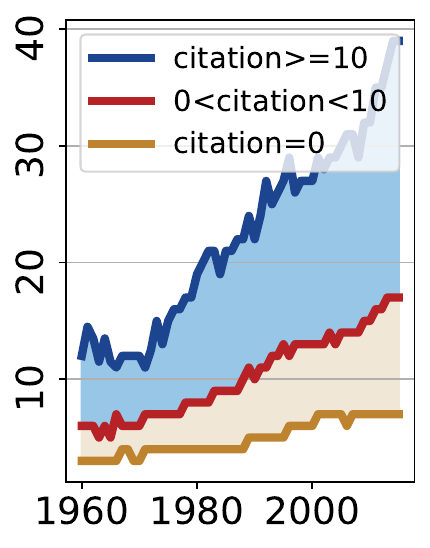}}
\quad
\subfigure[Age \label{sfig:refAgePhilosophy}]{\includegraphics[width=0.3605\textwidth]{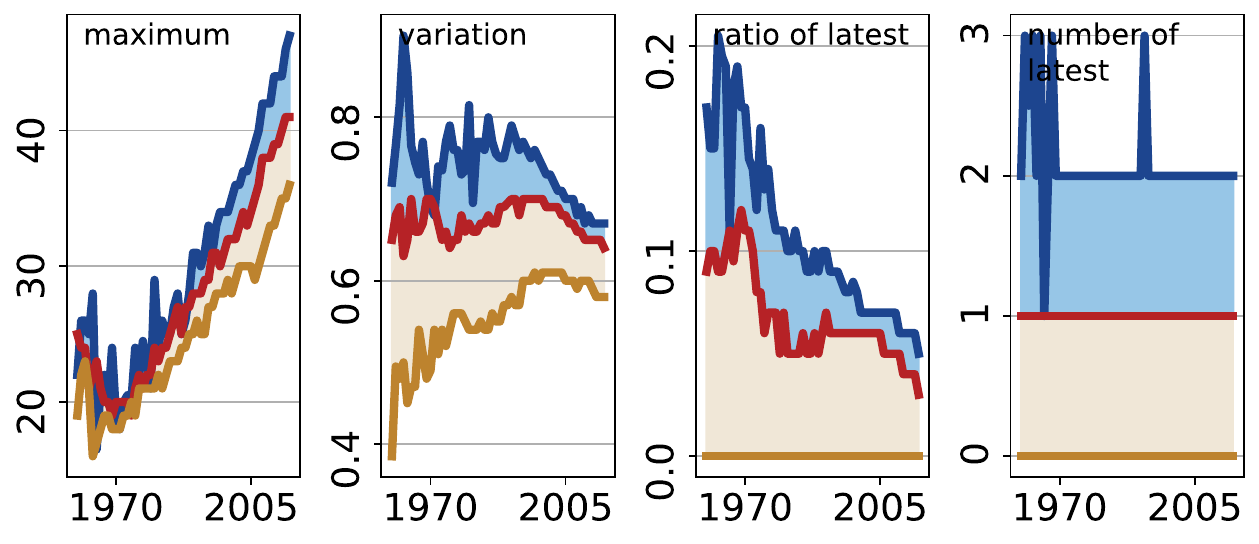}}
\quad
\subfigure[Citation count \label{sfig:refCitationPhilosophy}] {\includegraphics[width=0.45\textwidth]{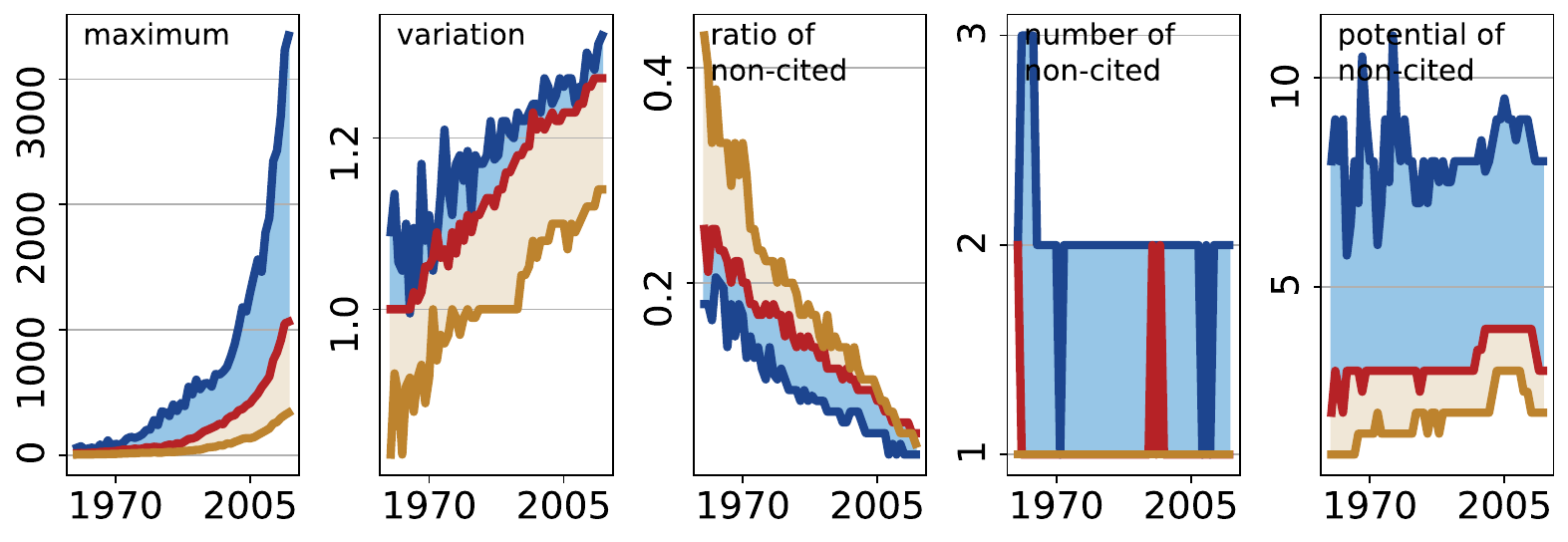}}
\caption{The indicators of reference pattern of Philosophy from 1960 to 2015.}
\label{fig:referencePatternPhilosophy}
\end{figure*}

\begin{figure*}[ht]
\centering
\subfigure[Number \label{sfig:refNumArt}] {\includegraphics[width=0.1233\textwidth]{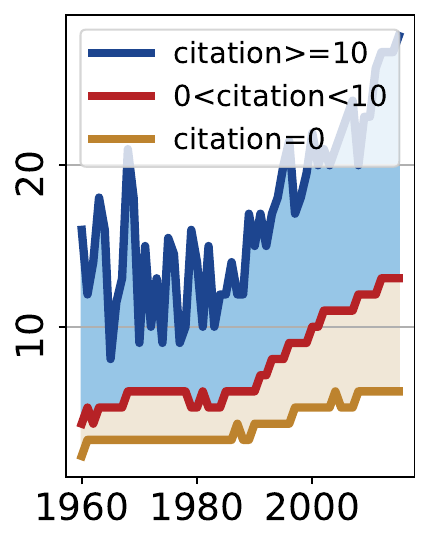}}
\quad
\subfigure[Age \label{sfig:refAgeArt}]{\includegraphics[width=0.3605\textwidth]{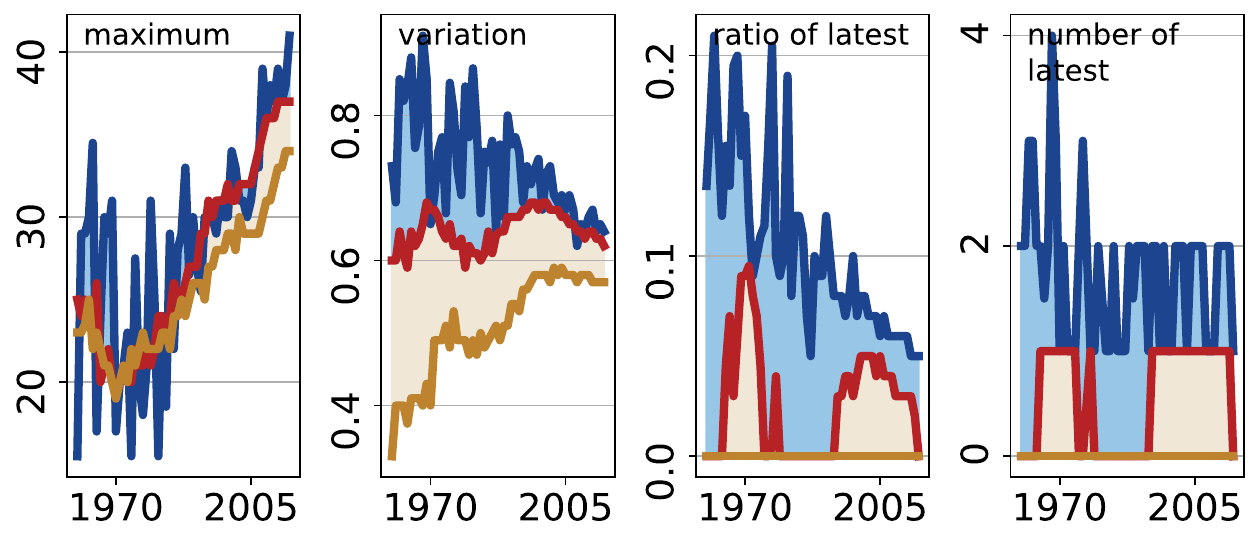}}
\quad
\subfigure[Citation count \label{sfig:refCitationArt}] {\includegraphics[width=0.45\textwidth]{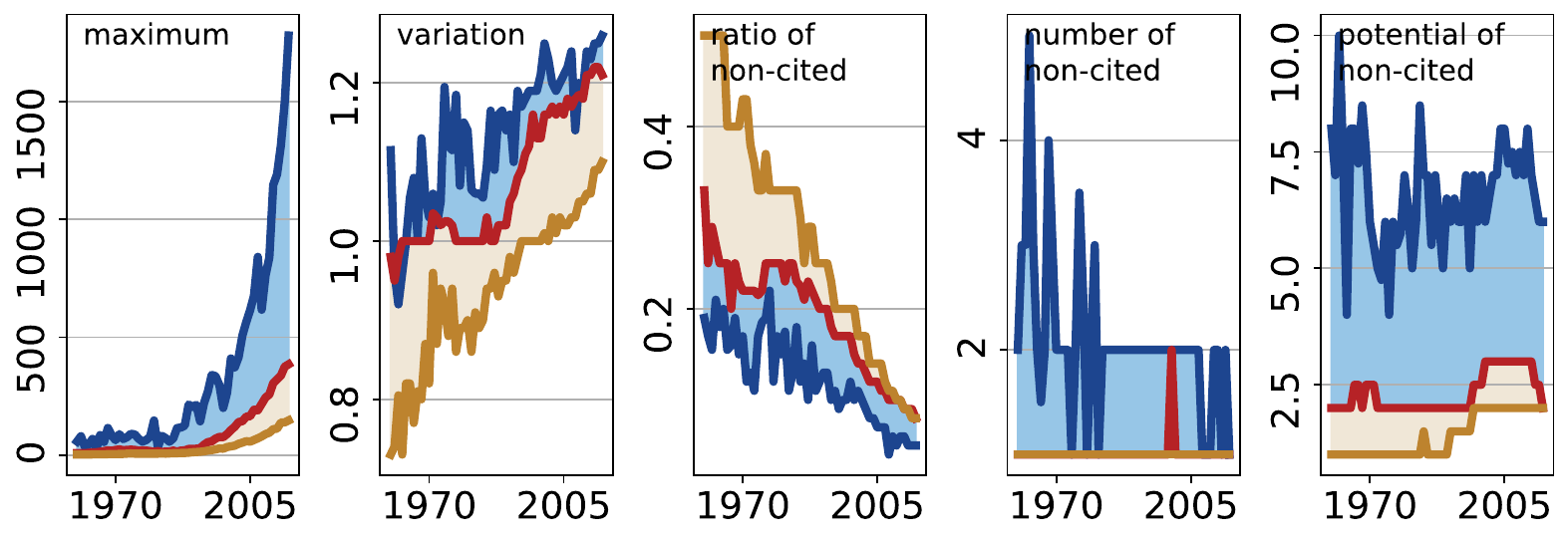}}
\caption{The indicators of reference pattern of Art from 1960 to 2015.}
\label{fig:referencePatternArt}
\end{figure*}

\begin{figure*}[ht]
\centering
\subfigure[Number \label{sfig:refNumHistory}] {\includegraphics[width=0.1233\textwidth]{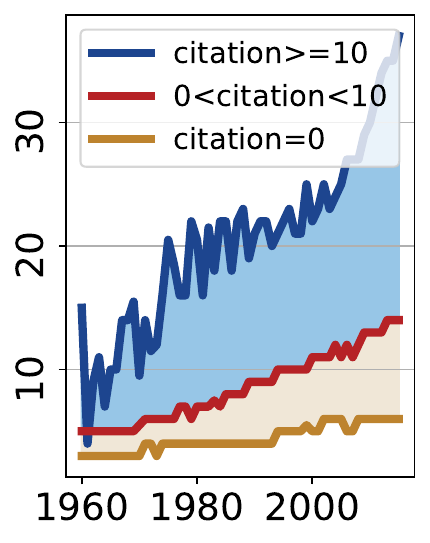}}
\quad
\subfigure[Age \label{sfig:refAgeHistory}]{\includegraphics[width=0.3605\textwidth]{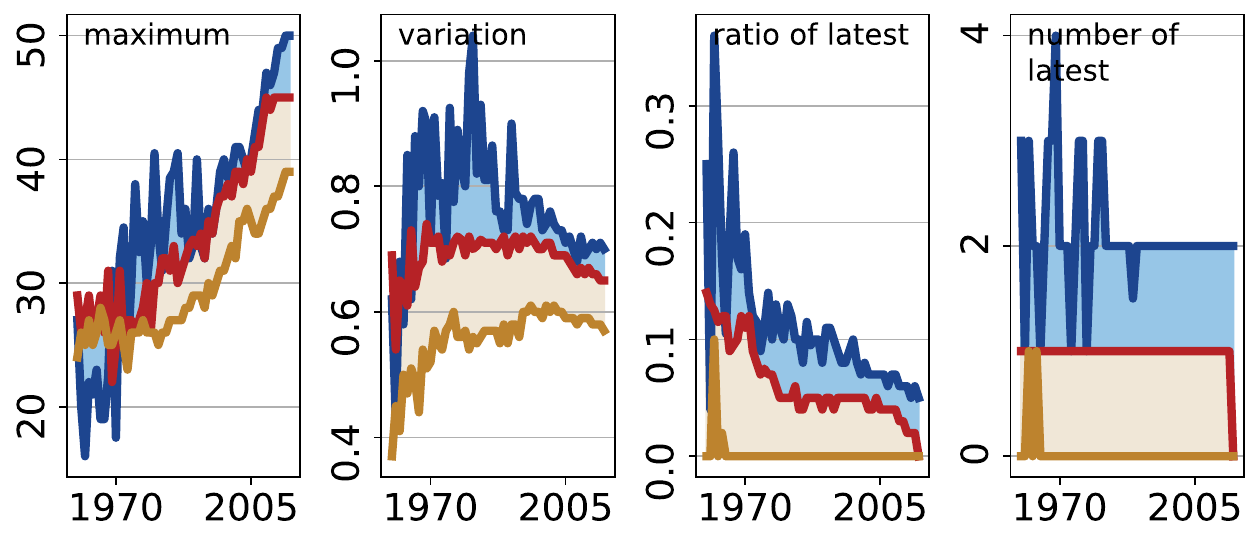}}
\quad
\subfigure[Citation count \label{sfig:refCitationHistory}] {\includegraphics[width=0.45\textwidth]{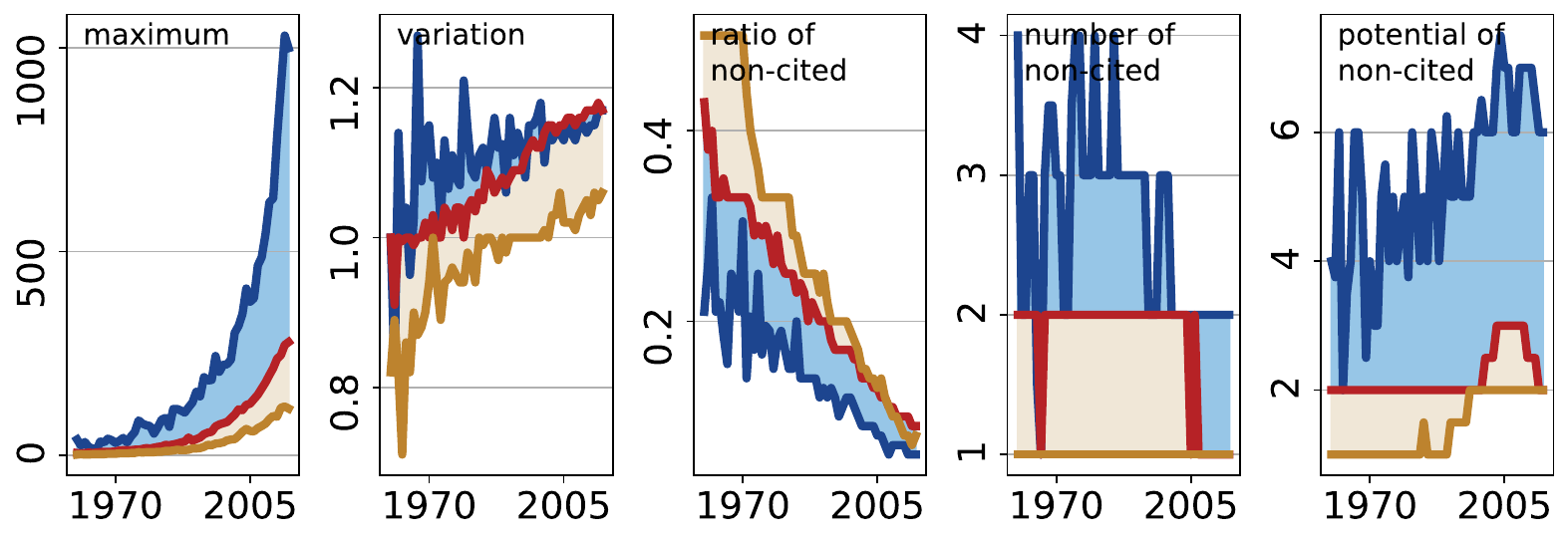}}
\caption{The indicators of reference pattern of History from 1960 to 2015.}
\label{fig:referencePatternHistory}
\end{figure*}

Figs. \ref{fig:referencePatternMedicine} to \ref{fig:referencePatternHistory} present reference pattern of each field of study, respectively. The reference pattern consists of the number of references, age features of references, and citation features of references. The figures are in the same order as the fields listed in Table 1 in the main text.

Table~\ref{tab:olsRegResAll} presents OLS regression results based on all papers. Tables \ref{tab:olsRegResChemistry} to \ref{tab:olsRegResBusiness} present OLS regression results based on papers in each field of study, respectively. The tables are in the same order as the fields listed in  Fig. 4 (a) in the main text. Each table includes list of independent variables and their coefficients. We also report standard error, t-Value, and p-Value of each coefficient. Moreover, $R^2$ and adjusted $R^2$ are employed to measure the proportion of the variance for a dependent variable that's explained by an independent variable in a regression model. The significance of the regression model is evaluated by $F$ test.

\begin{table*}[ht]
  \centering
  \caption{OLS regression results based on all papers.}
    \begin{tabular}{lrrrr}
    \toprule
    \multicolumn{1}{l|}{Variable} & \multicolumn{1}{l}{Coefficient} & \multicolumn{1}{l}{Std Err} & \multicolumn{1}{l}{t Value} & \multicolumn{1}{l}{p Value} \\
    \midrule
    \multicolumn{1}{l|}{Intercept} & 0.4617 & 0.001 & 356.886 & 0.000 \\
    \multicolumn{1}{l|}{\textbf{Number of accumulative publications} ($\bm{log_{10}}$)} & \textbf{-0.0279} & \textbf{0.000} & \textbf{-156.132} & \textbf{0.000} \\
    \multicolumn{1}{l|}{Reference number} & 0.0033 & 2.06E-06 & 1600.934 & 0.000 \\
    \multicolumn{1}{l|}{Maximum age of all references} & -0.0012 & 3.13E-06 & -378.115 & 0.000 \\
    \multicolumn{1}{l|}{Variation ratio of ages} & 0.0033 & 2.82E-05 & 117.310 & 0.000 \\
    \multicolumn{1}{l|}{Ratio of 1-year-old references} & 0.2787 & 0.000 & 796.039 & 0.000 \\
    \multicolumn{1}{l|}{Maximum of references' citation counts ($log_{10}$)} & 0.1350 & 8.69E-05 & 1554.214 & 0.000 \\
    \multicolumn{1}{l|}{Variation ratio of references' citation counts} & 0.0041 & 0.000     & 29.055 & 0.000 \\
    \multicolumn{1}{l|}{Ratio of non-cited references} & -0.2182 & 0.000     & -626.243 & 0.000 \\
    \midrule
    \multicolumn{1}{l|}{$R^2=0.184$, adjusted $R^2=0.184$.} & \multicolumn{4}{r}{$F=1.717E+06$, significance F: $p=0.00$.}  \\
    \bottomrule
    \end{tabular}%
  \label{tab:olsRegResAll}%
\end{table*}%

\begin{table*}[ht]
  \centering
  \caption{OLS regression results based on papers of Chemistry.}
    \begin{tabular}{lrrrr}
    \toprule
    \multicolumn{1}{l|}{Variable} & \multicolumn{1}{l}{Coefficient} & \multicolumn{1}{l}{Std Err} & \multicolumn{1}{l}{t Value} & \multicolumn{1}{l}{p Value} \\
    \midrule
    \multicolumn{1}{l|}{Intercept} & 0.0519 & 0.003 & 19.669 & 0.000 \\
    \multicolumn{1}{l|}{\textbf{Number of accumulative publications} ($\bm{log_{10}}$)} & \textbf{0.0329} & \textbf{0.000}     & \textbf{89.305} & \textbf{0.000} \\
    \multicolumn{1}{l|}{Reference number} & 0.0035 & 4.96E-06 & 705.231 & 0.000 \\
    \multicolumn{1}{l|}{Maximum age of all references} & -0.0005 & 8.24E-06 & -62.461 & 0.000 \\
    \multicolumn{1}{l|}{Variation ratio of ages} & 0.0034 & 5.75E-05 & 58.842 & 0.000 \\
    \multicolumn{1}{l|}{Ratio of 1-year-old references} & 0.2144 & 0.001 & 263.634 & 0.000 \\
    \multicolumn{1}{l|}{Maximum of references' citation counts ($log_{10}$)} & 0.1158 & 0.000 & 618.567 & 0.000 \\
    \multicolumn{1}{l|}{Variation ratio of references' citation counts} & 0.0654 & 0.000     & 203.215 & 0.000 \\
    \multicolumn{1}{l|}{Ratio of non-cited references} & -0.3245 & 0.001 & -381.242 & 0.000 \\
    \midrule
    \multicolumn{1}{l|}{$R^2=0.242$, adjusted $R^2=0.242$.} & \multicolumn{4}{r}{$F=4.326E+05$, significance F: $p=0.00$.}  \\
    \bottomrule
    \end{tabular}%
  \label{tab:olsRegResChemistry}%
\end{table*}%

\begin{table*}[ht]
  \centering
  \caption{OLS regression results based on papers of Materials science.}
    \begin{tabular}{lrrrr}
    \toprule
    \multicolumn{1}{l|}{Variable} & \multicolumn{1}{l}{Coefficient} & \multicolumn{1}{l}{Std Err} & \multicolumn{1}{l}{t Value} & \multicolumn{1}{l}{p Value} \\
    \midrule
    \multicolumn{1}{l|}{Intercept} & -0.0619 & 0.003 & -22.603 & 0.000 \\
    \multicolumn{1}{l|}{\textbf{Number of accumulative publications} ($\bm{log_{10}}$)} & \textbf{0.0327} & \textbf{0.000}     & \textbf{86.134} & \textbf{0.000} \\
    \multicolumn{1}{l|}{Reference number} & 0.0028 & 5.00E-06 & 555.138 & 0.000 \\
    \multicolumn{1}{l|}{Maximum age of all references} & -0.0003 & 8.19E-06 & -36.361 & 0.000 \\
    \multicolumn{1}{l|}{Variation ratio of ages} & 0.0092 & 0.000 & 74.439 & 0.000 \\
    \multicolumn{1}{l|}{Ratio of 1-year-old references} & 0.2692 & 0.001 & 362.972 & 0.000 \\
    \multicolumn{1}{l|}{Maximum of references' citation counts ($log_{10}$)} & 0.1643 & 0.000 & 773.663 & 0.000\\
    \multicolumn{1}{l|}{Variation ratio of references' citation counts} & -0.0003 & 0.000     & -1.024 & 0.306 \\
    \multicolumn{1}{l|}{Ratio of non-cited references} & -0.0794 & 0.001 & -120.904 & 0.000 \\
    \midrule
    \multicolumn{1}{l|}{$R^2=0.216$, adjusted $R^2=0.216$.} & \multicolumn{4}{r}{$F=3.424E+05$, significance F: $p=0.00$.}  \\
    \bottomrule
    \end{tabular}%
  \label{tab:olsRegResMaterialsScience}%
\end{table*}%

\begin{table*}[ht]
  \centering
  \caption{OLS regression results based on papers of Environmental science.}
    \begin{tabular}{lrrrr}
    \toprule
    \multicolumn{1}{l|}{Variable} & \multicolumn{1}{l}{Coefficient} & \multicolumn{1}{l}{Std Err} & \multicolumn{1}{l}{t Value} & \multicolumn{1}{l}{p Value} \\
    \midrule
    \multicolumn{1}{l|}{Intercept} & 0.0424 & 0.008 & 5.223 & 0.000 \\
    \multicolumn{1}{l|}{\textbf{Number of accumulative publications} ($\bm{log_{10}}$)} & \textbf{0.0227} & \textbf{0.001} & \textbf{20.429} & \textbf{0.000} \\
    \multicolumn{1}{l|}{Reference number} & 0.0045 & 1.46E-05 & 306.933 & 0.000 \\
    \multicolumn{1}{l|}{Maximum age of all references} & -0.0005 & 2.08E-05 & -25.129 & 0.000 \\
    \multicolumn{1}{l|}{Variation ratio of ages} & 0.0032 & 0.000 & 17.493 & 0.000 \\
    \multicolumn{1}{l|}{Ratio of 1-year-old references} & 0.2617 & 0.002 & 127.077 & 0.000 \\
    \multicolumn{1}{l|}{Maximum of references' citation counts ($log_{10}$)} & 0.1329 & 0.001 & 251.290 & 0.000 \\
    \multicolumn{1}{l|}{Variation ratio of references' citation counts} & -0.0051 & 0.001 & -6.441 & 0.000 \\
    \multicolumn{1}{l|}{Ratio of non-cited references} & -0.1162 & 0.002 & -63.225 & 0.000 \\
    \midrule
    \multicolumn{1}{l|}{$R^2=0.229$, adjusted $R^2=0.229$.} & \multicolumn{4}{r}{$F=6.129E+04$, significance F: $p=0.00$.}  \\
    \bottomrule
    \end{tabular}%
  \label{tab:olsRegResEnvironmentalScience}%
\end{table*}%

\begin{table*}[ht]
  \centering
  \caption{OLS regression results based on papers of Biology.}
    \begin{tabular}{lrrrr}
    \toprule
    \multicolumn{1}{l|}{Variable} & \multicolumn{1}{l}{Coefficient} & \multicolumn{1}{l}{Std Err} & \multicolumn{1}{l}{t Value} & \multicolumn{1}{l}{p Value} \\
    \midrule
    \multicolumn{1}{l|}{Intercept} & 0.3002 & 0.003 & 96.905 & 0.000 \\
    \multicolumn{1}{l|}{\textbf{Number of accumulative publications} ($\bm{log_{10}}$)} & \textbf{0.0147} & \textbf{0.000}     & \textbf{34.308} & \textbf{0.000} \\
    \multicolumn{1}{l|}{Reference number} & 0.0031 & 4.21E-06 & 729.619 & 0.000 \\
    \multicolumn{1}{l|}{Maximum age of all references} & -0.0013 & 7.17E-06 & -182.970 & 0.000 \\
    \multicolumn{1}{l|}{Variation ratio of ages} & 0.0012 & 4.69E-05 & 25.262 & 0.000 \\
    \multicolumn{1}{l|}{Ratio of 1-year-old references} & 0.6139 & 0.001 & 546.084 & 0.000 \\
    \multicolumn{1}{l|}{Maximum of references' citation counts ($log_{10}$)} & 0.1302 & 0.000 & 653.478 & 0.000 \\
    \multicolumn{1}{l|}{Variation ratio of references' citation counts} & -0.0148 & 0.000     & -37.927 & 0.000 \\
    \multicolumn{1}{l|}{Ratio of non-cited references} & -0.5276 & 0.001 & -435.058 & 0.000 \\
    \midrule
    \multicolumn{1}{l|}{$R^2=0.171$, adjusted $R^2=0.171$.} & \multicolumn{4}{r}{$F=3.025E+05$, significance F: $p=0.00$.}  \\
    \bottomrule
    \end{tabular}%
  \label{tab:olsRegResBiology}%
\end{table*}%

\begin{table*}[ht]
  \centering
  \caption{OLS regression results based on papers of Engineering.}
    \begin{tabular}{lrrrr}
    \toprule
    \multicolumn{1}{l|}{Variable} & \multicolumn{1}{l}{Coefficient} & \multicolumn{1}{l}{Std Err} & \multicolumn{1}{l}{t Value} & \multicolumn{1}{l}{p Value} \\
    \midrule
    \multicolumn{1}{l|}{Intercept} & 0.1863 & 0.003 & 67.531 & 0.000 \\
    \multicolumn{1}{l|}{\textbf{Number of accumulative publications} ($\bm{log_{10}}$)} & \textbf{0.0110} & \textbf{0.000}     & \textbf{28.781} & \textbf{0.000} \\
    \multicolumn{1}{l|}{Reference number} & 0.0024 & 3.95E-06 & 598.825 & 0.000 \\
    \multicolumn{1}{l|}{Maximum age of all references} & -0.0002 & 6.80E-06 & -22.380 & 0.000 \\
    \multicolumn{1}{l|}{Variation ratio of ages} & 0.0104 & 0.000 & 92.364 & 0.000 \\
    \multicolumn{1}{l|}{Ratio of 1-year-old references} & 0.2730 & 0.001 & 402.806 & 0.000 \\
    \multicolumn{1}{l|}{Maximum of references' citation counts ($log_{10}$)} & 0.1045 & 0.000 & 561.975 & 0.000 \\
    \multicolumn{1}{l|}{Variation ratio of references' citation counts} & 0.0105 & 0.000     & 37.970 & 0.000 \\
    \multicolumn{1}{l|}{Ratio of non-cited references} & -0.1666 & 0.001 & -281.394 & 0.000 \\
    \midrule
    \multicolumn{1}{l|}{$R^2=0.144$, adjusted $R^2=0.144$.} & \multicolumn{4}{r}{$F=2.602E+05$, significance $F=0.00$.}  \\
    \bottomrule
    \end{tabular}%
  \label{tab:olsRegResEngineering}%
\end{table*}%

\begin{table*}[ht]
  \centering
  \caption{OLS regression results based on papers of Physics.}
    \begin{tabular}{lrrrr}
    \toprule
    \multicolumn{1}{l|}{Variable} & \multicolumn{1}{l}{Coefficient} & \multicolumn{1}{l}{Std Err} & \multicolumn{1}{l}{t Value} & \multicolumn{1}{l}{p Value} \\
    \midrule
    \multicolumn{1}{l|}{Intercept} & 0.3012 & 0.003 & 98.272 & 0.000 \\
    \multicolumn{1}{l|}{\textbf{Number of accumulative publications} ($\bm{log_{10}}$)} & \textbf{-0.0107} & \textbf{0.000}     & \textbf{-24.981} & \textbf{0.000} \\
    \multicolumn{1}{l|}{Reference number} & 0.0031 & 5.37E-06 & 571.265 & 0.000 \\
    \multicolumn{1}{l|}{Maximum age of all references} & -0.0001 & 8.52E-06 & -12.950 & 0.000 \\
    \multicolumn{1}{l|}{Variation ratio of ages} & 0.0054 & 9.96E-05 & 54.699 & 0.000 \\
    \multicolumn{1}{l|}{Ratio of 1-year-old references} & 0.3488 & 0.001 & 407.016 & 0.000 \\
    \multicolumn{1}{l|}{Maximum of references' citation counts ($log_{10}$)} & 0.1225 & 0.000 & 508.121 & 0.000 \\
    \multicolumn{1}{l|}{Variation ratio of references' citation counts} & 0.0106 & 0.000     & 30.115 & 0.000 \\
    \multicolumn{1}{l|}{Ratio of non-cited references} & -0.1501 & 0.001 & -176.068 & 0.000 \\
    \midrule
    \multicolumn{1}{l|}{$R^2=0.160$, adjusted $R^2=0.160$.} & \multicolumn{4}{r}{$F=2.107E+05$, significance F: $p=0.00$.}  \\
    \bottomrule
    \end{tabular}%
  \label{tab:olsRegResPhysics}%
\end{table*}%

\begin{table*}[ht]
  \centering
  \caption{OLS regression results based on papers of Mathematics.}
    \begin{tabular}{lrrrr}
    \toprule
    \multicolumn{1}{l|}{Variable} & \multicolumn{1}{l}{Coefficient} & \multicolumn{1}{l}{Std Err} & \multicolumn{1}{l}{t Value} & \multicolumn{1}{l}{p Value} \\
    \midrule
    \multicolumn{1}{l|}{Intercept} & 0.4262 & 0.004 & 102.293 & 0.000 \\
    \multicolumn{1}{l|}{\textbf{Number of accumulative publications} ($\bm{log_{10}}$)} & \textbf{-0.0219} & \textbf{0.001} & \textbf{-37.718} & \textbf{0.000} \\
    \multicolumn{1}{l|}{Reference number} & 0.0055 & 9.82E-06 & 564.238 & 0.000 \\
    \multicolumn{1}{l|}{Maximum age of all references} & -0.0007 & 9.39E-06 & -72.352 & 0.000 \\
    \multicolumn{1}{l|}{Variation ratio of ages} & 0.0036 & 0.000 & 35.612 & 0.000 \\
    \multicolumn{1}{l|}{Ratio of 1-year-old references} & 0.3362 & 0.001 & 283.376 & 0.000 \\
    \multicolumn{1}{l|}{Maximum of references' citation counts ($log_{10}$)} & 0.0668 & 0.000 & 234.011 & 0.000 \\
    \multicolumn{1}{l|}{Variation ratio of references' citation counts} & 0.0206 & 0.000     & 47.926 & 0.000 \\
    \multicolumn{1}{l|}{Ratio of non-cited references} & -0.2142 & 0.001 & -184.396 & 0.000 \\
    \midrule
    \multicolumn{1}{l|}{$R^2=0.123$, adjusted $R^2=0.123$.} & \multicolumn{4}{r}{$F=1.068E+05$, significance F: $p=0.00$.}  \\
    \bottomrule
    \end{tabular}%
  \label{tab:olsRegResMathematics}%
\end{table*}%

\begin{table*}[ht]
  \centering
  \caption{OLS regression results based on papers of History.}
    \begin{tabular}{lrrrr}
    \toprule
    \multicolumn{1}{l|}{Variable} & \multicolumn{1}{l}{Coefficient} & \multicolumn{1}{l}{Std Err} & \multicolumn{1}{l}{t Value} & \multicolumn{1}{l}{p Value} \\
    \midrule
    \multicolumn{1}{l|}{Intercept} & 0.2692 & 0.010  & 26.726 & 0.000 \\
    \multicolumn{1}{l|}{\textbf{Number of accumulative publications} ($\bm{log_{10}}$)} & \textbf{-0.0250} & \textbf{0.001} & \textbf{-18.006} & \textbf{0.000} \\
    \multicolumn{1}{l|}{Reference number} & 0.0007 & 1.11E-05 & 66.107 & 0.000 \\
    \multicolumn{1}{l|}{Maximum age of all references} & -0.0002 & 1.28E-05 & -14.309 & 0.000 \\
    \multicolumn{1}{l|}{Variation ratio of ages} & 0.0009 & 0.000 & 6.026 & 0.000 \\
    \multicolumn{1}{l|}{Ratio of 1-year-old references} & 0.1385 & 0.002 & 56.064 & 0.000 \\
    \multicolumn{1}{l|}{Maximum of references' citation counts ($log_{10}$)} & 0.0859 & 0.001 & 118.788 & 0.000 \\
    \multicolumn{1}{l|}{Variation ratio of references' citation counts} & -0.0024 & 0.001 & -2.456 & 0.014 \\
    \multicolumn{1}{l|}{Ratio of non-cited references} & -0.0384 & 0.002 & -17.473 & 0.000 \\
    \midrule
    \multicolumn{1}{l|}{$R^2=0.075$, adjusted $R^2=0.075$.} & \multicolumn{4}{r}{$F=5842.0$, significance F: $p=0.00$.}  \\
    \bottomrule
    \end{tabular}%
  \label{tab:olsRegResHistory}%
\end{table*}%

\begin{table*}[ht]
  \centering
  \caption{OLS regression results based on papers of Geology.}
    \begin{tabular}{lrrrr}
    \toprule
    \multicolumn{1}{l|}{Variable} & \multicolumn{1}{l}{Coefficient} & \multicolumn{1}{l}{Std Err} & \multicolumn{1}{l}{t Value} & \multicolumn{1}{l}{p Value} \\
    \midrule
    \multicolumn{1}{l|}{Intercept} & 0.4405 & 0.005 & 82.026 & 0.000 \\
    \multicolumn{1}{l|}{\textbf{Number of accumulative publications} ($\bm{log_{10}}$)} & \textbf{-0.0374} & \textbf{0.001} & \textbf{-50.007} & \textbf{0.000}\\
    \multicolumn{1}{l|}{Reference number} & 0.0046 & 1.05E-05 & 439.216 & 0.000 \\
    \multicolumn{1}{l|}{Maximum age of all references} & -0.0003 & 1.22E-05 & -21.915 & 0.000 \\
    \multicolumn{1}{l|}{Variation ratio of ages} & 0.0024 & 0.000 & 22.841 & 0.000 \\
    \multicolumn{1}{l|}{Ratio of 1-year-old references} & 0.3315 & 0.002 & 204.050 & 0.000 \\
    \multicolumn{1}{l|}{Maximum of references' citation counts ($log_{10}$)} & 0.1695 & 0.000 & 420.271 & 0.000 \\
    \multicolumn{1}{l|}{Variation ratio of references' citation counts} & -0.0217 & 0.001 & -35.790 & 0.000 \\
    \multicolumn{1}{l|}{Ratio of non-cited references} & -0.1109 & 0.001 & -80.447 & 0.000 \\
    \midrule
    \multicolumn{1}{l|}{$R^2=0.288$, adjusted $R^2=0.288$.} & \multicolumn{4}{r}{$F=1.360E+05$, significance F: $p=0.00$.}  \\
    \bottomrule
    \end{tabular}%
  \label{tab:olsRegResGeology}%
\end{table*}%

\begin{table*}[ht]
  \centering
  \caption{OLS regression results based on papers of Political science.}
    \begin{tabular}{lrrrr}
    \toprule
    \multicolumn{1}{l|}{Variable} & \multicolumn{1}{l}{Coefficient} & \multicolumn{1}{l}{Std Err} & \multicolumn{1}{l}{t Value} & \multicolumn{1}{l}{p Value} \\
    \midrule
    \multicolumn{1}{l|}{Intercept} & 0.4766 & 0.011 & 44.403 & 0.000 \\
    \multicolumn{1}{l|}{\textbf{Number of accumulative publications} ($\bm{log_{10}}$)} & \textbf{-0.0383} & \textbf{0.001} & \textbf{-26.168} & \textbf{0.000} \\
    \multicolumn{1}{l|}{Reference number} & 0.0020 & 1.39E-05 & 140.619 & 0.000 \\
    \multicolumn{1}{l|}{Maximum age of all references} & -0.0012 & 1.59E-05 & -74.947 & 0.000 \\
    \multicolumn{1}{l|}{Variation ratio of ages} & 0.0024 & 0.000 & 12.978 & 0.000 \\
    \multicolumn{1}{l|}{Ratio of 1-year-old references} & 0.1928 & 0.002 & 92.205 & 0.000 \\
    \multicolumn{1}{l|}{Maximum of references' citation counts ($log_{10}$)} & 0.0779 & 0.001 & 130.669 & 0.000 \\
    \multicolumn{1}{l|}{Variation ratio of references' citation counts} & 0.0195 & 0.001 & 23.253 & 0.000 \\
    \multicolumn{1}{l|}{Ratio of non-cited references} & -0.1208 & 0.002 & -56.559 & 0.000 \\
    \midrule
    \multicolumn{1}{l|}{$R^2=0.079$, adjusted $R^2=0.079$.} & \multicolumn{4}{r}{$F=1.382E+04$, significance F: $p=0.00$.} \\
    \bottomrule
    \end{tabular}%
  \label{tab:olsRegResPoliticalScience}%
\end{table*}%

\begin{table*}[ht]
  \centering
  \caption{OLS regression results based on papers of Computer science.}
    \begin{tabular}{lrrrr}
    \toprule
    \multicolumn{1}{l|}{Variable} & \multicolumn{1}{l}{Coefficient} & \multicolumn{1}{l}{Std Err} & \multicolumn{1}{l}{t Value} & \multicolumn{1}{l}{p Value} \\
    \midrule
    \multicolumn{1}{l|}{Intercept} & 0.6372 & 0.004 & 154.222 & 0.000 \\
    \multicolumn{1}{l|}{\textbf{Number of accumulative publications} ($\bm{log_{10}}$)} & \textbf{-0.0441} & \textbf{0.001} & \textbf{-78.351} & \textbf{0.000} \\
    \multicolumn{1}{l|}{Reference number} & 0.0025 & 4.81E-06 & 517.196 & 0.000 \\
    \multicolumn{1}{l|}{Maximum age of all references} & -0.0006 & 9.95E-06 & -63.701 & 0.000 \\
    \multicolumn{1}{l|}{Variation ratio of ages} & 0.0085 & 0.000 & 73.089 & 0.000 \\
    \multicolumn{1}{l|}{Ratio of 1-year-old references} & 0.4017 & 0.001 & 474.500 & 0.000 \\
    \multicolumn{1}{l|}{Maximum of references' citation counts ($log_{10}$)} & 0.0768 & 0.000 & 342.102 & 0.000 \\
    \multicolumn{1}{l|}{Variation ratio of references' citation counts} & 0.0295 & 0.000 & 83.874 & 0.000 \\
    \multicolumn{1}{l|}{Ratio of non-cited references} & -0.2577 & 0.001 & -310.167 & 0.000 \\
    \midrule
    \multicolumn{1}{l|}{$R^2=0.117$, adjusted $R^2=0.117$.} & \multicolumn{4}{r}{$F=1.579E+05$, significance F: $p=0.00$.}  \\
    \bottomrule
    \end{tabular}%
  \label{tab:olsRegResComputerScience}%
\end{table*}%

\begin{table*}[ht]
  \centering
  \caption{OLS regression results based on papers of Medicine.}
    \begin{tabular}{lrrrr}
    \toprule
    \multicolumn{1}{l|}{Variable} & \multicolumn{1}{l}{Coefficient} & \multicolumn{1}{l}{Std Err} & \multicolumn{1}{l}{t Value} & \multicolumn{1}{l}{p Value} \\
    \midrule
    \multicolumn{1}{l|}{Intercept} & 0.6390 & 0.003 & 204.527 & 0.000 \\
    \multicolumn{1}{l|}{\textbf{Number of accumulative publications} ($\bm{log_{10}}$)} & \textbf{-0.0502} & \textbf{0.000}     & \textbf{-116.254} & \textbf{0.000} \\
    \multicolumn{1}{l|}{Reference number} & 0.0026 & 4.41E-06 & 598.096 & 0.000 \\
    \multicolumn{1}{l|}{Maximum age of all references} & -0.0011 & 7.73E-06 & -137.735 & 0.000 \\
    \multicolumn{1}{l|}{Variation ratio of ages} & 0.0014 & 5.05E-05 & 28.078 & 0.000 \\
    \multicolumn{1}{l|}{Ratio of 1-year-old references} & 0.3038 & 0.001 & 314.448 & 0.000 \\
    \multicolumn{1}{l|}{Maximum of references' citation counts ($log_{10}$)} & 0.1723 & 0.000 & 792.835 & 0.000 \\
    \multicolumn{1}{l|}{Variation ratio of references' citation counts} & -0.0147 & 0.000     & -38.817 & 0.000 \\
    \multicolumn{1}{l|}{Ratio of non-cited references} & -0.3260 & 0.001 & -299.930 & 0.000 \\
    \midrule
    \multicolumn{1}{l|}{$R^2=0.158$, adjusted $R^2=0.158$.} & \multicolumn{4}{r}{$F=3.083E+05$, significance F: $p=0.00$.}  \\
    \bottomrule
    \end{tabular}%
  \label{tab:olsRegResMedicine}%
\end{table*}%

\begin{table*}[ht]
  \centering
  \caption{OLS regression results based on papers of Geography.}
    \begin{tabular}{lrrrr}
    \toprule
    \multicolumn{1}{l|}{Variable} & \multicolumn{1}{l}{Coefficient} & \multicolumn{1}{l}{Std Err} & \multicolumn{1}{l}{t Value} & \multicolumn{1}{l}{p Value} \\
    \midrule
    \multicolumn{1}{l|}{Intercept} & 0.6695 & 0.009 & 71.026 & 0.000 \\
    \multicolumn{1}{l|}{\textbf{Number of accumulative publications} ($\bm{log_{10}}$)} & \textbf{-0.0592} & \textbf{0.001} & \textbf{-45.600} & \textbf{0.000} \\
    \multicolumn{1}{l|}{Reference number} & 0.0039 & 1.44E-05 & 268.849 & 0.000 \\
    \multicolumn{1}{l|}{Maximum age of all references} & -0.0008 & 1.61E-05 & -51.569 & 0.000 \\
    \multicolumn{1}{l|}{Variation ratio of ages} & 0.0024 & 0.000 & 15.782 & 0.000 \\
    \multicolumn{1}{l|}{Ratio of 1-year-old references} & 0.2669 & 0.002 & 118.332 & 0.000 \\
    \multicolumn{1}{l|}{Maximum of references' citation counts ($log_{10}$)} & 0.0962 & 0.001 & 157.641 & 0.000 \\
    \multicolumn{1}{l|}{Variation ratio of references' citation counts} & 0.0024 & 0.001 & 2.694 & 0.007 \\
    \multicolumn{1}{l|}{Ratio of non-cited references} & -0.2063 & 0.002 & -93.010 & 0.000 \\
    \midrule
    \multicolumn{1}{l|}{$R^2=0.134$, adjusted $R^2=0.134$.} & \multicolumn{4}{r}{$F=2.927E+04$, significance F: $p=0.00$.}  \\
    \bottomrule
    \end{tabular}%
  \label{tab:olsRegResGeography}%
\end{table*}%

\begin{table*}[ht]
  \centering
  \caption{OLS regression results based on papers of Art.}
    \begin{tabular}{lrrrr}
    \toprule
    \multicolumn{1}{l|}{Variable} & \multicolumn{1}{l}{Coefficient} & \multicolumn{1}{l}{Std Err} & \multicolumn{1}{l}{t Value} & \multicolumn{1}{l}{p Value} \\
    \midrule
    \multicolumn{1}{l|}{Intercept} & 0.5487 & 0.008 & 69.018 & 0.000 \\
    \multicolumn{1}{l|}{\textbf{Number of accumulative publications} ($\bm{log_{10}}$)} & \textbf{-0.0619} & \textbf{0.001} & \textbf{-56.798} & \textbf{0.000} \\
    \multicolumn{1}{l|}{Reference number} & 0.0013 & 1.34E-05 & 98.737 & 0.000 \\
    \multicolumn{1}{l|}{Maximum age of all references} & -0.0005 & 1.08E-05 & -47.349 & 0.000 \\
    \multicolumn{1}{l|}{Variation ratio of ages} & 0.0010 & 0.000 & 6.678 & 0.000 \\
    \multicolumn{1}{l|}{Ratio of 1-year-old references} & 0.1081 & 0.002 & 55.773 & 0.000 \\
    \multicolumn{1}{l|}{Maximum of references' citation counts ($log_{10}$)} & 0.0607 & 0.001 & 113.794 & 0.000 \\
    \multicolumn{1}{l|}{Variation ratio of references' citation counts} & 0.0114 & 0.001 & 15.379 & 0.000 \\
    \multicolumn{1}{l|}{Ratio of non-cited references} & -0.0545 & 0.002 & -32.769 & 0.000 \\
    \midrule
    \multicolumn{1}{l|}{$R^2=0.072$, adjusted $R^2=0.072$.} & \multicolumn{4}{r}{$F=7767.0$, significance F: $p=0.00$.}  \\
    \bottomrule
    \end{tabular}%
  \label{tab:olsRegResArt}%
\end{table*}%

\begin{table*}[ht]
  \centering
  \caption{OLS regression results based on papers of Population.}
    \begin{tabular}{lrrrr}
    \toprule
    \multicolumn{1}{l|}{Variable} & \multicolumn{1}{l}{Coefficient} & \multicolumn{1}{l}{Std Err} & \multicolumn{1}{l}{t Value} & \multicolumn{1}{l}{p Value} \\
    \midrule
    \multicolumn{1}{l|}{Intercept} & 0.9279 & 0.013 & 73.252 & 0.000 \\
    \multicolumn{1}{l|}{\textbf{Number of accumulative publications} ($\bm{log_{10}}$)} & \textbf{-0.0657} & \textbf{0.002} & \textbf{-37.441} & \textbf{0.000} \\
    \multicolumn{1}{l|}{Reference number} & 0.0039 & 1.78E-05 & 221.989 & 0.000 \\
    \multicolumn{1}{l|}{Maximum age of all references} & -0.0015 & 2.41E-05 & -60.212 & 0.000 \\
    \multicolumn{1}{l|}{Variation ratio of ages} & 0.0018 & 0.000 & 8.662 & 0.000 \\
    \multicolumn{1}{l|}{Ratio of 1-year-old references} & 0.5624 & 0.004 & 142.396 & 0.000 \\
    \multicolumn{1}{l|}{Maximum of references' citation counts ($log_{10}$)} & 0.1359 & 0.001 & 169.872 & 0.000\\
    \multicolumn{1}{l|}{Variation ratio of references' citation counts} & -0.0544 & 0.001 & -41.611 & 0.000 \\
    \multicolumn{1}{l|}{Ratio of non-cited references} & -0.5401 & 0.004 & -123.805 & 0.000 \\
    \midrule
    \multicolumn{1}{l|}{$R^2=0.146$, adjusted $R^2=0.146$.} & \multicolumn{4}{r}{$F=2.349E+04$, significance F: $p=0.00$.} \\
    \bottomrule
    \end{tabular}%
  \label{tab:olsRegResPopulation}%
\end{table*}%

\begin{table*}[ht]
  \centering
  \caption{OLS regression results based on papers of Economics.}
    \begin{tabular}{lrrrr}
    \toprule
    \multicolumn{1}{l|}{Variable} & \multicolumn{1}{l}{Coefficient} & \multicolumn{1}{l}{Std Err} & \multicolumn{1}{l}{t Value} & \multicolumn{1}{l}{p Value} \\
    \midrule
    \multicolumn{1}{l|}{Intercept} & 0.8813 & 0.008 & 107.592 & 0.000 \\
    \multicolumn{1}{l|}{\textbf{Number of accumulative publications} ($\bm{log_{10}}$)} & \textbf{-0.0860} & \textbf{0.001} & \textbf{-76.255} & \textbf{0.000} \\
    \multicolumn{1}{l|}{Reference number} & 0.0040 & 1.16E-05 & 345.906 & 0.000 \\
    \multicolumn{1}{l|}{Maximum age of all references} & -0.0012 & 1.40E-05 & -86.340 & 0.000 \\
    \multicolumn{1}{l|}{Variation ratio of ages} & 0.0024 & 0.000 & 18.963 & 0.000 \\
    \multicolumn{1}{l|}{Ratio of 1-year-old references} & 0.3095 & 0.002 & 170.853 & 0.000 \\
    \multicolumn{1}{l|}{Maximum of references' citation counts ($log_{10}$)} & 0.0807 & 0.000 & 166.750 & 0.000 \\
    \multicolumn{1}{l|}{Variation ratio of references' citation counts} & 0.0157 & 0.001 & 22.641 & 0.000 \\
    \multicolumn{1}{l|}{Ratio of non-cited references} & -0.2076 & 0.002 & -108.533 & 0.000 \\
    \midrule
    \multicolumn{1}{l|}{$R^2=0.106$, adjusted $R^2=0.106$.} & \multicolumn{4}{r}{$F=4.057E+04$, significance F: $p=0.00$.}  \\
    \bottomrule
    \end{tabular}%
  \label{tab:olsRegResEconomics}%
\end{table*}%

\begin{table*}[ht]
  \centering
  \caption{OLS regression results based on papers of Sociology.}
    \begin{tabular}{lrrrr}
    \toprule
    \multicolumn{1}{l|}{Variable} & \multicolumn{1}{l}{Coefficient} & \multicolumn{1}{l}{Std Err} & \multicolumn{1}{l}{t Value} & \multicolumn{1}{l}{p Value} \\
    \midrule
    \multicolumn{1}{l|}{Intercept} & 0.8839 & 0.009 & 101.302 & 0.000 \\
    \multicolumn{1}{l|}{\textbf{Number of accumulative publications} ($\bm{log_{10}}$)} & \textbf{-0.0891} & \textbf{0.001} & \textbf{-74.090} & \textbf{0.000} \\
    \multicolumn{1}{l|}{Reference number} & 0.0030 & 1.23E-05 & 246.832 & 0.000 \\
    \multicolumn{1}{l|}{Maximum age of all references} & -0.0017 & 1.30E-05 & -129.786 & 0.000 \\
    \multicolumn{1}{l|}{Variation ratio of ages} & 0.0022 & 0.000 & 15.603 & 0.000 \\
    \multicolumn{1}{l|}{Ratio of 1-year-old references} & 0.2559 & 0.002 & 116.238 & 0.000 \\
    \multicolumn{1}{l|}{Maximum of references' citation counts ($log_{10}$)} & 0.1187 & 0.001 & 220.712 & 0.000 \\
    \multicolumn{1}{l|}{Variation ratio of references' citation counts} & -0.0174 & 0.001 & -22.423 & 0.000 \\
    \multicolumn{1}{l|}{Ratio of non-cited references} & -0.2010 & 0.002 & -92.973 & 0.000 \\
    \midrule
    \multicolumn{1}{l|}{$R^2=0.117$, adjusted $R^2=0.117$.} & \multicolumn{4}{r}{$F=3.301E+04$, significance F: $p=0.00$.}  \\
    \bottomrule
    \end{tabular}%
  \label{tab:olsRegResSociology}%
\end{table*}%

\begin{table*}[ht]
  \centering
  \caption{OLS regression results based on papers of Philosophy.}
    \begin{tabular}{lrrrr}
    \toprule
    \multicolumn{1}{l|}{Variable} & \multicolumn{1}{l}{Coefficient} & \multicolumn{1}{l}{Std Err} & \multicolumn{1}{l}{t Value} & \multicolumn{1}{l}{p Value} \\
    \midrule
    \multicolumn{1}{l|}{Intercept} & 0.7884 & 0.009 & 88.471 & 0.000 \\
    \multicolumn{1}{l|}{\textbf{Number of accumulative publications} ($\bm{log_{10}}$)} & \textbf{-0.0959} & \textbf{0.001} & \textbf{-78.273} & \textbf{0.000} \\
    \multicolumn{1}{l|}{Reference number} & 0.0027 & 1.43E-05 & 186.489 & 0.000 \\
    \multicolumn{1}{l|}{Maximum age of all references} & -0.0009 & 1.27E-05 & -72.229 & 0.000 \\
    \multicolumn{1}{l|}{Variation ratio of ages} & 0.0014 & 0.000 & 8.721 & 0.000 \\
    \multicolumn{1}{l|}{Ratio of 1-year-old references} & 0.1703 & 0.002 & 74.522 & 0.000 \\
    \multicolumn{1}{l|}{Maximum of references' citation counts ($log_{10}$)} & 0.1033 & 0.001 & 171.390 & 0.000 \\
    \multicolumn{1}{l|}{Variation ratio of references' citation counts} & 0.0028 & 0.001 & 3.197 & 0.001 \\
    \multicolumn{1}{l|}{Ratio of non-cited references} & -0.0764 & 0.002 & -35.399 & 0.000 \\
    \midrule
    \multicolumn{1}{l|}{$R^2=0.128$, adjusted $R^2=0.128$.} & \multicolumn{4}{r}{$F=1.940E+04$, significance F: $p=0.00$.}  \\
    \bottomrule
    \end{tabular}%
  \label{tab:olsRegResPhilosophy}%
\end{table*}%

\begin{table*}[ht]
  \centering
  \caption{OLS regression results based on papers of Psychology.}
    \begin{tabular}{lrrrr}
    \toprule
    \multicolumn{1}{l|}{Variable} & \multicolumn{1}{l}{Coefficient} & \multicolumn{1}{l}{Std Err} & \multicolumn{1}{l}{t Value} & \multicolumn{1}{l}{p Value} \\
    \midrule
    \multicolumn{1}{l|}{Intercept} & 1.1415 & 0.006 & 186.616 & 0.000 \\
    \multicolumn{1}{l|}{\textbf{Number of accumulative publications} ($\bm{log_{10}}$)} & \textbf{-0.1245} & \textbf{0.001} & \textbf{-144.398} & \textbf{0.000} \\
    \multicolumn{1}{l|}{Reference number} & 0.0030 & 7.79E-06 & 386.534 & 0.000 \\
    \multicolumn{1}{l|}{Maximum age of all references} & -0.0015 & 1.29E-05 & -117.833 & 0.000 \\
    \multicolumn{1}{l|}{Variation ratio of ages} & 0.0020 & 0.000 & 17.800  & 0.000 \\
    \multicolumn{1}{l|}{Ratio of 1-year-old references} & 0.3750 & 0.002 & 182.681 & 0.000 \\
    \multicolumn{1}{l|}{Maximum of references' citation counts ($log_{10}$)} & 0.1714 & 0.000 & 385.446 & 0.000 \\
    \multicolumn{1}{l|}{Variation ratio of references' citation counts} & -0.0603 & 0.001 & -87.833 & 0.000 \\
    \multicolumn{1}{l|}{Ratio of non-cited references} & -0.3288 & 0.002 & -151.060 & 0.000 \\
    \midrule
    \multicolumn{1}{l|}{$R^2=0.155$, adjusted $R^2=0.155$.} & \multicolumn{4}{r}{$F=8.334E+04$, significance F: $p=0.00$.} \\
    \bottomrule
    \end{tabular}%
  \label{tab:olsRegResPsychology}%
\end{table*}%

\begin{table*}[ht] 
  \centering
  \caption{OLS regression results based on papers of Business.}
    \begin{tabular}{lrrrr}
    \toprule
    \multicolumn{1}{l|}{Variable} & \multicolumn{1}{l}{Coefficient} & \multicolumn{1}{l}{Std Err} & \multicolumn{1}{l}{t Value} & \multicolumn{1}{l}{p Value} \\
    \midrule
    \multicolumn{1}{l|}{Intercept} & 1.3884 & 0.012 & 111.895 & 0.000 \\
    \multicolumn{1}{l|}{\textbf{Number of accumulative publications} ($\bm{log_{10}}$)} & \textbf{-0.1514} & \textbf{0.002} & \textbf{-89.751} & \textbf{0.000} \\
    \multicolumn{1}{l|}{Reference number} & 0.0041 & 1.46E-05 & 283.576 & 0.000 \\
    \multicolumn{1}{l|}{Maximum age of all references} & -0.0011 & 2.28E-05 & -47.260 & 0.000 \\
    \multicolumn{1}{l|}{Variation ratio of ages} & 0.0044 & 0.000 & 20.511 & 0.000 \\
    \multicolumn{1}{l|}{Ratio of 1-year-old references} & 0.2930 & 0.002 & 128.569 & 0.000 \\
    \multicolumn{1}{l|}{Maximum of references' citation counts ($log_{10}$)} & 0.0618 & 0.001 & 98.849 & 0.000 \\
    \multicolumn{1}{l|}{Variation ratio of references' citation counts} & 0.0324 & 0.001 & 36.305 & 0.000 \\
    \multicolumn{1}{l|}{Ratio of non-cited references} & -0.2342 & 0.002 & -98.783 & 0.000 \\
    \midrule
    \multicolumn{1}{l|}{$R^2=0.114$, adjusted $R^2=0.114$.} & \multicolumn{4}{r}{$F=2.509E+04$, significance F: $p=0.00$.} \\
    \bottomrule
    \end{tabular}%
  \label{tab:olsRegResBusiness}%
\end{table*}%

\end{document}